\documentclass[journal,twoside]{IEEEtran}
\usepackage{multirow}
\usepackage[driverfallback=dvipdfm]{hyperref}
\usepackage[usenames,dvipsnames]{pstricks}

\usepackage{amsmath,amssymb,amsfonts,bm}
\usepackage{algorithmic}
\usepackage{algorithm}

\usepackage{graphicx}
\usepackage{textcomp}

\usepackage{multirow}
\graphicspath{{figures/}}
\usepackage[usenames,dvipsnames]{pstricks}
\usepackage{colortbl}

\newcommand{\sign}{\ensuremath{\text{\rm sign}}}

\newcommand{\subfigimg}[3][,]{%
  \setbox1=\hbox{\includegraphics[#1]{#3}}
  \leavevmode\rlap{\usebox1}
  \rlap{\hspace*{1pt}\raisebox{\dimexpr\ht1-0.6\baselineskip}{#2}}
  \phantom{\usebox1}
}
\definecolor{gbypink}{rgb}{0.98, 0.65, 0.8} 
\definecolor{gbyblue}{rgb}{0.0, 0.78, 0.98} 
\definecolor{gbygreen}{rgb}{0.67, 0.88, 0.69}

\newcommand{\myfontNew}{\fontsize{9pt}{\baselineskip}\selectfont}

\begin{document}
\title{MetaInv-Net: Meta Inversion Network for Sparse View CT Image Reconstruction}
\author{Haimiao Zhang, Baodong Liu, Hengyong Yu, \IEEEmembership{Senior Member, IEEE}, and Bin Dong
\thanks{
This work was supported 
in part by the China Postdoctoral Science Foundation under Grant 2018M641056, 
in part by the National Key R\&D Program of China under Grant 2017YFF0107201, 
in part by the CAS Interdisciplinary Innovation Team under Project No. JCTD-2019-02,
in part by the National Natural Science Foundation of China (NSFC) under Grant 11831002, 
in part by the Beijing Natural Science Foundation under Grant 180001, 
and in part by the Beijing Academy of Artificial Intelligence (BAAI).
(Corresponding authors: Baodong Liu; Bin Dong.)
}
\thanks{
H. Zhang and B. Dong are with the Beijing International Center for Mathematical Research, Peking University, Beijing, 100871, P. R. China (e-mail: hmzhang@pku.edu.cn; dongbin@math.pku.edu.cn). }
\thanks{B. Liu is with the Beijing Engineering Research Center of Radiographic Techniques and Equipment, Institute of High Energy Physics, Chinese Academy of Sciences, Beijing 100049, China. School of Nuclear Science and Technology, University of Chinese Academy of Sciences, Beijing 100049, China (e-mail: liubd@ihep.ac.cn). }
\thanks{H. Yu is with the Department of Electrical and Computer Engineering, University of Massachusetts Lowell, Lowell, MA 01854, USA (e-mail: hengyong-yu@ieee.org).}
}
\maketitle

\begin{abstract}
X-ray Computed Tomography (CT) is widely used in clinical applications such as diagnosis and image-guided interventions. In this paper, we propose a new deep learning based model for CT image reconstruction with the backbone network architecture built by unrolling an iterative algorithm. However, unlike the existing strategy to include as many data-adaptive components in the unrolled dynamics model as possible, we find that it is enough to only learn the parts where traditional designs mostly rely on intuitions and experience. More specifically, we propose to learn an initializer for the conjugate gradient (CG) algorithm that involved in one of the subproblems of the backbone model. Other components, such as image priors and hyperparameters, are kept as the original design. Since a hypernetwork is introduced to inference on the initialization of the CG module, it makes the proposed model a certain meta-learning model. Therefore, we shall call the proposed model the meta-inversion network (MetaInv-Net). The proposed MetaInv-Net can be designed with much less trainable parameters while still preserves its superior image reconstruction performance than some state-of-the-art deep models in CT imaging. In simulated and real data experiments, MetaInv-Net performs very well and can be generalized beyond the training setting, i.e., to other scanning settings, noise levels, and data sets.
\end{abstract}


\section{Introduction}
\IEEEPARstart{C}{omputed} tomography (CT) is one of the most important diagnostic imaging techniques. In clinical applications, sparse view CT is adopted to decrease the radiation dose and reduce the scanning time. However, this inevitably leads to ill-posed inverse problems \cite{davison1983the,louis1986incomplete} and gives rise to numerous new and exciting image reconstruction models and algorithms. 

\subsection{Trends of CT Models and Algorithms}
Monochromatic energy CT imaging can be formulated as the following linear inverse problem
$$ \bm{Y}=\bm{P}\bm{u}+\bm{\eta},$$
where $\bm{Y}$ is the measured sinogram (projection data) after correction and log transform, $\bm{u}$ is the unknown image, and $\bm{P}$ is an operator that models the imaging system. In the 2D continuum case, $\bm{P}$ is equivalent to the Radon transform for parallel beam imaging geometry. The parameter $\bm{\eta}$ is the measurement error or additive noise (e.g., electronic noise).  In incomplete data CT (e.g., sparse view CT, limited-angle CT, and interior/exterior CT), the dominated noise in $\bm{Y}$ is the Poisson noise and the operator $\bm{P}$ is noninvertable. Thus, the above inverse problem is highly ill-posed. 

Classical CT image reconstruction methods include the well-known Filtered Backprojection (FBP), algebraic reconstruction technique (ART), etc. \cite{natterer2001mathematics}. However, these methods are relatively sensitive to noise $\bm{\eta}$. This is a manifestation of the ill-posed nature of the problem. To solve this issue, regularization based models have been widely adopted for the past few decades. Typical regularization based models take the following form \cite{aubert2006mathematical}
\begin{equation}\label{l2+g}
\min_{\bm{u}} \frac{1}{2}\|\bm{P}\bm{u}-\bm{Y}\|^{2}+\lambda R(\bm{u}),
\end{equation}
where the first term is the discrepancy between the estimated $\bm{u}$ and measured data $\bm{Y}$, and $R(\bm{u})$ is the regularization term that incorporates our prior knowledge on the image to be reconstructed. 
Successful examples for CT image reconstruction models include 1) piecewise constant/smoothness constrained models, i.e., total variation (TV) based models \cite{sidky2006accurate,sidky2008image,yang2010high,mahmood2018adaptive}, 2) nonlocal/patch similarity based models \cite{zeng2015spectral}, 3) applied harmonic analysis models, i.e., wavelets and wavelet frame models \cite{rantala2006wavelet,Dong2013,choi2016limited,zhang2018reweighted} and convolutional sparse coding \cite{bao2019convolutional}, 4) low-rank models \cite{gao2011robust,kim2014sparse,semerci2014tensor}, 5) dictionary learning models \cite{zhou2013adaptive,zhao2012dual,tai2016multiscale,zhan2016ct,zhang2016tensor,bai2017z,dong2018joint,kong2020spectral}, etc.

Recent years, the rapid development of machine learning, especially deep learning, has lead to a paradigm shift of modeling and algorithmic design in computer vision and medical imaging \cite{deep-learning-book-2016,wang2016perspective,wang2017machine,Unser2017review,wang2018image,monga2019algorithm,zhang2020review,zhou2020review}. Deep learning based models (or deep models for short) are able to leverage large image datasets to learn better image representations and produce better image reconstruction results than traditional methods \cite{Jin2017FBPConvNet,chen2017low,kang2017deep,zhang2018sparse,zhu2018image,yang2018low,yoo2020deepDOT}. 
Some earlier work on deep models for CT imaging, i.e., FBPConv-Net, focused on post-processing where the FBP reconstructed image is processed by the deep neural network to remove artifacts induced by low tube current and incomplete projections \cite{Jin2017FBPConvNet}. A residual encoder-decoder CNN is adopted in \cite{chen2017low,zhang2018sparse} to learn the map between the degraded and clean image. Authors in \cite{yang2018low} proposed to use the generative adversarial network with Wasserstein distance and perceptual loss to restore the subtle structures from the FBP reconstructed CT image. To improve the intepretability and reliability of deep models, a recent trend of deep modeling is to combine deep learning with traditional iterative reconstruction methods. This emerging new approach is often known as the unrolled dynamics (UD) approach, which firstly unrolls an iterative algorithm (e.g. an optimization algorithm associated to a regularization model or a partial differential equation) to form the backbone network architecture and then replaces parts of its key ingredients, such as image representations, model parameters, matrix inversions, proximal operators, etc., by deep (convolutional) neural networks. The UD approach started with the seminal work of \cite{gregor2010learning} and has lead to many exciting developments in signal/image processing \cite{SCN-SR-2015,schuler2016learning,zhang2017learning,zhang2018ista-net,dong2019denoising,li2020efficient} and medical imaging \cite{yang2016deep,yang2017admm,mardani2018neural,gupta2018cnn,adler2018learned,zhang2019JSRNet,solomon2020UnfoldRPCA,yang2020admm-csnet}.

In \cite{gregor2010learning}, the authors proposed to unroll the iterative soft-thresholding algorithm (ISTA) \cite{daubechies2004iterative} to improve model efficiency for the sparse coding problem. Later, many other popular iterative algorithms were unrolled to generate deep models for various ill-posed inverse problems in medical image reconstruction, such as the UD model from the primal-dual hybrid gradient (PDHG) algorithm \cite{adler2018learned,Adler2017Solving}, the algorithm of alternating direction method of multipliers (ADMM) \cite{yang2016deep,yang2017admm,yang2020admm-csnet,he2019optimizing}, the projected/proximal gradient descent (PGD) \cite{gupta2018cnn,Meinhardt_2017learn_prox_op,ding2020low}, etc. Now, we review these UD methods in a collective fashion. 
Consider a general object functional as
$$F_{\lambda,\bm{\gamma}}(\bm{u},\bm{z};\bm{\beta})=\mathcal{D}(\bm{u},\bm{Y})+\bm{\gamma}\mathcal{C}(\bm{u},\bm{z},\bm{\beta})+\lambda \mathcal{R}_{\bm{W}}(\bm{z}),$$
where $ \mathcal{D}(\bm{u},\bm{Y})$ is the data discrepancy term, $\mathcal{C}(\bm{u},\bm{z},\bm{\beta})$ is a constraint term that links the primal variable $\bm{u}$, the auxiliary variable $\bm{z}$ and the dual variable $\bm{\beta}$, and $\mathcal{R}_{\bm{W}}(\bm{z})$ is the regularization term with variable $\bm{z}$ and a certain sparsifying transformation $\bm{W}$ which represents the image prior used by the model. Then, the aforementioned deep UD models use the following iterative algorithm (at iteration $k+1$) as their backbone network architecture

\begin{align}\label{eq:abstract-BCD}
\bm{u}^{k+1}&=\arg\min_{\bm{u}}F_{\lambda,\bm{\gamma}}(\bm{u},\bm{z}^{k};\bm{\beta}^{k}),\notag\\
\bm{z}^{k+1}&=\arg\min_{\bm{z}}F_{\lambda,\bm{\gamma}}(\bm{u}^{k+1},\bm{z};\bm{\beta}^{k}),\notag\\
\bm{\beta}^{k+1}&= \arg\max_{\bm{\beta}}F_{\lambda,\bm{\gamma}}(\bm{u}^{k+1},\bm{z}^{k+1};\bm{\beta}).
\end{align} 
In these deep UD models, the $\bm{u}$-subproblem is the image reconstruction layer, the $\bm{z}$-subproblem is the image denoising module, and $\lambda$ and $\bm{\gamma}$ are either hyperparameters or trainable parameters. The update of $\bm{\beta}^{k+1}$ is commonly a residual term. Different deep UD model has its own unique design on the $\bm{u}$- and $\bm{z}$-layer, and makes its own choice on handcrafting and learning in every module of the model. Next, we shall provide some details on the design of a few deep UD models. A summary is given in Table \ref{table:diff-UDs-models}.

\begin{table*}[t]
  \myfontNew
  \caption{ The learned issue of different UD models.  The existence of learnable parameters is indicated by Yes or No within the bracket.
  }  \label{table:diff-UDs-models}
\centering
  \begin{tabular}{|c|c|c|c|c|}
  \hline
  \multirow{2}{*}{ Models } & \multicolumn{3}{c|}{Learned Issue}\\
\cline{2-4}
 & \multicolumn{1}{c|}{$\bm{u}$-Layer (Param.) } & \multicolumn{1}{c|}{$\bm{z}$-Layer (Param.)  } & Image Prior ($\bm{W}$) \\
  \cline{1-4}
  DUBLID\cite{li2020efficient}, ADMM-Net\cite{yang2016deep} & Analytic formula (Yes) & Soft-threshold(Yes) & Learned  \\
  \cline{1-4}
  DnCNN\cite{zhang2017learning}, ADMM-CSNet\cite{yang2020admm-csnet} & Analytic formula (Yes) & CNN (Yes) & Learned \\
  \cline{1-4}
  ISTA-Net\cite{zhang2018ista-net}, DPDNN\cite{dong2019denoising} & Gradient descent (Yes) & CNN (Yes)  &Learned \\
  \cline{1-4}
  PD-Net\cite{adler2018learned} & CNN (Yes) & CNN (Yes) & Learned \\
  \cline{1-4}
  MetaInv-Net (Alg.\ref{alg:meta-HQS}) & CG with learned initialization (Yes) & Soft-threshold (No) &Handcrafted \\
  \cline{1-4}
\hline
\end{tabular}
\end{table*}

In DUBLID\cite{li2020efficient} and ADMM-Net \cite{yang2016deep}, $\bm{u}$-layer was a handcrafted analytic inversion formula with trainable parameters. For $\bm{z}$-layer, a soft-threshold operation was implemented with learnable parameters. The sparsifying transformation $\bm{W}$ was learned from the dataset in an end-to-end manner. 

In DnCNN \cite{zhang2017learning} and ADMM-CSNet \cite{yang2020admm-csnet}, the $\bm{u}$-layer was solved by a handcrafted analytic formula with trainable parameters. For $\bm{z}$-layer, a convolutional neural network was implemented along with the learnable sparsifying transformation $\bm{W}$ as an image denoiser.

In ISTA-Net \cite{zhang2018ista-net} and DPDNN  \cite{dong2019denoising}, $\bm{u}$-layer was implemented by a one-step gradient descent with trainable step size. For $\bm{z}$-layer, convolutional neural network was implemented along with the learnable sparsifying transformation $\bm{W}$ as a image denoiser.

In PD-Net \cite{adler2018learned}, the authors replaced the proximal operators in both $\bm{u}$-layer and $\bm{z}$-layer by multilayer convolutional neural networks. Model parameters and sparsifying transformation $\bm{W}$ were learned from the dataset. 

\subsection{Motivation}
The general strategy of designing deep UD models is to first select an appropriate iterative algorithm as the backbone network architecture and then decide which components of the UD need to be data-aware. As arbitrary as the design of deep UD models may sound, our general rule of thumb is that it is unnecessary to learn the knowledge or principles that are certain to play an important role; instead, we should focus on learning the parts of the unrolled dynamics where traditional designs mostly rely on oversimplification (e.g., images being piecewise constant), inaccurate intuitions (e.g., convenient initialization of the $\bm{u}$-subproblem of \eqref{eq:abstract-BCD}) or trails-and-errors (e.g., hyperparameter selection). This is partially supported by \cite{liu2019alista,chen2018theoretical}, where the authors showed that we do not need to rely too much on learning to achieve an asymptotic linear convergence rate of the learned ISTA. However, the choice of the trainable components in the unrolled dynamical model heavily depends on the underlying task. For example, for image reconstruction, the quality of the reconstructed image is more important than linear convergence rate of the underlying algorithm. Thus, merely learning the step size of the optimization algorithm as in \cite{liu2019alista,chen2018theoretical} may not be enough nor necessary for image reconstruction. The empirical results in Section \ref{sec:numerical-simul} show that the initialization of the conjugate gradient (CG) algorithm in the $\bm{u}$-subproblem of the UD model is important and needs to be learned from the dataset for better generalization of the whole UD model.

As can be seen from \eqref{eq:abstract-BCD}, the backbone architecture of unrolled dynamics can be obtained from most popular optimization algorithms such as the half-quadratic splitting (HQS) algorithm \cite{geman1995HQS}, ADMM, PGD, and PDHG, etc. The multiplier/auxiliary variable update layer $\beta^{k}$ is naively treated in the unrolled dynamics. The $\bm{u}$-subproblem is often a least square problem that can be solved by CG algorithm as adopted by ADMM and HQS, while PDHG and PGD use one step of gradient descent instead. The $\bm{z}$-subproblem can be expressed as a proximal operator associated to the regularization $\mathcal{R}_{\bm{W}}(z)$.

In this paper, we propose to unroll the HQS algorithm to form the backbone network. We shall call the HQS algorithm with the $\bm{u}$-subproblem solved by CG as HQS-CG. At iteration $k$, the common practice is to initialize the CG algorithm (with  $L$ iterations) using $\bm{u}^{k-1,L}$, the approximated solution $\bm{u}$ from the iteration $k-1$. Ideally, if $\bm{u}^{k-1,L}$ gets close to the fixed point of the HQS-CG algorithm quickly, the entire algorithm will terminate quickly and output a high-quality CT image. However, it is hard to design a good initialization for the CG step, since it can be as difficult as finding the solution of the inverse problem itself. In this work, we adopt a neural network to predict the initial value at iteration $k$ as
$$\bm{u}^{k,0}=\mathcal{N}(\bm{u}^{k-1,L};\Theta^k),$$
where $\Theta^{k}$ collects the trainable parameters. 
For the $\bm{z}$-subproblem (i.e., the image denoising layer), we observe that the sparsifying transformation $\bm{W}$ can be fixed. In the numerical experiment, we choose $\bm{W}$ as the highpass components of the piecewise linear tight wavelet frame transform. 

Unlike most of the existing deep CT image reconstruction models, we only use a trainable CNN to infer an initialization for CG algorithm while keeping all the other components of the HQS-CG algorithm unchanged. Other than the initialization for each CG step, the hyperparameters $\lambda$ and $\bm{\gamma}$ are also important and difficult to tune in practice. However, as shown in our numerical experiments that using the neural network to approximate $\lambda$ and $\bm{\gamma}$ along with $\bm{u}^{k,0}$ does not bring noticeable overall benefit, whereas it inevitably introduces more trainable parameters to the proposed model. Therefore, we shall manually fix $\lambda$ and $\bm{\gamma}$ in all of our experiments without tuning them for each test case.   

To utilize the intermediate reconstruction results in the iterative reconstruction procedure, the graph total variation (GTV) based CT image reconstruction model in \cite{mahmood2018adaptive} proposed to learn the "non-local patch similarity" image prior from the intermediate reconstructions. In contrast, the proposed model in this work is to learn the initialization of the $\bm{u}$-subproblem from the intermediate reconstructions rather than to estimate an optimal image prior.

\begin{figure}[t]
\centering
\includegraphics[scale=0.35]{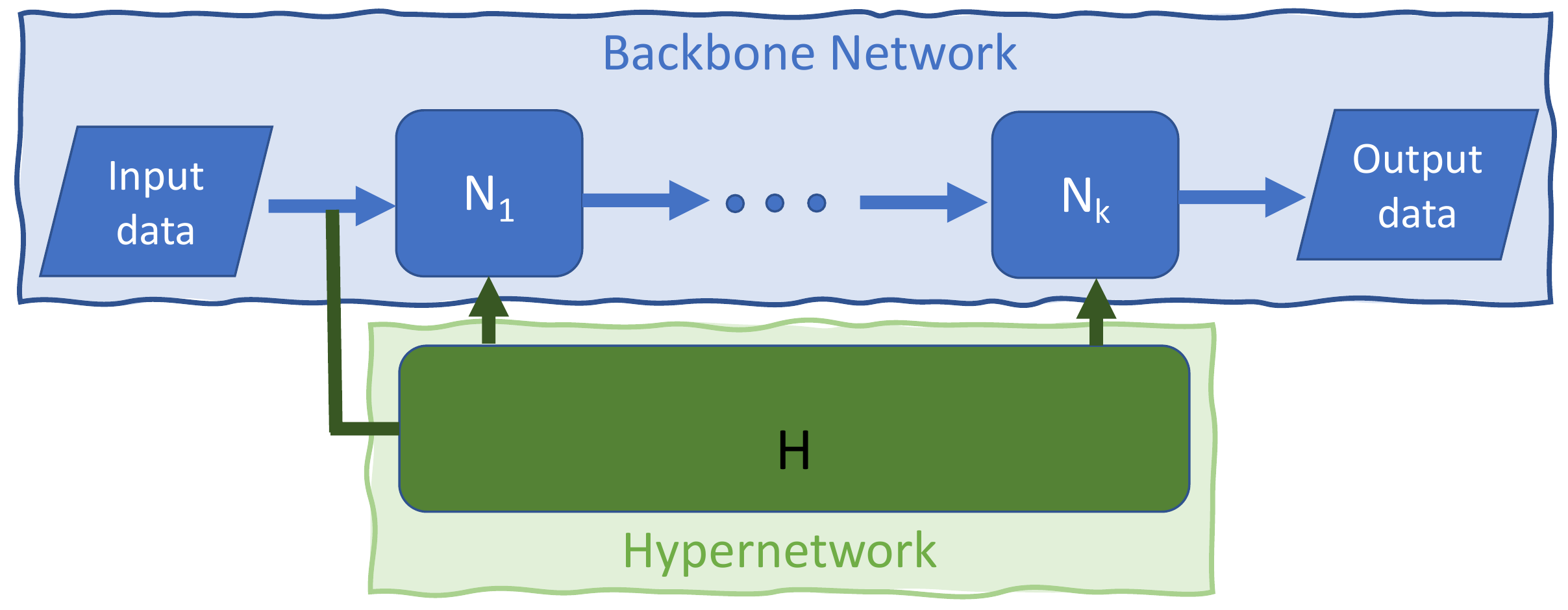}\label{HyperNetFlow}
\caption{A standard hypernetwork.}
\label{fig:HyperNetFlow}
\end{figure}

We would argue that the proposed deep UD model based on HQS-CG can be interpreted as a certain meta-learning model. Meta-learning is a branch of machine learning that exploits intrinsic common knowledge between different tasks to effectively solve new tasks. Meta-learning can be applied for hyperparameter initialization \cite{nichol2018first,finn2017model}, multitask learning \cite{Meyerson2019NIPS-MTL}, weight pruning \cite{liu2019metapruning}, neural architecture search \cite{brock2018smash,zhang2019graph}, Bayesian neural networks \cite{pawlowski2017implicit}, and hyperparameter optimization \cite{lorraine2018stochastic}. A specific form of meta-learning models adopts a hypernetwork to predict hyperparameters or weights in the backbone neural network \cite{ha2016hypernetworks}. Fig. \ref{fig:HyperNetFlow} shows a common architecture of the hypernetwork based deep model. If we write the image reconstruction model as
$$T(\bm{Y},\bm{\theta}) \to \bm{u},$$
where $\bm{\theta}$ represents the model (or algorithm) parameters, $\bm{Y}$ is the given measured data, and $\bm{u}$ is the reconstructed image. In Hypernetwork, the model parameters $\bm{\theta}$ can be predicted by another neural network $H_{\bm{\phi}}(\cdot)$ rather than being learned directly from a dataset. Hence, the image reconstruction model can be rewritten as 
$$T(\bm{Y},H_{\bm{\phi}}(\bm{Y}))\to \bm{u},$$
with $\bm{\phi}$ collects the weight of Hypernetwork and is learned by end-to-end training on a dataset. Hypernetwork helps to compress the trainable parameters and simplify the training process while maintaining its meta-learning property.

\begin{figure}[t]
\centering
\includegraphics[scale=0.35]{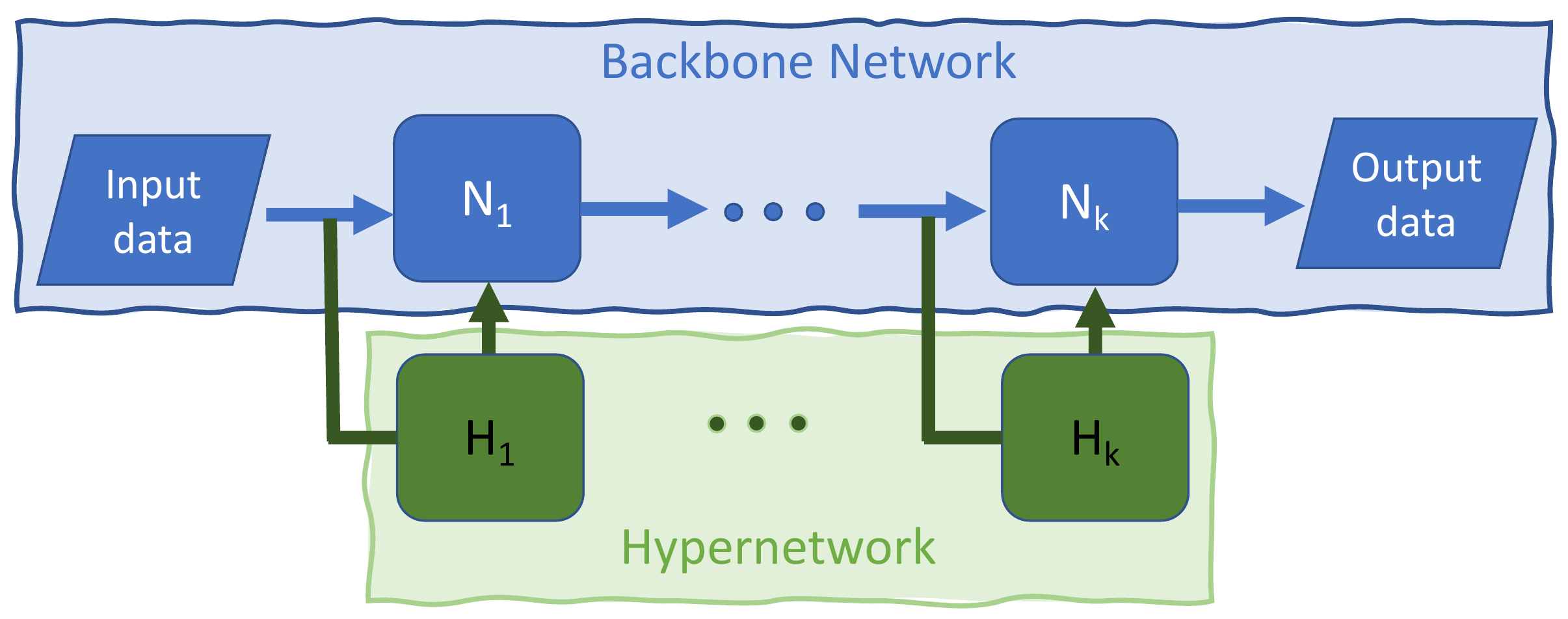}\label{Meta-Inv-abs-flow}
\caption{MetaInv-Net.}
\label{fig:Meta-Inv-abs-flow}
\end{figure}

For our proposed neural network model, the $\bm{u}$-subproblem that is solved by the CG algorithm is initialized by a predicted value from a CNN with the residual structure taking the current state (i.e., the approximation of image $\bm{u}^{k-1,L}$) as input. This CNN is a Hypernetwork of the full model, which is illustrated in Fig. \ref{fig:Meta-Inv-abs-flow}. We denote our model as Meta-Inversion Network (MetaInv-Net) in the rest of the paper. Note that there are not any trainable parameters in the backbone network. All trainable parameters are in the hypernetwork.

\subsection{Our Contribution}
The contributions of this work are summarized as follows:
\begin{itemize}
  \item New UD model. We propose a new deep model, MetaInv-Net. It adopts a CNN module as the hypernetwork to predict initialization for $\bm{u}$-subproblem, which is the only trainable component. This makes the proposed MetaInv-Net distinctively different from most existing UD models where more components of the underlying dynamics are set to be trainable. 
  \item Better generalization. The proposed MetaInv-Net generalizes well beyond the training setting (i.e., to other scanning settings, noise levels, and data sets).
\end{itemize}

The remaining part of this paper is organized as follows. In Section \ref{sec-methods}, we present details on the design of the MetaInv-Net. Numerical experiments are presented in Section \ref{sec:numerical-simul}. Finally, the remarks on conclusions and future works are presented in Section \ref{sec:conclusion}.

\section{Methods}\label{sec-methods}
In this section, we introduce full details on the proposed MetaInv-Net. We first introduce the backbone model used by MetaInv-Net, followed by a detailed description of the Hypernetwork and the loss function design.

\subsection{Half Quadratic Splitting (HQS)-CG Algorithm}
The backbone model is the unrolled HQS algorithm \cite{geman1995HQS} solving the following optimization problem
\begin{equation}\label{variable-sep-model}
\min_{\bm{u},\bm{z}} \frac{1}{2}\|\bm{P}\bm{u}-\bm{Y}\|^{2}+\lambda\|\bm{z}\|_{1}+\frac{1}{2}\sum_{i=1}^{M} \bm{\gamma}_{i}\|\bm{W}_{i}\bm{u}-\bm{z}_{i}\|^2,
\end{equation}
where $\bm{W}=(\bm{W}_{1},...,\bm{W}_{M})$ is a $M$ channnel operator, $\bm{z}=(\bm{z}_{1},...,\bm{z}_{M})$, $\lambda>0$ and $\bm{\gamma}=(\bm{\gamma}_{1},...,\bm{\gamma}_{M})$ with $\bm{\gamma}_{i}>0, i=1,...,M$. The operator $\bm{W}$ is chosen as the highpass components of the piecewise linear tight wavelet frame transform \cite{ron1997affine,Dong2010IASNotes}.

Now, the variables in optimization problem \eqref{variable-sep-model} are splitted into two blocks that can be updated alternatively
\begin{small}
\begin{align}
 \bm{u}^{k+1}&=\arg\min_{\bm{u}}\|\bm{P}\bm{u}-\bm{Y}\|^{2}+\sum_{i=1}^{M} \bm{\gamma}_{i}\|\bm{W}_{i}\bm{u}-\bm{z}^{k}_{i}\|^2,\notag\\
 \bm{z}^{k+1}&=\arg\min_{\bm{z}}\lambda \|\bm{z}\|_{1}+\frac{1}{2}\sum_{i=1}^{M} \bm{\gamma}_{i}\|\bm{W}_{i}\bm{u}^{k+1}-\bm{z}_{i}\|^2,
 \end{align}
\end{small}
with proper initialization $\bm{u}_{0}$ and $\bm{z}_{0}$. Solution to each of the two subproblem takes the form
\begin{small}
\begin{align}\label{exact-solution}
 \bm{u}^{k+1}&=\left(\bm{P}^{\top}\bm{P}+\sum_{i=1}^{M}\bm{\gamma}_{i}\bm{W}_{i}^{\top}\bm{W}_{i}\right)^{-1}\left[ \bm{P}^{\top}\bm{Y}+\sum_{i=1}^{M}\bm{\gamma}_{i}\bm{W}_{i}^{\top}\bm{z}_{i}^{k}\right],\notag\\
 \bm{z}^{k+1}&=\mathcal{T}_{\lambda/\bm{\gamma}}(\bm{W}\bm{u}^{k+1}).
\end{align}
\end{small}

The linear system in the $\bm{u}$-subproblem of \eqref{exact-solution} is solved approximately by the CG algorithm. Rewrite this large scale linear system as 
$$\bm{\mathcal{A}}\bm{u}=\bm{\mathcal{B}}^{k},$$ 
where $\bm{\mathcal{A}}=\bm{P}^{\top}\bm{P}+\sum_{i=1}^{M}\bm{\gamma}_{i}\bm{W}_{i}^{\top}\bm{W}_{i}$ and $\bm{\mathcal{B}}^{k}=\bm{P}^{\top}\bm{Y}+\sum_{i=1}^{M}\bm{\gamma}_{i}\bm{W}_{i}^{\top}\bm{z}_{i}^{k}$. 
Commonly, no hyperparameters are necessarily being provided to the CG algorithm except the stopping criteria ($L$ iterations). For $(k+1)$-th out-loop of HQS-CG algorithm, the initial variable $\bm{u}^{k+1,0}$ for CG algorithm is usually set to $0$ or the former output $\bm{u}^{k,L}$. 

The proximal operator $\mathcal{T}_{\lambda}(\cdot)$ is the anisotropic soft-thresholding operator defined by
$\mathcal{T}_{\lambda}(\bm{x})=\sign{(\bm{x})}\max\{|\bm{x}|-\lambda,0\}$.
The summary of the HQS-CG method is shown in Algorithm \ref{alg:HQSalg}.

\begin{algorithm}[t]
\caption{HQS-CG algorithm}
\label{alg:HQSalg}
\begin{algorithmic}[1]
\STATE 
\small{Initialization:} $\bm{u}^{0,L}=\bm{u}_{\mbox{\tiny FBP}},\bm{z}^{0}=\bm{W}\bm{u}^{0,L}$
\FOR{\texorpdfstring{$k=0:K-1$}{} }
\STATE
$\bm{u}^{k+1,0}=\bm{u}^{k,L} \;\mbox{or}\; \bm{0}$\\
$\bm{u}^{k+1,L}=\mbox{CG}(\bm{u}^{k+1,0},\bm{z}^{k},\bm{\gamma})$ \\
$\bm{z}^{k+1}=\mathcal{T}_{\lambda/\bm{\gamma}}(\bm{W}\bm{u}^{k+1,L})$
\ENDFOR
\STATE Output: $\bm{u}^{K,L}$
\end{algorithmic}
\end{algorithm}

Note that, the hyperparameters $\lambda$ and $\bm{\gamma}$ in Algorithm \ref{alg:HQSalg} can be manually selected with extensive trials-and-errors. To make the algorithm more practically applicable, data-adaptive hyperparameter estimation methods are often adopted, which include the discrepancy principle \cite{morozov1966solution,morozov2012methods}, cross validation, L-curve \cite{morozov1966solution} and the quasi-optimality criterion \cite{morozov1966solution}. Recently, deep learning approaches are able to determine pixel-wise, data-adaptive and task-dependent hyperparameters and operators in the model through end-to-end training on the given dataset \cite{yang2016deep,adler2018learned,Meinhardt_2017learn_prox_op,gregor2010learning}. Recent development of deep learning based hyperparameter optimization includes hypernetwork based methods \cite{lorraine2018stochastic,liu2019metapruning,zhang2019graph} and gradient descent-based approaches \cite{bengio2000gradient,maclaurin2015gradient,luketina2016scalable,andrychowicz2016learning,franceschi2017forward}. 

Other components in Algorithm \ref{alg:HQSalg}, such as the sparsifying transform $\bm{W}$ and the proximal operator $\mathcal{T}(\cdot)$ can also be adaptively learned from the dataset as what has been done by the earlier works listed in Table \ref{table:diff-UDs-models}. In this work, however, we fix the hyperparameters, $\lambda, \bm{\gamma}$, and $\bm{W}$, throughout our experiments. We observe that these hyperparameters are less critical than the initial values for the CG algorithm in the backbone model. Our numerical experiments confirm that although using a hypernetwork to estimate all the aforementioned hyperparameters may lead to better results for some cases (but not all as shown in Fig. \ref{fig:learn-diff-hyper-parameters-metainv-net} and Fig. \ref{fig:diff-denoiser-metainv-net}), it also leads to more trainable parameters for the model and out-weights its benefit in general.

\subsection{Meta-Inversion Network (MetaInv-Net)}
The backbone architecture of MetaInv-Net is built by unrolling the HQS-CG Algorithm (Algorithm \ref{alg:HQSalg}) with $K$ iterations forming a $K$-layer feed-forward network.

\subsubsection{Image Reconstruction Module}

\begin{figure}[t]
\centering
\includegraphics[scale=0.4]{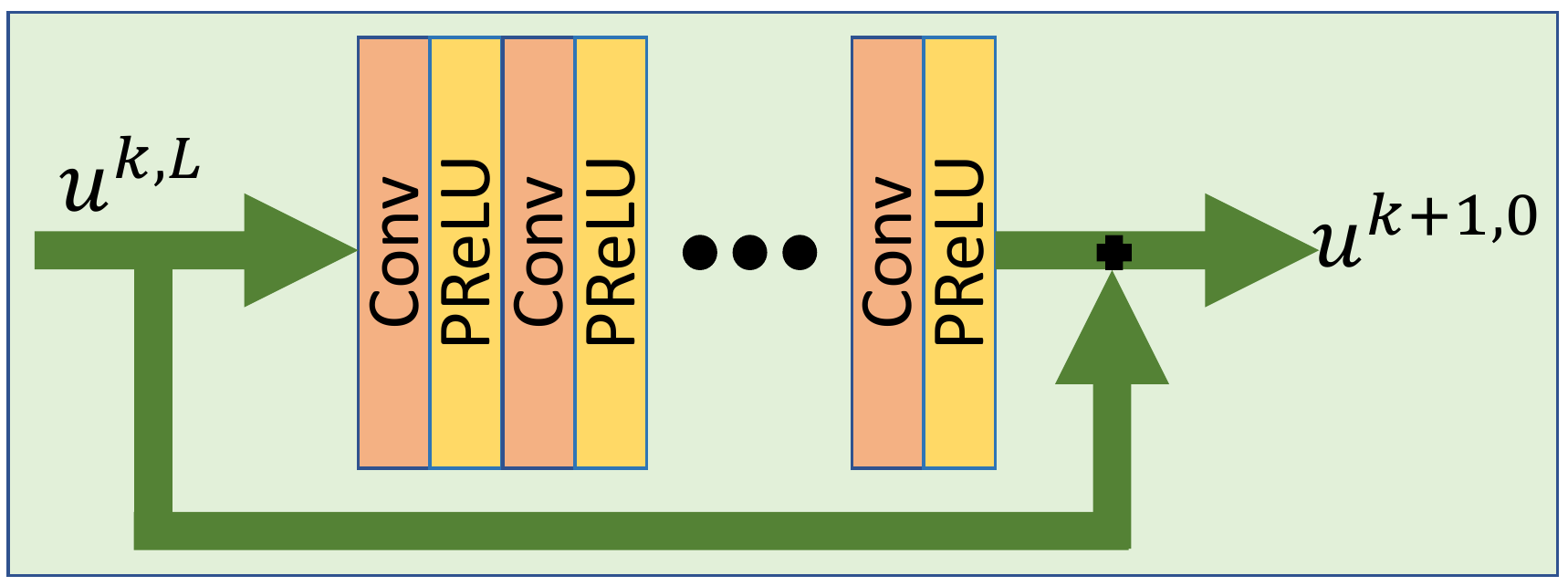}
  \caption{Light weight CG initializer module with vanilla CNN.}
  \label{fig:CG-init-CNN}
\end{figure}

\begin{figure}[t]
  \centering
  \includegraphics[scale=0.4]{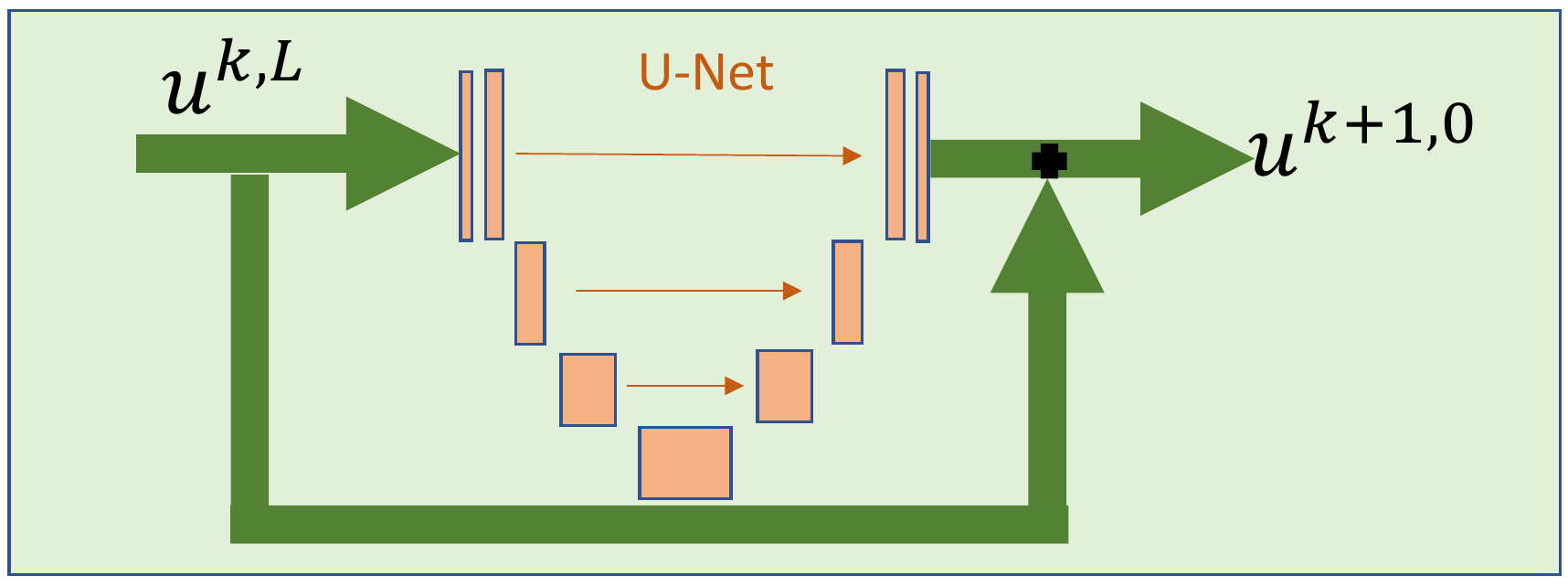}
  \caption{Heavy weight CG initializer module with U-Net.}
  \label{fig:CG-init-CNN-H}
\end{figure}

In this paper, a CNN-based initializer is adopted to providing a better initialization variable $\bm{u}^{k+1,0}$ for CG algorithm. The CG initializer with trainable parameters $\bm{\Theta}^{k+1}$ and past reconstruction $\bm{u}^{k,L}$ as input is built as
\begin{equation}\label{CNN-module}
\bm{u}^{k+1,0}=\bm{u}^{k,L}+\mbox{CNN}(\bm{u}^{k,L};\bm{\Theta}^{k+1}).
\end{equation}
Here, the skip connection helps to avoid gradient vanishing/exploding during training. It also admits a dynamic system interpretation \cite{lu18beyond,weinan2017proposal,chen2018neural}. The detailed architecture of CNN in \eqref{CNN-module} can be designed in various forms. Fig. \ref{fig:CG-init-CNN} shows a vanilla CNN that is composed of $S$ convolutional (Conv) layer and Parametric ReLU (PReLU) activation. This CNN architecture is rather light-weighted. Fig. \ref{fig:CG-init-CNN-H} shows a variant CG initializer (CG-Init) that is built by the U-Net architecture \cite{Ronneberger2015U-net} whereas the parameters are shared across  external layers. The number of model parameters from U-Net is much larger than vanilla CNN CG-Init.

\subsubsection{Denoiser Module}
\begin{figure}[ht]
\centering
\includegraphics[scale=0.5]{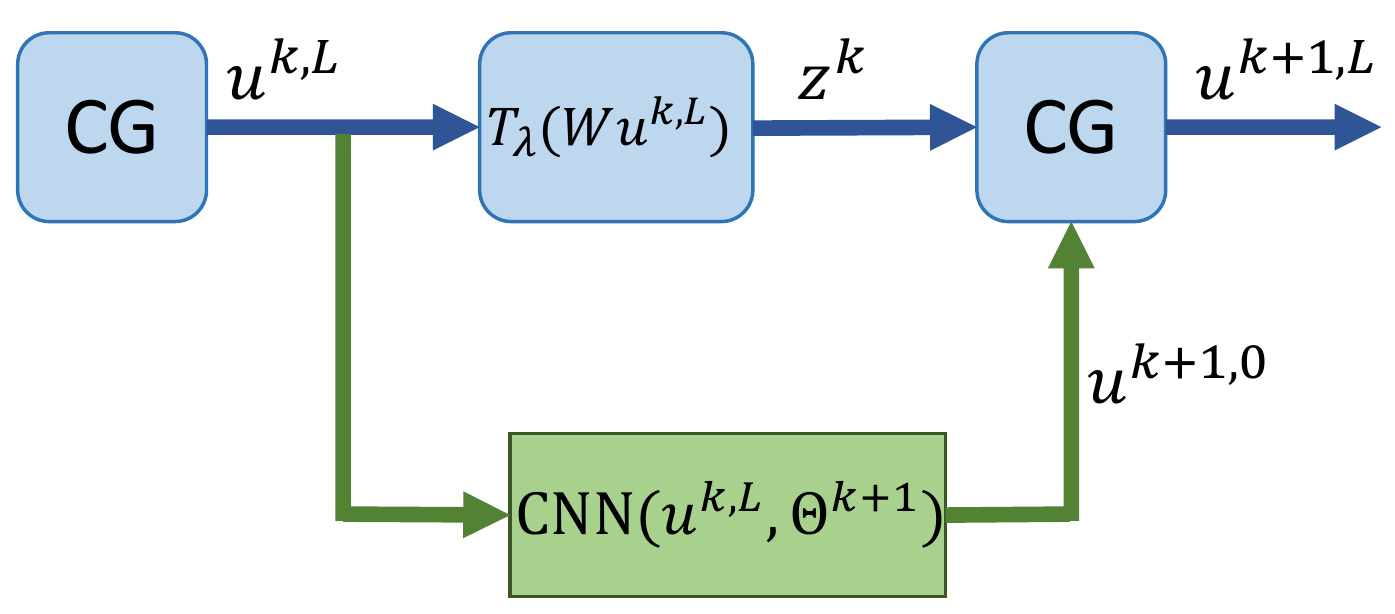}\label{Meta-Inversion}
\caption{MetaInv-Net building block.}
\label{fig:Meta-inversion-block}
\end{figure}

\begin{figure*}[ht]
\centering
\includegraphics[scale=0.35]{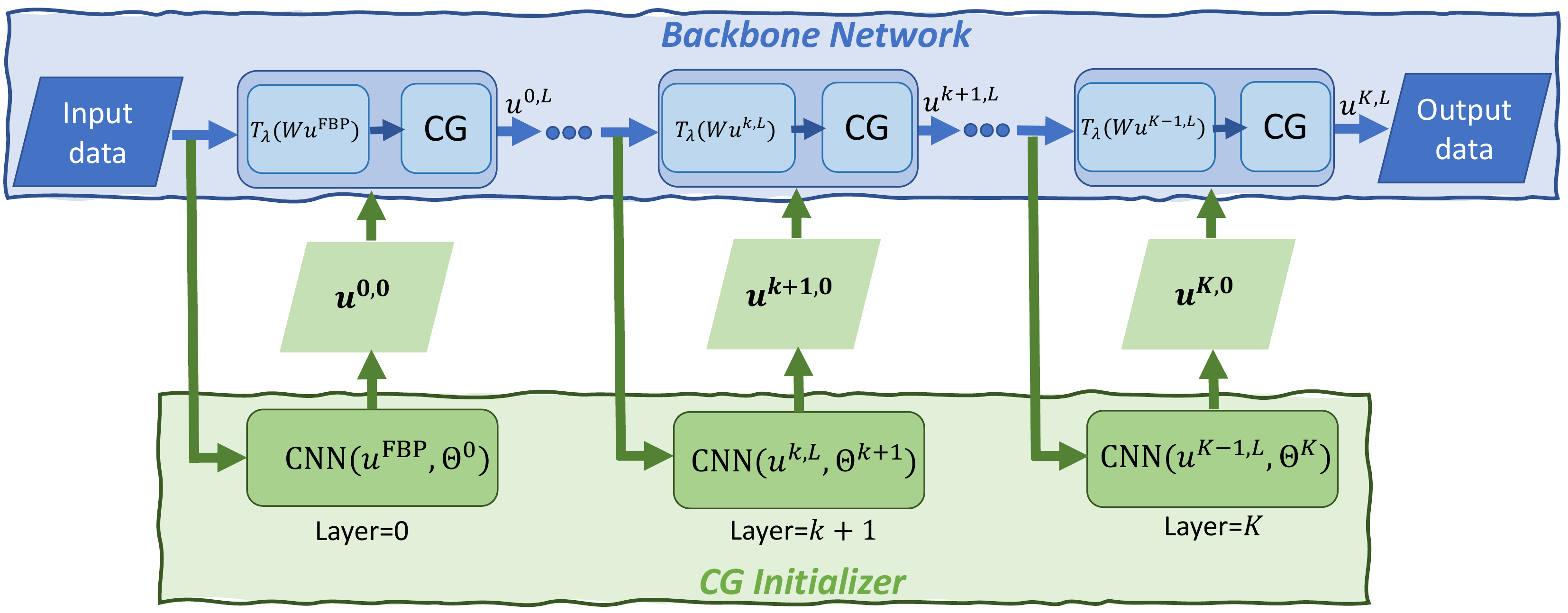}\label{Meta-Inv-Network}
\caption{MetaInv-Net.}
\label{fig:Meta-Inv-Network}
\end{figure*}

There is no trainable parameter in this module. We choose the operator $\bm{W}$ as the tight wavelet frame transform. The proximal operator $\mathcal{T}_{\lambda}(\cdot)$ is a soft-thresholding. The empirical result shows that this handcrafted image denoiser does not explicitly influence the quality of final reconstruction. In conclusion, a handcrafted denoiser has some benefits 1) the handcrafted $\mathcal{T}_{\lambda}(\cdot)$ has no training parameters, thus reduces the total number of parameters in MetaInv-Net, 2) $\bm{W}$ is dataset independent and thus should have better generalization property. Fig. \ref{fig:Meta-inversion-block} shows one block of the MetaInv-Net. The summary of MetaInv-Net is presented in Algorithm \ref{alg:meta-HQS}, and its architecture is presented in Fig. \ref{fig:Meta-Inv-Network}.

\begin{algorithm}[ht]
\caption{ MetaInv-Net}
\label{alg:meta-HQS}
\begin{algorithmic}[1]
\STATE 
\small{Initialization:} $\bm{u}^{0,L}=\bm{u}^{\mbox{\tiny FBP}},\bm{z}^{0}=W\bm{u}^{0,L}$
\\
\FOR{ \texorpdfstring{$k=0:K-1$}{} }
\STATE 
 $\bm{u}^{k+1,0}=\bm{u}^{k,L}+\mbox{CNN}(\bm{u}^{k,L};\bm{\Theta}^{k+1})$\\
\STATE
 $\bm{u}^{k+1,L}= \mbox{CG}(\bm{u}^{k+1,0},\bm{z}^{k},\bm{\gamma})$\\
\STATE
 $\bm{z}^{k+1}=\mathcal{T}_{\lambda/\bm{\gamma}}(\bm{W}\bm{u}^{k+1,L})$
\ENDFOR
\STATE Output: $\bm{u}^{K,L}$.
\end{algorithmic}
\end{algorithm}

\subsubsection{Loss Function}

Loss function in MetaInv-Net is defined by
\begin{multline*}
\mathcal{L}(\bm{u}^{1,L},...,\bm{u}^{K,L};\bm{u}^{gt})=\cr \sum_{k=1}^{K}\left(\mu_{1} \mathcal{L}_{2}(\bm{u}^{k,L},\bm{u}^{gt})+\mu_{2}\mathcal{L}_{ssim}(\bm{u}^{k,L},\bm{u}^{gt})\right),\notag
\end{multline*} 
where $\mathcal{L}_{2}(\bm{x},\bm{y})$ represents the $\ell_{2}$-norm of $\bm{x}-\bm{y}$ and $\mathcal{L}_{ssim}$ is the SSIM loss. Similar to the value function design in reinforcement learning, constants $\mu_{1}$ and $\mu_{2}$ can be viewed as discount factors for $\ell_{2}$ loss $\mathcal{L}_{2}$ and SSIM loss $\mathcal{L}_{ssim}$, respectively.

\section{Numerical Experiments}\label{sec:numerical-simul}
In this section, we validate the performance of MetaInv-Net on sparse view CT image reconstruction beyond the training setting (e.g., to different scanning settings, noise levels, and data sets). MetaInv-Net with vanilla CNN and U-Net based CG initializer that trained end-to-end on the training set are denoted by MetaInv-Net (L) and MetaInv-Net (H), respectively.

\subsection{Comparison Methods}
The FBP with ``Ram-Lak" filter is adopted to provide initialization for HQS-CG and compared deep models. The HQS-CG Algorithm (Algorithm \ref{alg:HQSalg}) is an iterative reconstruction method. Its stopping criterion is set to be the relative error $\frac{\|\bm{u}^{k+1}-\bm{u}^{k}\|}{\|\bm{u}^{k+1}\|}\le 3\times 10^{-4}$. PWLS-CSCGR \cite{bao2019convolutional} is a state-of-the-art iterative reconstruction method. The MetaInv-Net (L) and MetaInv-Net (H) are truncated version (step K=6) of HQS-CG. Three state-of-the-art deep UD models, i.e., PD-Net \cite{adler2018learned}, JSR-Net \cite{zhang2019JSRNet}, and PFBS-AIR \cite{ding2020low}, are adopted in our comparison with initialization $\bm{u}^{0}=\bm{u}^{\mbox{\tiny FBP}}$. In PD-Net and JSR-Net models, the SS2 loss function \cite{zhang2019JSRNet} is adopted, which is a weighted sum of mean SSIM index loss, semantic segmentation loss and mean square error (MSE) loss. The PFBS-AIR adopts the MSE loss as in the original work \cite{ding2020low}. FBPConv-Net \cite{Jin2017FBPConvNet} is a post-processing method that adopts a U-Net \cite{Ronneberger2015U-net} to reduce artifacts in FBP reconstruction. Another variant of MetaInv-Net is a specific form of MetaInv-Net (H) that adopts the pre-trained FBPConv-Net as CG initializer. We denote this model by MetaInv-Net (P).

\subsection{Training Setting}
The ``2016 NIH-AAPM-Mayo Clinic Low Dose CT Grand Challenge" data set has ten patients' normal dose CT images \cite{mccollough2016tu}. We adopt five patients' data as the training set. The contained 1684 slices of high-resolution images ($512\times 512$ pixels) are used as the ground truth. An additional one patient's data from the same data set containing 366 slices image is used as the validation set. The scanning geometry is a fan beam X-ray source with $180$ scanning views equally distributed around $360^{\circ}$ and 800 detector elements. The normal dose Mayo data were scanned from the chest to the abdomen under the protocol 120 kVp and 235 effective mAs (500mA/0.47s). In the numerical experiments, the simulated sinogram is contaminated by Poisson noise (with different noise level indicated by photon intensity $I_{0}$) and electronic system noise (with standard deviation $\sigma$). The electronic noise level is regarded to follow a zero-mean normal distribution and is assumed to be stable for a commercial CT scanner. In our simulation, the noise is controlled by the incident photon intensity $I_{0}$ per detector bin. More clinic realistic noisy sinogram can be obtained following the procedure of low dose CT imaging data synthesis at different scanning protocol (i.e., the X-ray tube output parameter with unit mAs that reflects the photon flux/intensity) in \cite{zeng2015simple,won2014realistic}.

To have a fair comparison, we set the hyperparameters of each deep model properly, so that they contain a similar number of training parameters. CG initializer of MetaInv-Net (L) has depth $S=6$ and channel width $8$. Model parameters are not shared in each layer of MetaInv-Net (L). For MetaInv-Net (H) and MetaInv-Net (P), we adopt the same U-Net architecture from FBPConv-Net as a CG initializer. Since the number of model parameters is large in U-Net, they are shared in each layer of MetaInv-Net (H) and MetaInv-Net (P). Table \ref{table:train-setting} summarizes the details of training setting. \textit{All the deep models are only trained under the setting of sparse view CT reconstruction as described in Table \ref{table:train-setting}}. In addition, all the deep models except MetaInv-Net (P) are trained end-to-end on the training set.

\begin{table}[ht]
\tiny
  \caption{ Training setting for compared deep models.}\label{table:train-setting}
\centering
  \begin{tabular}{|c|c|c|c|c|c|}
  \hline
 \multirow{2}{*}{ Items } & \multicolumn{5}{c|}{Models}\\
\cline{2-6}
 & \multicolumn{1}{c|}{PD-Net} & \multicolumn{1}{c|}{JSR-Net} &
 \multicolumn{1}{c|}{PFBS-AIR} &
 \multicolumn{1}{c|}{MetaInv-Net (L)} & \multicolumn{1}{c|}{MetaInv-Net (H)}\\
  \cline{1-6}
   Training set size & 1683 & 1683 & 1683  & 1683 & 1683 \\
  \cline{1-6}
  $\# \mbox{Param.}$ & 253220 & 225160 & 1152980 & 17703 & 13394177 \\
  \cline{1-6}
  Unrolled Layers & 10 & 6 & 6 & 6 & 10 \\
  \cline{1-6}
  \textcolor{blue}{Sinogram Size} & 180$\times$800 & 180$\times$800 &180$\times$800 &180$\times$800 &180$\times$800 \\
  \cline{1-6}
  Image Size & 512$\times$512 & 512$\times$512 & 512$\times$512 & 512$\times$512 & 512$\times$512 \\
  \cline{1-6}
  Batch Size & 1 & 1 &4 &1 &2\\
  \cline{1-6}
  Training Epochs & 10 & 10 & 10 &10 &10\\
  \cline{1-6}
  \textcolor{blue}{Noise Level($I_{0}$)} & 5e6 & 5e6 & 5e6 & 5e6 & 5e6 \\
  \cline{1-6}
\hline
\end{tabular}
\end{table}

For simplicity, we replace the original parameter representation $\lambda/\bm{\gamma}$ by $\lambda$. In each layer of MetaInv-Net, $\lambda_k$ and $\bm{\gamma}_k$ are set by $\lambda_k=\lambda_{k-1}-\delta_{\lambda}$ and $\bm{\gamma}_k=\bm{\gamma}_{k-1}-\delta_{\gamma}$. We empirically set $\lambda_0=0.005$, $\delta_{\lambda}=0.0008$, $\bm{\gamma}_0=0.01$ and $\delta_{\gamma}=0.02$. The discount factors $\mu_{1}$ and $\mu_{2}$ in the definition of loss function are set to $1.1$ and $1$, respectively.

The training is conducted on PyTorch 1.3.1 backend with an NVIDIA Titan Xp GPU with memory 10.75G. The ADAM optimizer \cite{kingma2015adam} with the learning rate, $10^{-3}$, is adopted in the training phase.

\begin{table*}[ht]
\tiny
\caption{ 
Sparse view CT image reconstruction on AAPM-Test set. \colorbox[rgb]{0.67, 0.88, 0.69}{Green} rows (\#views) indicate different number of measured projection views. The sinogram is corrupted by different noise level, i.e., Noise-1 (Low),  Noise-2,  Noise-3, and Noise-4 (High) corresponding to photon intensity $I_0=1\times 10^7, 5\times 10^6, 5\times 10^5$ and $1\times 10^5$, respectively.
\colorbox{gbypink}{Red} and \colorbox{gbyblue}{Blue} 
indicate the Rank-1 and Rank-2 mean value of SSIM/PSNR, respectively. The last column shows the average inference time (in seconds).  }
\label{table:sparse-view-CT-Poisson-AAPM}
\centering
  \begin{tabular}{|c|c|c|c|c|c|c|c|c|c|}
  \hline
\multirow{3}{*}{Models} & \multicolumn{8}{c|}{Quality measure}& \multirow{3}{*}{Time (sec.)}\\
  \cline{2-9}
& \multicolumn{2}{c|}{ Noise-1 (Low) }& \multicolumn{2}{c|}{ Noise-2 } 
& \multicolumn{2}{c|}{ Noise-3 }& \multicolumn{2}{c|}{ Noise-4 (High) }& \\
  \cline{2-9}
& \multicolumn{1}{c|}{SSIM } & \multicolumn{1}{c|}{PSNR } &\multicolumn{1}{c|}{SSIM } & \multicolumn{1}{c|}{PSNR }
& \multicolumn{1}{c|}{SSIM } & \multicolumn{1}{c|}{PSNR } &\multicolumn{1}{c|}{SSIM } & \multicolumn{1}{c|}{PSNR } & \\
  \cline{1-10}

 \multicolumn{10}{|c|}{\cellcolor{gbygreen}\# views=15} \\
 \cline{1-10}
FBP 
&0.4082(0.0360) &16.5251(1.6878) 
&0.4079(0.0359) &16.5121(1.6879) 
&0.4015(0.0350) &16.2694(1.6625) 
&0.3781(0.0322) &15.3687(1.5945) & - \\
    \cline{1-10}
HQS-CG 
 &\cellcolor{gbyblue} 0.8101(0.0367) &\cellcolor{gbypink} 25.7443(1.8202) 
 &\cellcolor{gbyblue}0.8098(0.0365) &\cellcolor{gbypink}25.7365(1.8159) 
 &\cellcolor{gbyblue}0.8054(0.0376) &\cellcolor{gbypink}25.6542(1.8201) 
 &\cellcolor{gbyblue}0.7892(0.0406) &\cellcolor{gbypink}25.3639(1.8256) &15.6952 \\
    \cline{1-10}
PWLS-CSCGR
  &0.7639(0.0619) &\cellcolor{gbyblue}24.9305(2.2790) 
  &0.7637(0.0620) &\cellcolor{gbyblue}24.9252(2.2799)
  &0.7611(0.0625) &\cellcolor{gbyblue}24.8588(2.2703) 
  &0.7539(0.0625) &\cellcolor{gbyblue}24.6148(2.2043) &664.5800 \\
    \cline{1-10}
FBPConv-Net
&0.3837(0.0418) &16.4643(1.9660) 
&0.3831(0.0416) &16.4496(1.9642) 
&0.3745(0.0393) &16.2013(1.9409) 
&0.3455(0.0316) &15.2017(1.8446) &0.0332 \\
    \cline{1-10}
PFBS-AIR 
&- &- &- &- &- &- &- &- &-\\
    \cline{1-10}
PD-Net 
&0.4785(0.0779) &17.1390(2.0992) 
&0.4783(0.0777) &17.1267(2.0995) 
&0.4737(0.0774) &16.9208(2.0668) 
&0.4545(0.0761) &16.1902(2.0220) &0.1195 \\
  \cline{1-10}
JSR-Net 
&0.4059(0.0476) &16.6454(1.9448) 
&0.4057(0.0476) &16.6405(1.9433) 
&0.4033(0.0473) &16.5588(1.9289) 
&0.3927(0.0456) &16.1938(1.8552) &0.1786 \\
  \cline{1-10}
MetaInv-Net (P) 
&\cellcolor{gbypink}0.8115(0.0260) &24.5116(1.9003)
&\cellcolor{gbypink}0.8114(0.0261) &24.5110(1.9133) 
&\cellcolor{gbypink}0.8062(0.0277) &24.4832(1.9039) 
&\cellcolor{gbypink}0.7986(0.0319) &24.3738(1.9676) &0.6388\\
  \cline{1-10}
MetaInv-Net (L)
 &0.3234(0.1606) &8.8102(5.5876) 
 &0.3139(0.1604) &8.3742(5.6837) 
 &0.2746(0.1488) &6.8801(5.5527) 
 &0.1679(0.0982) &3.5315(4.8865) &0.5203\\
  \cline{1-10}
MetaInv-Net (H) 
&0.7530(0.0443) &24.5627(1.8439) 
&0.7529(0.0442) &24.5593(1.8446) 
&0.7518(0.0423) &24.5241(1.8468) 
&0.7454(0.0390) &24.3780(1.8452) &0.5863\\
 \cline{1-10}

 \multicolumn{10}{|c|}{\cellcolor{gbygreen}\# views=30} \\
  \cline{1-10}
FBP 
&0.5461(0.0405) &19.8849(1.7178) 
&0.5457(0.0405) &19.8684(1.7163) 
&0.5374(0.0370) &19.5782(1.7008) 
&0.5124(0.0382) &18.5002(1.6595) &- \\
  \cline{1-10}
HQS-CG 
&\cellcolor{gbyblue}0.8928(0.0245) &\cellcolor{gbypink}28.5001(1.9123) 
&\cellcolor{gbypink}0.8926(0.0245) &\cellcolor{gbypink}28.4958(1.9113) 
&\cellcolor{gbypink}0.8888(0.0257) &\cellcolor{gbypink}28.3871(1.9129) \
&\cellcolor{gbyblue}0.8713(0.0312) &\cellcolor{gbypink}27.9315(1.9261) &11.5589 \\
  \cline{1-10}
PWLS-CSCGR
&0.8501(0.0480) &27.5853(2.3455) 
&0.8497(0.0481) &27.5691(2.3421)
&0.8451(0.0486) &27.3308(2.3024) 
&0.8270(0.0492) &26.3593(2.1128) &769.5800\\
    \cline{1-10}
FBPConv-Net 
&0.5673(0.0524) &21.1052(2.0655) 
&0.5666(0.0525) &21.0846(2.0659) 
&0.5559(0.0509) &20.7814(2.0314) 
&0.5183(0.0466) &19.5723(1.9555) &0.0318 \\
    \cline{1-10}
PFBS-AIR 
&- &- &- &- &- &- &- &- &-\\
    \cline{1-10}
PD-Net 
&0.6625(0.0776) &21.4266(2.5030) 
&0.6621(0.0777) &21.4046(2.4976) 
&0.6555(0.0774) &21.0271(2.4362) 
&0.6295(0.0771) &19.8095(2.2767) &0.1327\\
  \cline{1-10}
JSR-Net 
&0.6023(0.0666) &21.5884(2.2440) 
&0.6021(0.0665) &21.5798(2.2425) 
&0.5981(0.0663) &21.4199(2.2190) 
&0.5830(0.0658) &20.7977(2.1484) &0.1871\\
  \cline{1-10}
MetaInv-Net (P) 
  &\cellcolor{gbypink}0.8997(0.0275) &28.0763(1.9416) 
  &\cellcolor{gbyblue}0.8899(0.0318) &28.0710(1.9420)
  &\cellcolor{gbyblue}0.8881(0.0325) &28.0094(1.9692) 
  &\cellcolor{gbypink}0.8782(0.0359) &27.6790(2.0190) &0.6476\\
    \cline{1-10}
MetaInv-Net (L)
&0.6874(0.1253) &21.6817(4.7651) 
&0.6689(0.1343) &20.9490(5.0416) 
&0.6531(0.1353) &20.6462(4.8600) 
&0.5786(0.1265) &18.5290(4.2655) &0.5252 \\
  \cline{1-10}
MetaInv-Net (H) 
&0.8680(0.0336) &\cellcolor{gbyblue}28.3096(2.0272) 
&0.8659(0.0288) &\cellcolor{gbyblue}28.3076(2.0290) 
&0.8650(0.0301) &\cellcolor{gbyblue}28.2561(2.0379) 
&0.8610(0.0297) &\cellcolor{gbyblue}27.9216(1.9994) &0.5945\\
  \cline{1-9}

 \multicolumn{10}{|c|}{\cellcolor{gbygreen}\# views=60} \\
  \cline{1-10}
 FBP
&0.6957(0.0409) &23.5904(1.6945) 
&0.6953(0.0408) &23.5703(1.6931) 
&0.6880(0.0405) &23.2190(1.6823) 
&0.6597(0.0406) &21.9475(1.6759) &-\\
  \cline{1-10}
 HQS-CG 
&0.9410(0.0151) &30.9317(1.8999) 
&0.9410(0.0151) &30.9287(1.9000) 
&0.9396(0.0155) &30.8706(1.9035) 
&0.9326(0.0181) &30.6016(1.9215) &8.0788\\
  \cline{1-10}
PWLS-CSCGR 
&0.9283(0.0288) &31.2204(2.6553) 
&0.9277(0.0289) &31.1556(2.6438) 
&0.9162(0.0309) &30.0369(2.4445) 
&0.8740(0.0357) &26.6885(2.1354) &1144.0700\\
    \cline{1-10}
FBPConv-Net 
&0.7826(0.0508) &26.8272(2.3432) 
&0.7819(0.0510) &26.8030(2.3465) 
&0.7713(0.0520) &26.3855(2.3189) 
&0.7292(0.0550) &24.7066(2.2295) &0.0311\\
    \cline{1-10}
 PFBS-AIR 
&- &- &- &- &- &- &- &- &-\\
    \cline{1-10}
 PD-Net 
&0.8351(0.0554) &27.1657(2.6494) 
&0.8346(0.0555) &27.1270(2.6473) 
&0.8254(0.0565) &26.4614(2.6435) 
&0.7930(0.0601) &24.3375(2.5690) &0.1455\\
  \cline{1-10}
  JSR-Net
&0.7953(0.0594) &26.7365(2.4171) 
&0.7951(0.0594) &26.7230(2.4145) 
&0.7902(0.0597) &26.4516(2.3849) 
&0.7703(0.0612) &25.4008(2.3133) &0.2129\\
  \cline{1-10}
  MetaInv-Net (P)
&\cellcolor{gbyblue}0.9480(0.0163) &\cellcolor{gbyblue}31.5918(1.7928) 
&\cellcolor{gbyblue}0.9476(0.0157) &\cellcolor{gbyblue}31.5855(1.7924) 
&\cellcolor{gbyblue}0.9458(0.0165) &\cellcolor{gbyblue}31.4332(1.8320) 
&\cellcolor{gbypink}0.9447(0.0116) &\cellcolor{gbyblue}30.7730(1.9376) & 0.7057\\
 \cline{1-10}
MetaInv-Net (L) 
&0.9265(0.0256) &31.1223(2.4271) 
&0.9259(0.0263) &31.0893(2.4317) 
&0.9214(0.0268) &30.8072(2.3409) 
&0.8900(0.0378) &29.2109(2.4673) &0.5717\\
  \cline{1-10}
  MetaInv-Net (H) 
&\cellcolor{gbypink}0.9528(0.0165) &\cellcolor{gbypink}32.9784(2.1852) 
&\cellcolor{gbypink}0.9519(0.0159) &\cellcolor{gbypink}32.9658(2.1842) 
&\cellcolor{gbypink}0.9487(0.0151) &\cellcolor{gbypink}32.7204(2.1572) 
&\cellcolor{gbyblue}0.9360(0.0183) &\cellcolor{gbypink}31.3990(2.0836) &0.5911\\
 \cline{1-10}

 \multicolumn{10}{|c|}{\cellcolor{gbygreen}\# views=120} \\
  \cline{1-10}
FBP
&0.8401(0.0304) &26.9864(1.8169) 
&0.8397(0.0305) &26.9637(1.8164) 
&0.8321(0.0308) &26.5664(1.8053) 
&0.8023(0.0333) &25.1634(1.7939) &-\\
  \cline{1-10}
 HQS-CG 
&0.9693(0.0092) &33.9859(1.7956) 
&0.9692(0.0092) &33.9797(1.7959) 
&0.9674(0.0098) &33.8241(1.8080) 
&0.9566(0.0136) &\cellcolor{gbyblue}33.0693(1.8955) &5.6082\\
    \cline{1-10}
PWLS-CSCGR 
&0.9764(0.0099) &35.3358(2.6822) 
&0.9747(0.0104) &34.9593(2.6366) 
&0.9488(0.0173) &30.7337(2.3359) 
&0.8740(0.0326) &24.4759(2.2454) &1611.1800\\
    \cline{1-10}
 FBPConv-Net 
&0.9500(0.0183) &32.3030(2.8179) 
&0.9497(0.0184) &32.2792(2.8153) 
&0.9441(0.0201) &31.8446(2.7632) 
&0.9162(0.0277) &29.9024(2.5648) &0.0333\\
    \cline{1-10}
 PFBS-AIR 
&0.1113(0.0337) &-1.2418(3.2403) 
&0.1110(0.0334) &-1.2564(3.2326) 
&0.1075(0.0317) &-1.5061(3.1497) 
&0.0932(0.0254) &-2.6090(2.8855) &0.3371\\
    \cline{1-10}
  PD-Net
&0.9639(0.0150) &34.0681(2.3871) 
&0.9632(0.0153) &33.9898(2.4024) 
&0.9524(0.0196) &32.7167(2.5802) 
&0.9125(0.0321) &29.0437(2.7483) &0.1758\\
  \cline{1-10}
  JSR-Net
&0.9561(0.0179) &33.1083(2.4801) 
&0.9558(0.0180) &33.0768(2.4804) 
&0.9504(0.0196) &32.5668(2.4731) 
&0.9284(0.0257) &30.7716(2.4706) &0.2581\\
  \cline{1-10}
  MetaInv-Net (P)
&0.9760(0.0062) &33.5379(1.7293) 
&0.9707(0.0095) &33.5340(1.7281) 
&0.9675(0.0148) &33.4103(1.7597)
&\cellcolor{gbypink}0.9570(0.0212) &32.6454(1.8821) &0.7578\\
  \cline{1-10}
MetaInv-Net (L) 
&\cellcolor{gbyblue}0.9765(0.0082) &\cellcolor{gbyblue}35.7060(1.9659) 
&\cellcolor{gbyblue}0.9762(0.0083) &\cellcolor{gbyblue}35.6596(1.9631) 
&\cellcolor{gbyblue}0.9710(0.0098) &\cellcolor{gbyblue}34.8902(1.9700) 
&0.9463(0.0176) &32.3087(2.0382) &0.6236\\
   \cline{1-10}
 MetaInv-Net (H)
&\cellcolor{gbypink}0.9792(0.0075) &\cellcolor{gbypink}36.4048(2.0861) 
&\cellcolor{gbypink}0.9784(0.0078) &\cellcolor{gbypink}36.3677(2.0860) 
&\cellcolor{gbypink}0.9743(0.0086) &\cellcolor{gbypink}35.7150(2.0754) 
&\cellcolor{gbyblue}0.9569(0.0138) &\cellcolor{gbypink}33.1037(2.1027) &0.6813\\
   \cline{1-10}

 \multicolumn{10}{|c|}{\cellcolor{gbygreen}\# views=180} \\
  \cline{1-10}
FBP
&0.9085(0.0199) &29.5282(1.6358) 
&0.9074(0.0199) &29.5006(1.6362) 
&0.8997(0.0202) &29.0138(1.6385) 
&0.8715(0.0251) &27.3736(1.6722) &-\\
  \cline{1-10}
 HQS-CG
&0.9815(0.0060) &36.2793(1.8663) 
&0.9812(0.0061) &36.2377(1.8649) 
&0.9760(0.0075) &35.5099(1.8595) 
&0.9473(0.0164) &32.7485(2.0585) &5.7709\\
  \cline{1-10}
PWLS-CSCGR 
&0.9841(0.0061) &36.2639(2.6258)
&0.9843(0.0058) &36.1489(2.4681) 
&0.9475(0.0152) &29.4591(2.0957) 
&0.8563(0.0359) &22.6454(2.1609) &2130.3300\\
    \cline{1-10}
 FBPConv-Net 
&0.9725(0.0100) &35.2701(1.8857) 
&0.9725(0.0100) &35.2588(1.8860) 
&0.9714(0.0098) &34.9693(1.8958) 
&0.9600(0.0131) &33.1696(2.0366) &0.0326\\
    \cline{1-10}
 PFBS-AIR
&0.9814(0.0060) &36.8503(1.9515) 
&0.9809(0.0062) &36.7700(1.9436) 
&0.9724(0.0080) &35.5657(1.8591) 
&0.9397(0.0151) &32.4312(1.8216) &0.4259\\
    \cline{1-10}
PD-Net
&\cellcolor{gbypink}0.9877(0.0038) &\cellcolor{gbyblue}37.3635(1.6844) 
&\cellcolor{gbypink}0.9874(0.0039) &\cellcolor{gbyblue}37.2891(1.7018) 
&0.9808(0.0067) &35.8432(2.0235) 
&0.9489(0.0193) &31.3517(2.5932) &0.2201\\
  \cline{1-10}
  JSR-Net
&\cellcolor{gbyblue}0.9870(0.0043) &36.8654(1.8133) 
&\cellcolor{gbyblue}0.9868(0.0044) &36.8219(1.8201) 
&\cellcolor{gbypink}0.9826(0.0059) &36.0825(1.9181) 
&\cellcolor{gbyblue}0.9646(0.0123) &\cellcolor{gbyblue}33.7080(2.1778) &0.3057\\
  \cline{1-10}
  MetaInv-Net (P)
&0.9752(0.0104) &34.1345(1.7729)
&0.9749(0.0110) &34.1308(1.7729) 
&0.9741(0.0079) &34.0279(1.7953) 
&\cellcolor{gbypink}0.9665(0.0102) &33.3867(1.8938) &0.8682\\
  \cline{1-10}
MetaInv-Net (L) 
&0.9835(0.0058) &37.1915(1.9723)
&0.9824(0.0066) &37.1393(1.9688) 
&0.9801(0.0067) &\cellcolor{gbyblue}36.2534(1.9385) 
&0.9594(0.0131) &33.3991(2.0124) &0.6832\\
   \cline{1-10}
 MetaInv-Net (H)
&0.9852(0.0056) &\cellcolor{gbypink}37.5739(2.0790) 
&0.9850(0.0055) &\cellcolor{gbypink}37.5256(2.0788)
&\cellcolor{gbyblue}0.9811(0.0068) &\cellcolor{gbypink}36.7087(2.0695) 
&0.9642(0.0118) &\cellcolor{gbypink}33.9069(2.1356) &0.7296\\
   \cline{1-10}
\hline
\end{tabular}
\end{table*}

\begin{table*}[ht]
\tiny
\caption{ 
Sparse view CT image reconstruction on Pancreas-Test set. \colorbox[rgb]{0.67, 0.88, 0.69}{Green} rows (\#views) indicate different number of measured projection views. The sinogram is corrupted by different noise level, i.e., Noise-1 (Low),  Noise-2,  Noise-3, and Noise-4 (High) corresponding to photon intensity $I_0=1\times 10^7, 5\times 10^6, 5\times 10^5$ and $1\times 10^5$, respectively.
\colorbox{gbypink}{Red} and \colorbox{gbyblue}{Blue} 
indicate the Rank 1 and Rank 2 mean value of SSIM/PSNR, respectively. The last column shows the average inference time (in seconds).   }
\label{table:sparse-view-CT-Poisson-Panc}
\centering
  \begin{tabular}{|c|c|c|c|c|c|c|c|c|c|}
  \hline
\multirow{3}{*}{Models} &\multicolumn{8}{c|}{Quality measure} &\multirow{3}{*}{Time (sec.)}\\
  \cline{2-9}
& \multicolumn{2}{c|}{ Noise-1 (Low) }& \multicolumn{2}{c|}{ Noise-2 } 
& \multicolumn{2}{c|}{ Noise-3 }& \multicolumn{2}{c|}{ Noise-4 (High) } & \\
  \cline{2-9}
& \multicolumn{1}{c|}{SSIM } & \multicolumn{1}{c|}{PSNR } &\multicolumn{1}{c|}{SSIM } & \multicolumn{1}{c|}{PSNR }
& \multicolumn{1}{c|}{SSIM } & \multicolumn{1}{c|}{PSNR } &\multicolumn{1}{c|}{SSIM } & \multicolumn{1}{c|}{PSNR } &\\
  \cline{1-10}

 \multicolumn{10}{|c|}{\cellcolor{gbygreen}\# views=15} \\
  \cline{1-10}
  FBP 
 &0.4516(0.0296) &17.0171(1.1072) 
 &0.4503(0.0293) &16.9901(1.0989) 
 &0.4327(0.0242) &16.5536(0.9695)
 &0.3867(0.0199) &15.1176(0.8105) &-\\
  \cline{1-10}
 HQS-CG 
&\cellcolor{gbypink}0.8436(0.0239) &\cellcolor{gbypink}25.8317(1.5215) 
&\cellcolor{gbypink}0.8434(0.0238) &\cellcolor{gbypink}25.8271(1.5198) 
&\cellcolor{gbypink}0.8376(0.0240) &\cellcolor{gbypink}25.7169(1.4901) 
&\cellcolor{gbypink}0.8117(0.0258) &\cellcolor{gbypink}25.2377(1.3734) &13.1581\\
  \cline{1-10}
PWLS-CSCGR 
&\cellcolor{gbyblue}0.8205(0.0331) &\cellcolor{gbyblue}25.4258(1.5663) 
&\cellcolor{gbyblue}0.8202(0.0332) &\cellcolor{gbyblue}25.4194(1.5673) 
&\cellcolor{gbyblue}0.8164(0.0329) &\cellcolor{gbyblue}25.3147(1.5299) 
&\cellcolor{gbyblue}0.8057(0.0321) &\cellcolor{gbyblue}24.9378(1.4108) &665.2000\\
    \cline{1-10}
 FBPConv-Net 
&0.4187(0.0451) &16.5333(1.3220) 
&0.4173(0.0445) &16.5125(1.3169) 
&0.3958(0.0364) &16.1488(1.2173) 
&0.3416(0.0219) &14.7228(0.9659) &0.0313\\
    \cline{1-10}
PFBS-AIR 
   &- &- &- &- &- &- &- &- &-\\
    \cline{1-10}
 PD-Net 
&0.5663(0.0566) &19.1668(1.5238) 
&0.5655(0.0567) &19.1327(1.5208) 
&0.5523(0.0567) &18.5542(1.4183) 
&0.5027(0.0564) &16.8135(1.4095) &0.1170\\
  \cline{1-10}
JSR-Net 
&0.4820(0.0399) &18.0415(1.4340) 
&0.4814(0.0397) &18.0286(1.4309) 
&0.4735(0.0402) &17.8211(1.4019) 
&0.4492(0.0425) &17.0509(1.3412) &0.1700\\
  \cline{1-10}
  MetaInv-Net (P)
&0.8113(0.0342) &24.5356(1.5088) 
&0.8108(0.0430) &24.5360(1.4997) 
&0.8085(0.0449) &24.5108(1.4967) 
&0.7924(0.0396) &24.3885(1.4589) &0.6457\\
  \cline{1-10}
MetaInv-Net (L) 
&0.4115(0.1277) &11.3040(4.2423) 
&0.3919(0.1048) &10.4769(3.6548) 
&0.3312(0.1021) &8.7673(3.3889) 
&0.1571(0.0771) &2.9709(3.9229) &0.4660\\  
  \cline{1-10}
 MetaInv-Net (H) 
&0.7667(0.0368) &24.4583(1.4115) 
&0.7646(0.0329) &24.4608(1.4124) 
&0.7643(0.0325) &24.4345(1.3935) 
&0.7591(0.0312) &24.2419(1.3279) &0.6075\\
  \cline{1-10}

 \multicolumn{10}{|c|}{\cellcolor{gbygreen}\# views=30} \\
  \cline{1-10}
FBP 
&0.5842(0.0334) &20.4992(1.1063) 
&0.5830(0.0332) &20.4695(1.0957)
&0.5654(0.0278) &19.9718(0.9639)
&0.5152(0.0217) &18.3368(0.8571) &-\\
  \cline{1-10}
 HQS-CG 
&\cellcolor{gbypink}0.9256(0.0122) &\cellcolor{gbypink}29.1793(1.4924) 
&\cellcolor{gbypink}0.9253(0.0122) &\cellcolor{gbypink}29.1695(1.4913) 
&\cellcolor{gbypink}0.9198(0.0127) &\cellcolor{gbypink}28.9884(1.4621) 
&\cellcolor{gbypink}0.8943(0.0159) &\cellcolor{gbypink}28.2461(1.3325) &8.7918\\
  \cline{1-10}
PWLS-CSCGR 
&\cellcolor{gbyblue}0.9068(0.0202) &\cellcolor{gbyblue}29.0386(1.6164) 
&\cellcolor{gbyblue}0.9059(0.0203) &\cellcolor{gbyblue}29.0082(1.6055) 
&0.8976(0.0198) &\cellcolor{gbyblue}28.5308(1.4306) 
&0.8693(0.0203) &26.9269(1.0783) &800.6500\\
    \cline{1-10}
 FBPConv-Net 
&0.5934(0.0561) &21.2793(1.5511) 
&0.5919(0.0556) &21.2517(1.5429) 
&0.5692(0.0477) &20.7918(1.3869) 
&0.5016(0.0294) &19.0348(1.0452) &0.0301\\
    \cline{1-10}
 PFBS-AIR 
&- &- &- &- &- &- &- &- &-\\
    \cline{1-10}
 PD-Net 
&0.7624(0.0532) &24.2574(1.6450) 
&0.7616(0.0533) &24.2145(1.6439) 
&0.7478(0.0555) &23.4603(1.6040) 
&0.6968(0.0624) &21.1395(1.6813) &0.1307\\
  \cline{1-10}
JSR-Net 
&0.6905(0.0478) &23.0992(1.5911)
&0.6898(0.0479) &23.0808(1.5894)
&0.6797(0.0486) &22.7710(1.5446) 
&0.6484(0.0524) &21.5855(1.5021) &0.1743\\
  \cline{1-10}
  MetaInv-Net (P)
&0.9063(0.0264) &28.6021(1.5336)
&0.9050(0.0274) &28.5877(1.5292)
&\cellcolor{gbyblue}0.9024(0.0265) &28.4636(1.5109) 
&\cellcolor{gbyblue}0.8883(0.0258) &\cellcolor{gbyblue}27.9636(1.3941) &0.6904\\
  \cline{1-10}
MetaInv-Net (L) 
&0.7742(0.0889) &23.8102(3.3368) 
&0.7677(0.0760) &23.4768(3.0160) 
&0.7454(0.0756) &22.9318(2.9937) 
&0.5789(0.0915) &18.2322(2.9399) &0.4945\\
  \cline{1-10}
MetaInv-Net (H)
&0.8819(0.0300) &28.3993(1.6209) 
&0.8805(0.0270) &28.4123(1.6188) 
&0.8800(0.0257) &28.3889(1.5736) 
&0.8715(0.0225) &27.9061(1.3718) &0.6238\\
  \cline{1-10}

 \multicolumn{10}{|c|}{\cellcolor{gbygreen}\# views=60} \\
  \cline{1-10}
 FBP
&0.7296(0.0282) &24.0254(0.8833) 
&0.7254(0.0293) &23.9902(0.8733) 
&0.7113(0.0218) &23.4161(0.7709)
&0.6562(0.0171) &21.6016(0.8183) &-\\
  \cline{1-10}
 HQS-CG 
&\cellcolor{gbyblue}0.9756(0.0035) &34.3479(0.9856) 
&\cellcolor{gbypink}0.9756(0.0035) &34.3411(0.9858) 
&\cellcolor{gbypink}0.9738(0.0037) &\cellcolor{gbyblue}34.1746(0.9730) 
&\cellcolor{gbypink}0.9641(0.0062) &\cellcolor{gbypink}33.4346(0.9200) &8.9038\\
  \cline{1-10}
PWLS-CSCGR 
&0.9687(0.0098) &\cellcolor{gbyblue}34.8211(1.8090) 
&0.9672(0.0099) &\cellcolor{gbyblue}34.6028(1.7249) 
&0.9501(0.0112) &32.0859(1.2265) 
&0.8864(0.0215) &26.9394(1.3748) &1073.7100\\
    \cline{1-10}
 FBPConv-Net 
&0.8009(0.0479) &27.4394(1.6627)
&0.7997(0.0477) &27.4002(1.6496)
&0.7804(0.0428) &26.6932(1.4159) 
&0.7086(0.0281) &24.1474(1.0237) &0.0309\\
    \cline{1-10}
 PFBS-AIR 
&- &- &- &- &- &- &- &- &-\\
    \cline{1-10}
 PD-Net 
&0.9092(0.0261) &30.6645(1.6190)
&0.9083(0.0262) &30.5828(1.6118)
&0.8930(0.0298) &29.2236(1.6061) 
&0.8446(0.0433) &25.6502(1.9381) &0.1415\\
  \cline{1-10}
  JSR-Net
&0.8712(0.0267) &29.1897(1.4972) 
&0.8705(0.0267) &29.1524(1.4904) 
&0.8592(0.0282) &28.5108(1.4321) 
&0.8236(0.0367) &26.4357(1.6568) &0.2018\\
  \cline{1-10}
  MetaInv-Net (P)
&\cellcolor{gbypink}0.9764(0.0045) &34.1475(1.3494)
&\cellcolor{gbyblue}0.9706(0.0068) &34.1361(1.3520) 
&\cellcolor{gbyblue}0.9677(0.0077) &33.8695(1.4005) 
&\cellcolor{gbyblue}0.9525(0.0116) &32.4288(1.2921) &0.7192\\
  \cline{1-10}
MetaInv-Net (L)  
&0.9624(0.0118) &34.2580(1.6995) 
&0.9628(0.0106) &34.2810(1.5937) 
&0.9542(0.0136) &33.3820(1.6055) 
&0.9107(0.0255) &30.3115(1.6876) &0.4647\\
  \cline{1-10}
  MetaInv-Net (H) 
&0.9687(0.0114) &\cellcolor{gbypink}35.3443(1.6520) 
&0.9668(0.0102) &\cellcolor{gbypink}35.3593(1.6374) 
&0.9656(0.0079) &\cellcolor{gbypink}35.0658(1.3024) 
&0.9447(0.0080) &\cellcolor{gbyblue}32.5760(1.1662) &0.6442\\
  \cline{1-10}

 \multicolumn{10}{|c|}{\cellcolor{gbygreen}\# views=120} \\
  \cline{1-10}
FBP
&0.8612(0.0232) &27.2495(0.7731) 
&0.8602(0.0229) &27.2110(0.7686) 
&0.8446(0.0179) &26.5897(0.7370) 
&0.7925(0.0160) &24.6635(0.9180) &-\\
  \cline{1-10}
 HQS-CG
&0.9908(0.0013) &39.3226(0.8928) 
&0.9907(0.0013) &39.2890(0.8877) 
&\cellcolor{gbypink}0.9882(0.0017) &\cellcolor{gbyblue}38.6940(0.8084) 
&\cellcolor{gbypink}0.9710(0.0084) &\cellcolor{gbypink}36.2080(0.9603) &4.8941\\
  \cline{1-10}
PWLS-CSCGR 
&\cellcolor{gbyblue}0.9916(0.0031) &40.8751(1.3077) 
&0.9895(0.0034) &39.5576(1.2323)
&0.9549(0.0119) &31.5463(1.6547)
&0.8589(0.0325) &24.1316(1.7994) &1580.4300\\
    \cline{1-10}
 FBPConv-Net 
&0.9628(0.0169) &35.3363(1.3810) 
&0.9625(0.0169) &35.2898(1.3773) 
&0.9533(0.0166) &34.2064(1.2560) 
&0.9038(0.0177) &30.1637(1.2320) &0.0313\\
    \cline{1-10}
 PFBS-AIR 
&0.1867(0.0691) &2.2811(4.6148) 
&0.1861(0.0686) &2.2601(4.5781) 
&0.1723(0.0578) &1.6445(4.2386) 
&0.1321(0.0346) &-0.6975(3.2446) &0.3484\\
    \cline{1-10}
 PD-Net
&0.9847(0.0055) &38.3296(1.4376) 
&0.9840(0.0056) &38.1844(1.4212) 
&0.9711(0.0099) &35.5081(1.4918) 
&0.9245(0.0269) &29.6941(2.1431) &0.1756\\
  \cline{1-10}
  JSR-Net
&0.9790(0.0070) &36.9827(1.1474) 
&0.9784(0.0071) &36.8703(1.1233) 
&0.9692(0.0096) &35.2859(1.2078) 
&0.9375(0.0224) &31.6307(2.0329) &0.2443\\
  \cline{1-10}
  MetaInv-Net (P)
&0.9882(0.0017) &36.1685(1.2710)
&0.9863(0.0038) &36.1539(1.2842) 
&0.9836(0.0043) &35.9100(1.2658)
&\cellcolor{gbyblue}0.9646(0.0132) &34.1008(1.2977) &0.7537\\
  \cline{1-10}
MetaInv-Net (L)
&\cellcolor{gbyblue}0.9916(0.0027) &\cellcolor{gbyblue}40.9295(1.0356) 
&\cellcolor{gbyblue}0.9913(0.0022) &\cellcolor{gbyblue}40.7613(1.0097) 
&0.9822(0.0126) &38.2993(1.0203) 
&0.9497(0.0150) &33.2050(1.4287) &0.5475\\
   \cline{1-10}
MetaInv-Net (H)
&\cellcolor{gbypink}0.9924(0.0012) &\cellcolor{gbypink}42.0034(1.0408) 
&\cellcolor{gbypink}0.9921(0.0012) &\cellcolor{gbypink}41.8687(1.0025) 
&\cellcolor{gbyblue}0.9862(0.0029) &\cellcolor{gbypink}39.4843(0.9146) 
&0.9593(0.0113) &\cellcolor{gbyblue}34.1313(1.5777) &0.7143\\
   \cline{1-10}

\multicolumn{10}{|c|}{\cellcolor{gbygreen}\# views=180} \\
  \cline{1-10}
FBP
&0.9232(0.0156) &28.9723(0.5548)
&0.9223(0.0153) &28.9340(0.5563) 
&0.9074(0.0111) &28.3004(0.6208) 
&0.8581(0.0131) &26.3417(0.9302) &-\\
  \cline{1-10}
 HQS-CG
&0.9942(0.0008) &42.0735(0.9346) 
&0.9938(0.0008) &41.8752(0.8826) 
&0.9858(0.0029) &39.2351(0.7430) 
&0.9393(0.0182) &33.2973(1.5542) &5.3342\\
  \cline{1-10}
PWLS-CSCGR 
&0.9927(0.0020) &40.4991(1.0592) 
&0.9909(0.0024) &39.3250(1.3162)
&0.9460(0.0157) &29.7683(1.8690)
&0.8429(0.0396) &22.1592(2.0043) &2064.5500\\
    \cline{1-10}
 FBPConv-Net 
&0.9867(0.0018) &37.5854(0.4184) 
&0.9867(0.0019) &37.5690(0.4247) 
&0.9847(0.0021) &37.0423(0.5287)
&0.9607(0.0087) &33.4007(1.2857) &0.0304\\
    \cline{1-10}
 PFBS-AIR 
&0.9927(0.0010) &41.9570(0.8870) 
&0.9918(0.0011) &41.6115(0.7966) 
&0.9779(0.0044) &37.9591(0.8004)
&0.9289(0.0164) &32.5821(1.2432) &0.4165\\
    \cline{1-10}
  PD-Net
&\cellcolor{gbyblue}0.9950(0.0012) &41.0472(1.2566) 
&0.9946(0.0013) &40.9784(1.2237) 
&0.9865(0.0050) &38.3362(1.3871) 
&0.9498(0.0205) &31.6755(2.2069) &0.2067\\
  \cline{1-10}
JSR-Net
&0.9938(0.0012) &40.1419(0.5713)
&0.9934(0.0013) &39.9953(0.5065) 
&0.9864(0.0047) &38.0498(0.9324) 
&0.9603(0.0170) &33.8541(2.1250) &0.2855\\
  \cline{1-10}
  MetaInv-Net (P)
&0.9905(0.0014) &36.7996(1.3177)
&0.9895(0.0073) &36.7971(1.3183)
&0.9879(0.0017) &36.5808(1.3171) 
&\cellcolor{gbypink}0.9741(0.0068) &\cellcolor{gbypink}34.9201(1.3093) &0.8479\\
  \cline{1-10}
MetaInv-Net (L) 
&\cellcolor{gbypink}0.9954(0.0010) &\cellcolor{gbyblue}43.0075(0.9122) 
&\cellcolor{gbypink}0.9950(0.0011) &\cellcolor{gbyblue}42.7803(0.8787) 
&\cellcolor{gbyblue}0.9887(0.0030) &\cellcolor{gbyblue}39.6737(1.0566) 
&0.9569(0.0138) &33.9335(1.5614) &0.6505\\
   \cline{1-10}
 MetaInv-Net (H)
&\cellcolor{gbyblue}0.9950(0.0012) &\cellcolor{gbypink}43.7750(0.9621) 
&\cellcolor{gbyblue}0.9947(0.0012) &\cellcolor{gbypink}43.5942(0.9356) 
&\cellcolor{gbypink}0.9891(0.0029) &\cellcolor{gbypink}40.7332(1.0429)
&\cellcolor{gbyblue}0.9643(0.0106) &\cellcolor{gbyblue}34.8208(1.6459) &0.7888\\
   \cline{1-9}
\hline
\end{tabular}
\end{table*}

\begin{figure*}[ht]
  \centering
  \begin{tabular}{@{}p{0.12\linewidth}@{}p{0.12\linewidth}@{}p{0.12\linewidth}@{}p{0.12\linewidth}@{}p{0.12\linewidth}@{}p{0.12\linewidth}@{}p{0.12\linewidth}@{}p{0.12\linewidth}}
    \subfigimg[width=\linewidth]{}{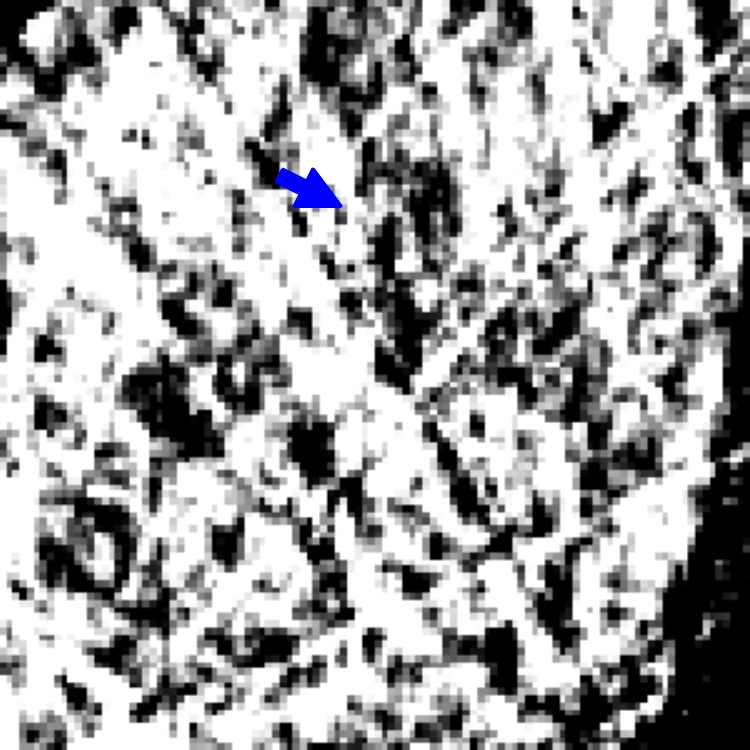} &
    \subfigimg[width=\linewidth]{}{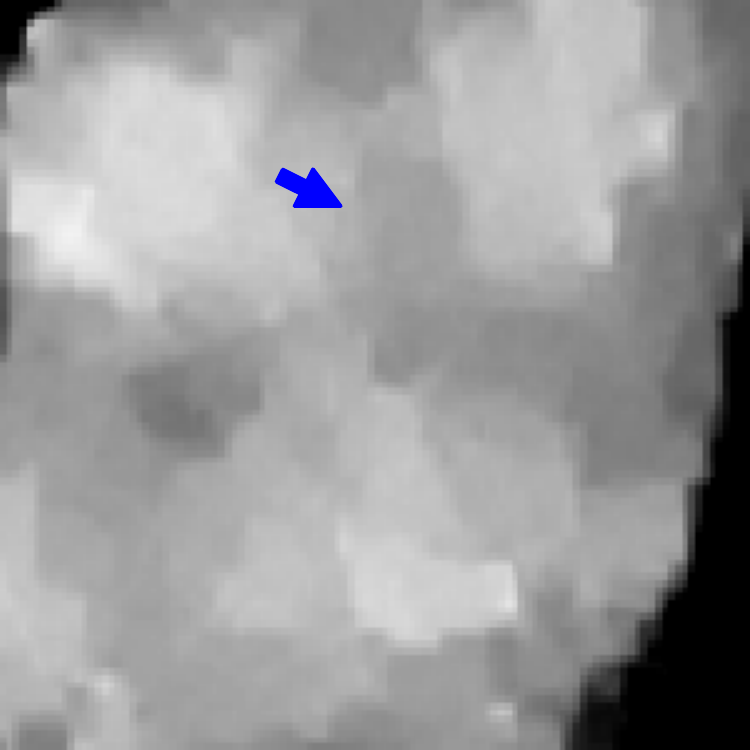} &
    \subfigimg[width=\linewidth]{}{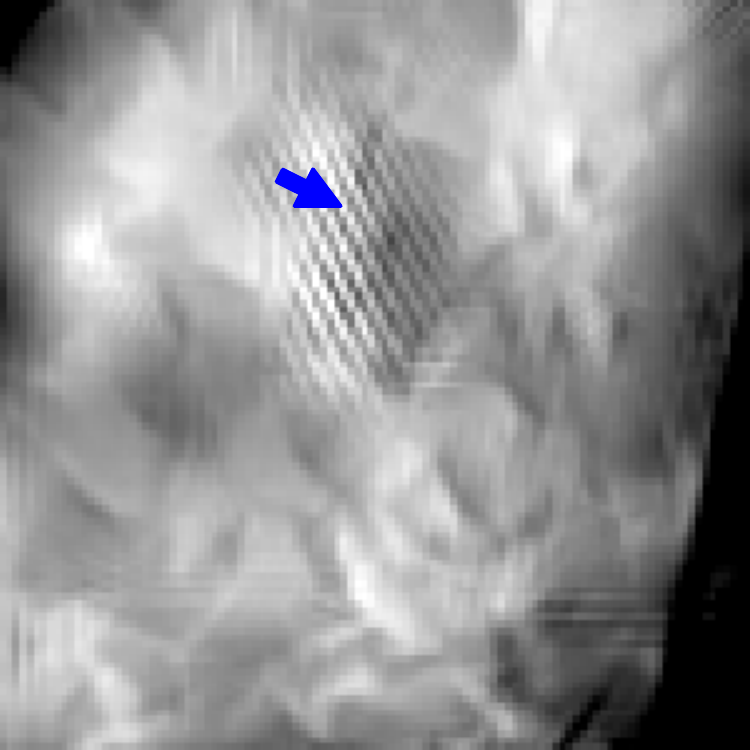} &
    \subfigimg[width=\linewidth]{}{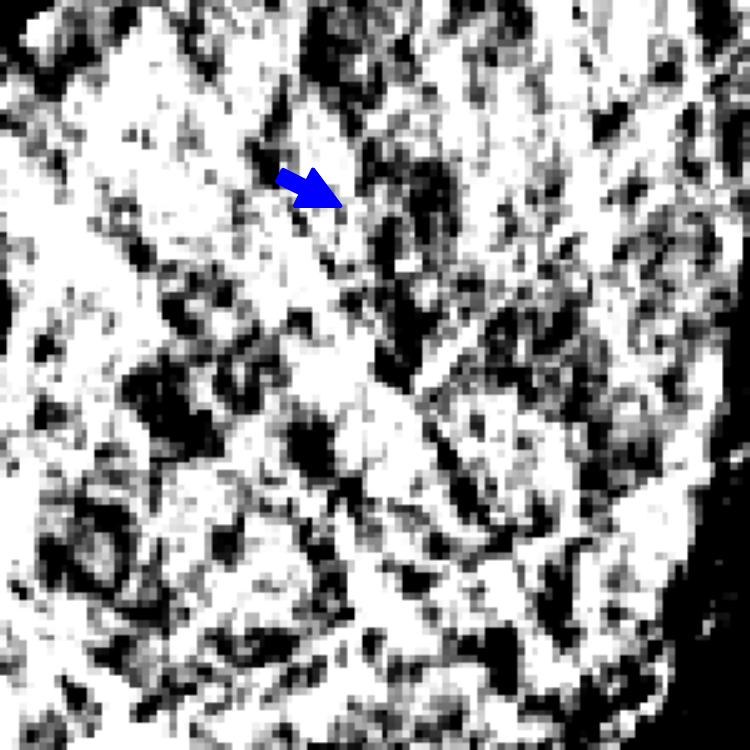} &
    \subfigimg[width=\linewidth]{}{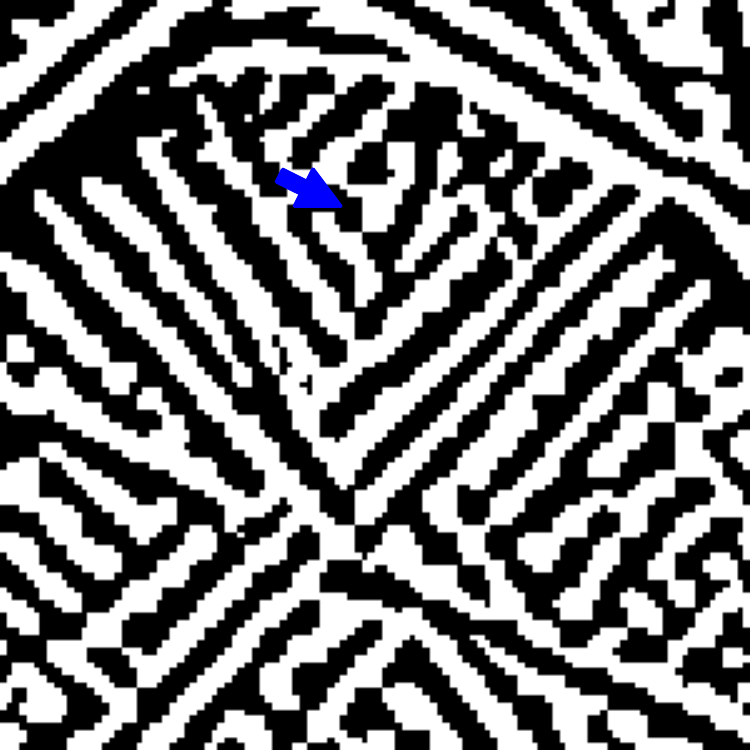} &
    \subfigimg[width=\linewidth]{}{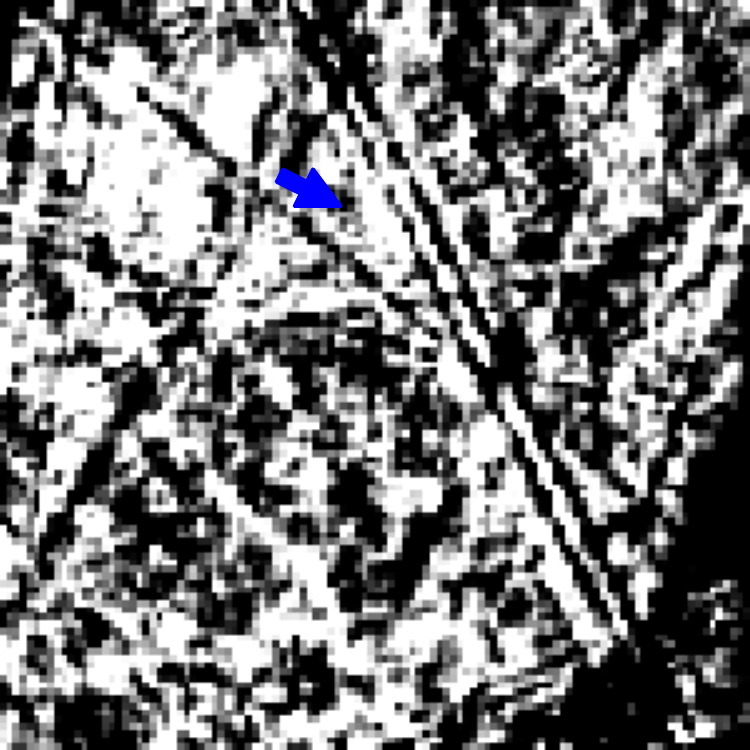} &
    \subfigimg[width=\linewidth]{}{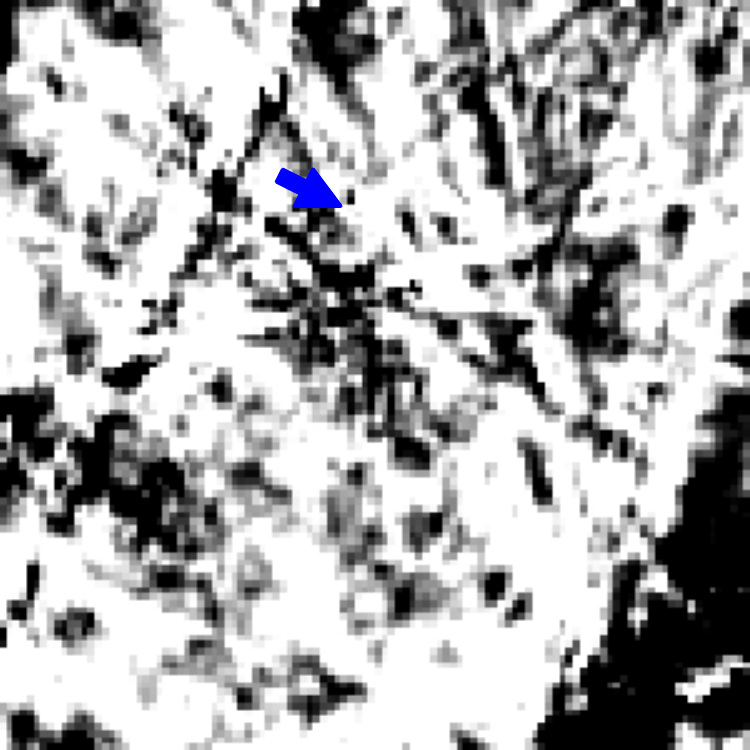} &
    \subfigimg[width=\linewidth]{}{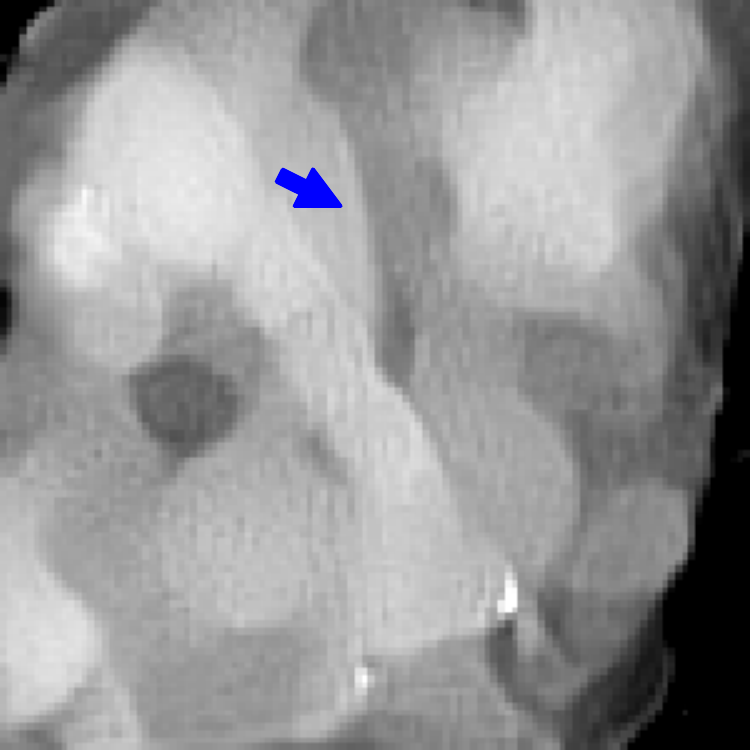} \\

    \subfigimg[width=\linewidth]{}{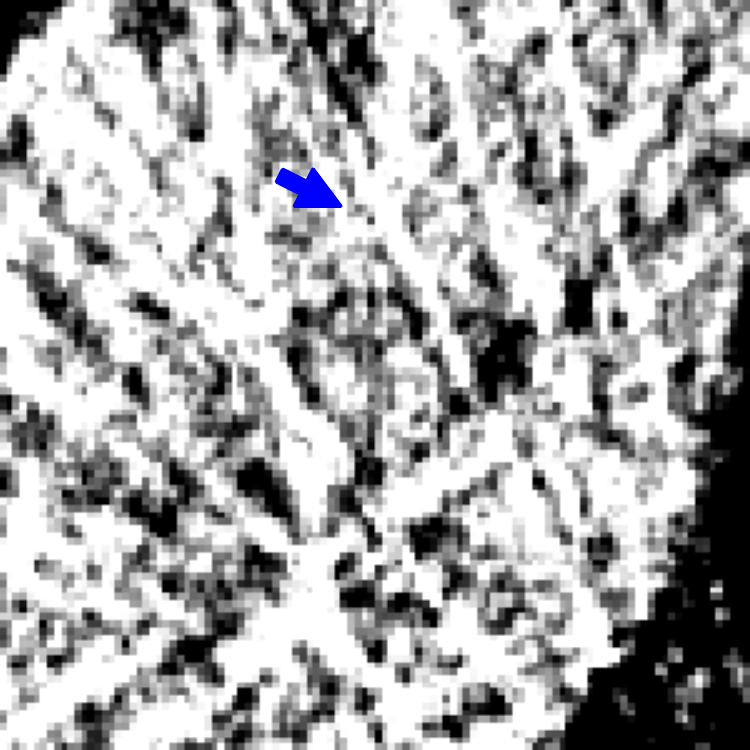} &
    \subfigimg[width=\linewidth]{}{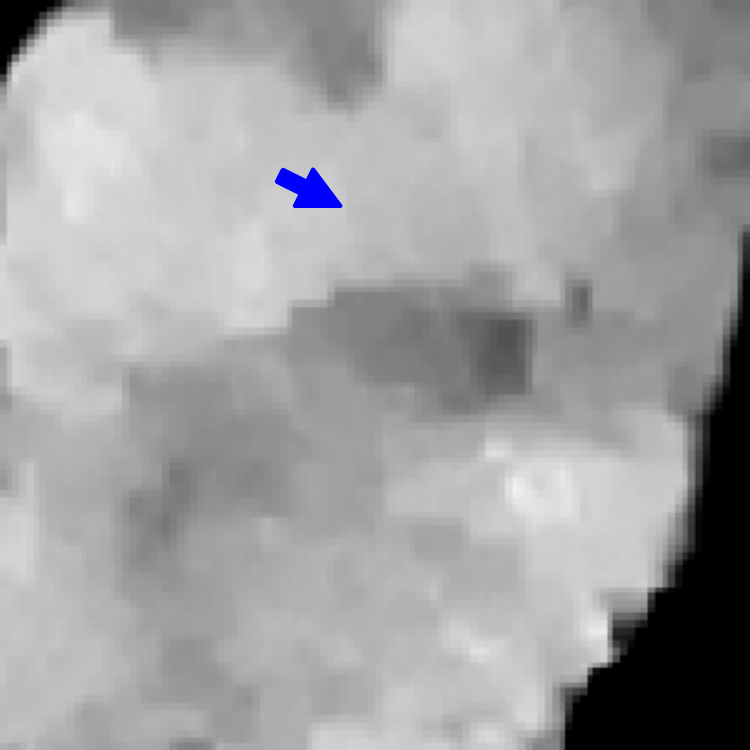} &
    \subfigimg[width=\linewidth]{}{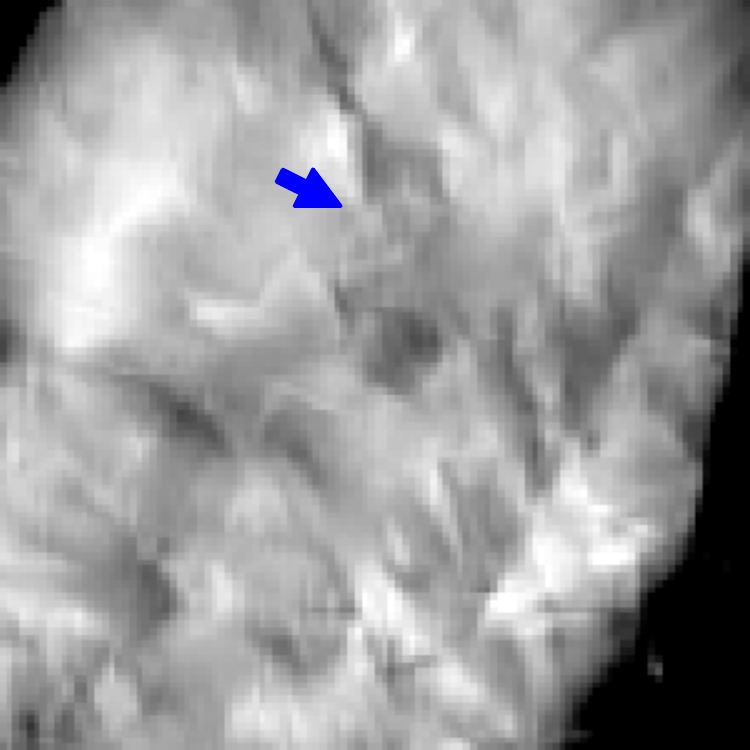} &
    \subfigimg[width=\linewidth]{}{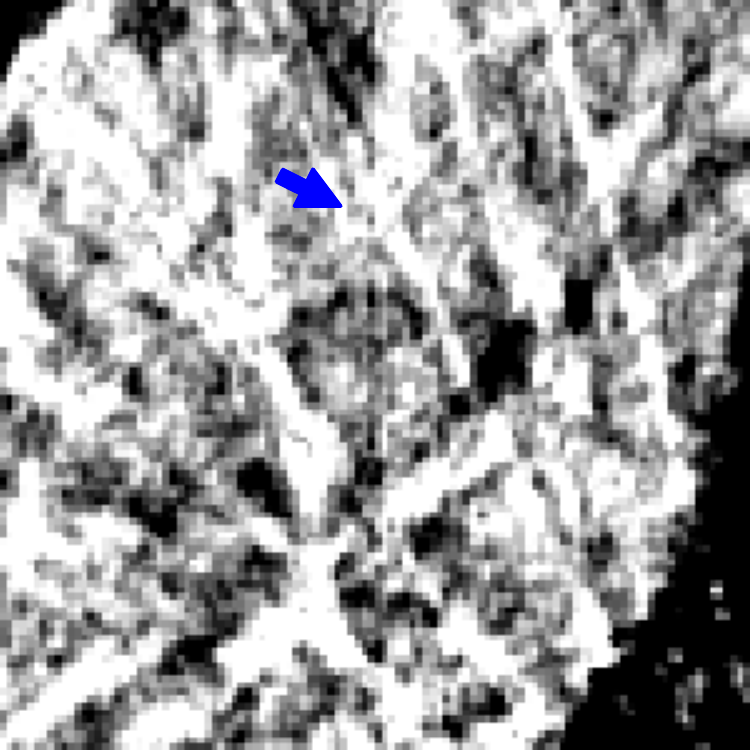} &
    \subfigimg[width=\linewidth]{}{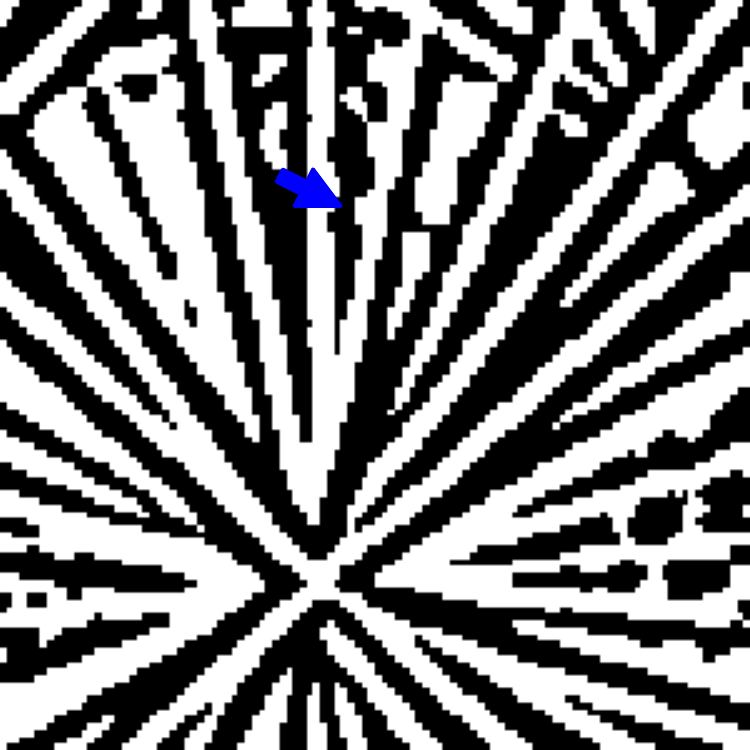} &
    \subfigimg[width=\linewidth]{}{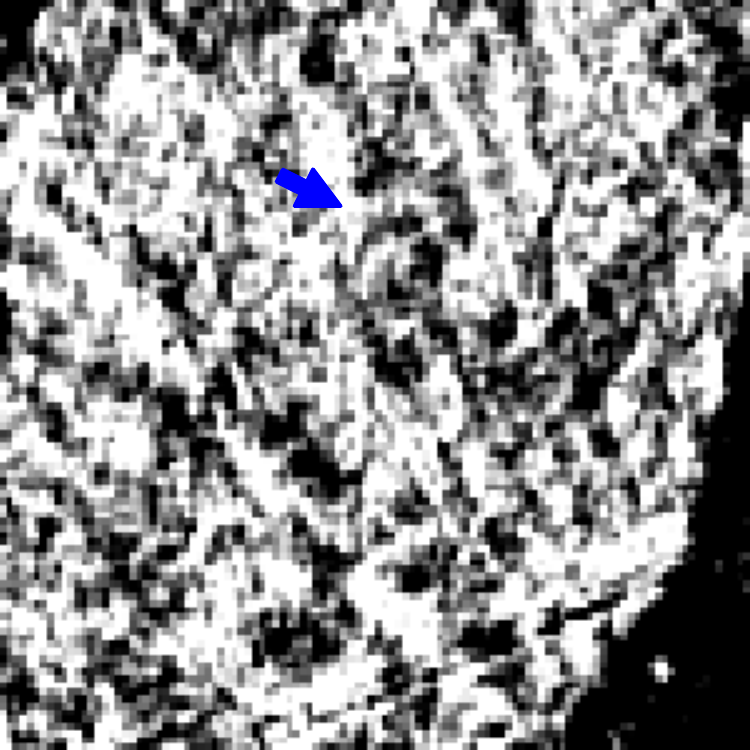} &
    \subfigimg[width=\linewidth]{}{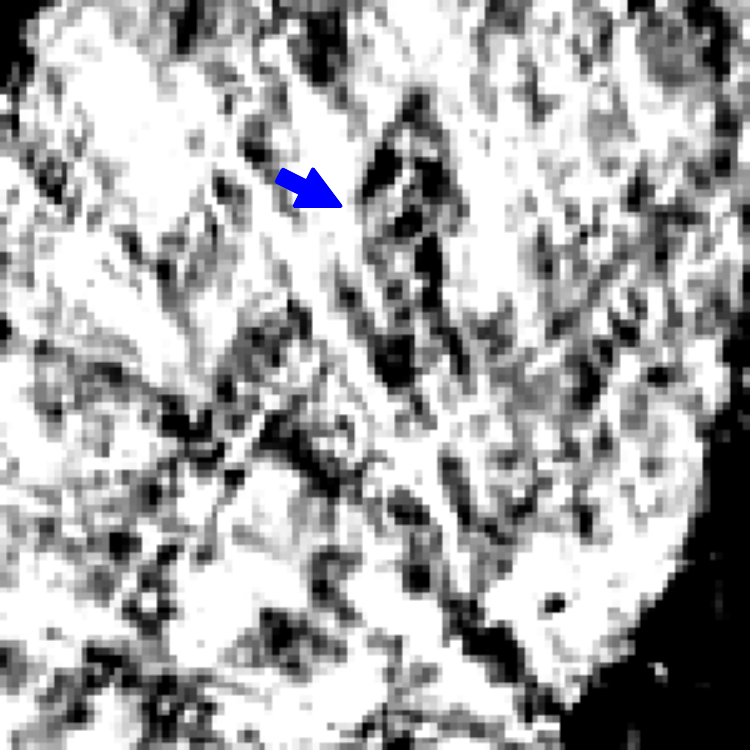} &
    \subfigimg[width=\linewidth]{}{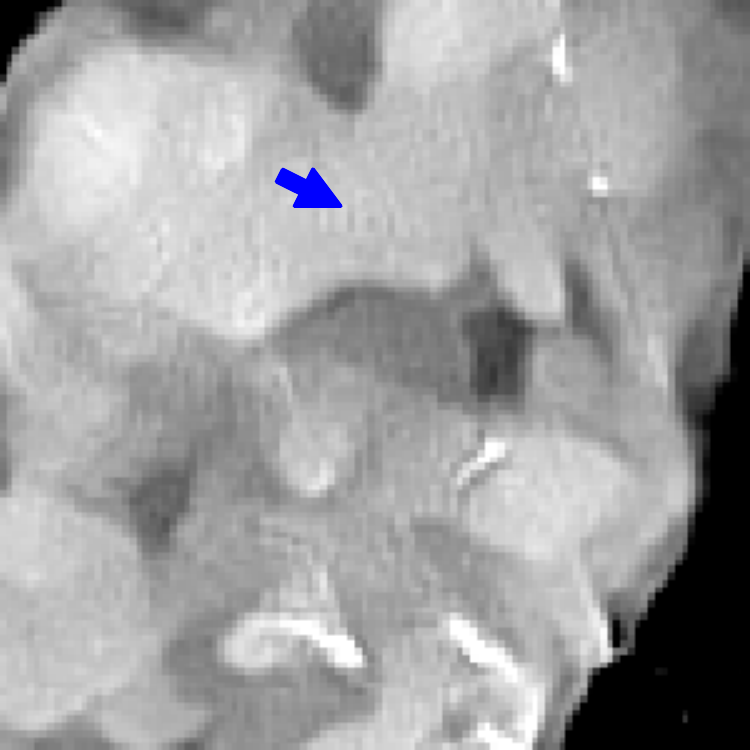} \\
    
    \subfigimg[width=\linewidth]{}{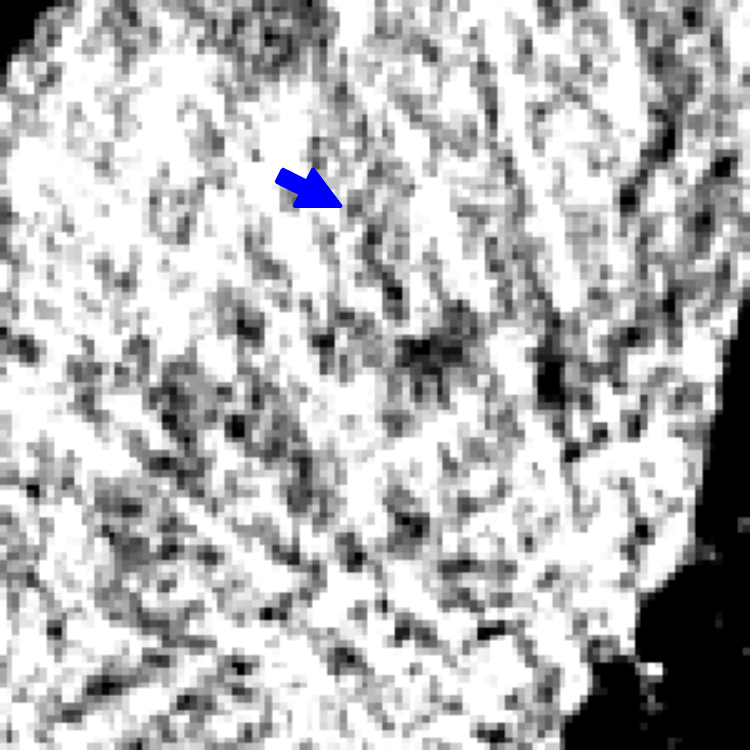} &
    \subfigimg[width=\linewidth]{}{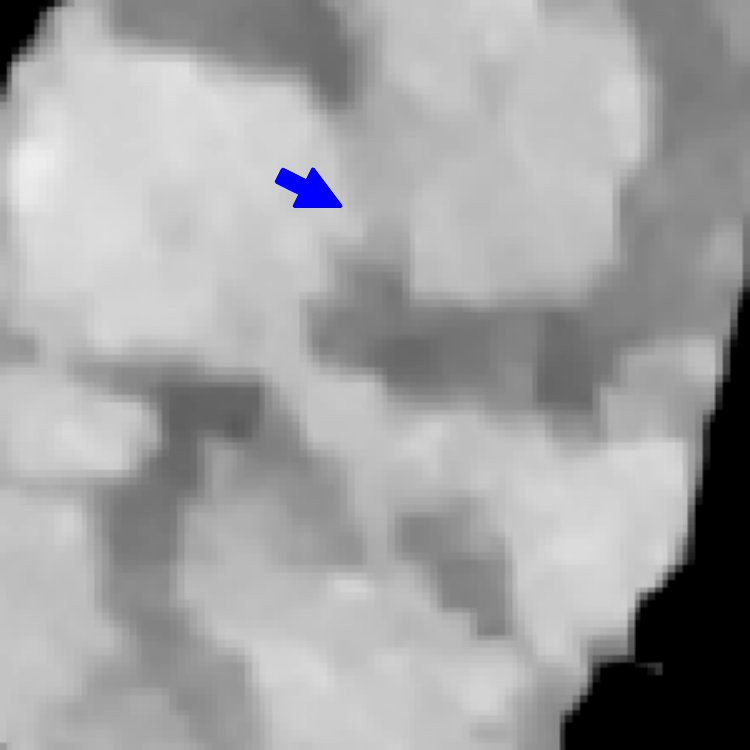} &
    \subfigimg[width=\linewidth]{}{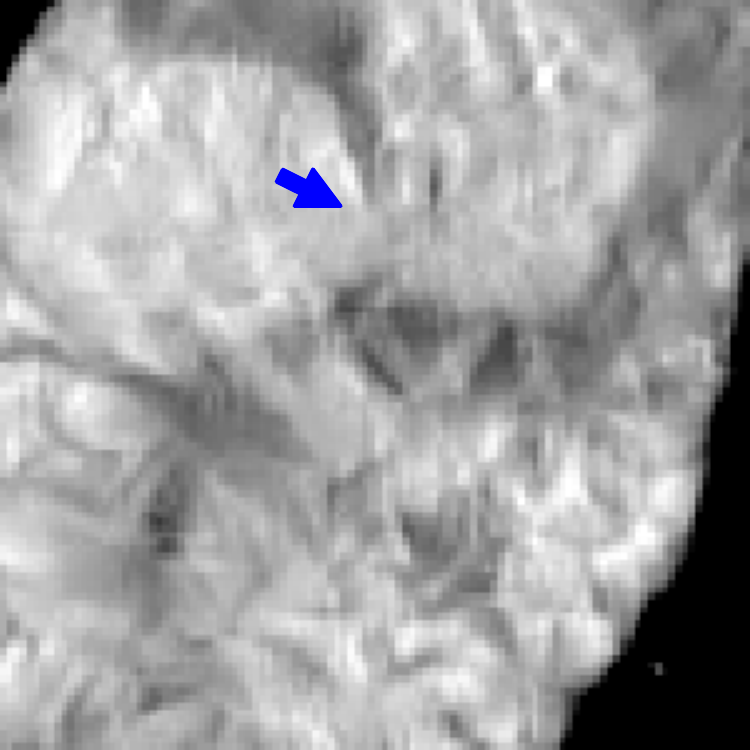} &
    \subfigimg[width=\linewidth]{}{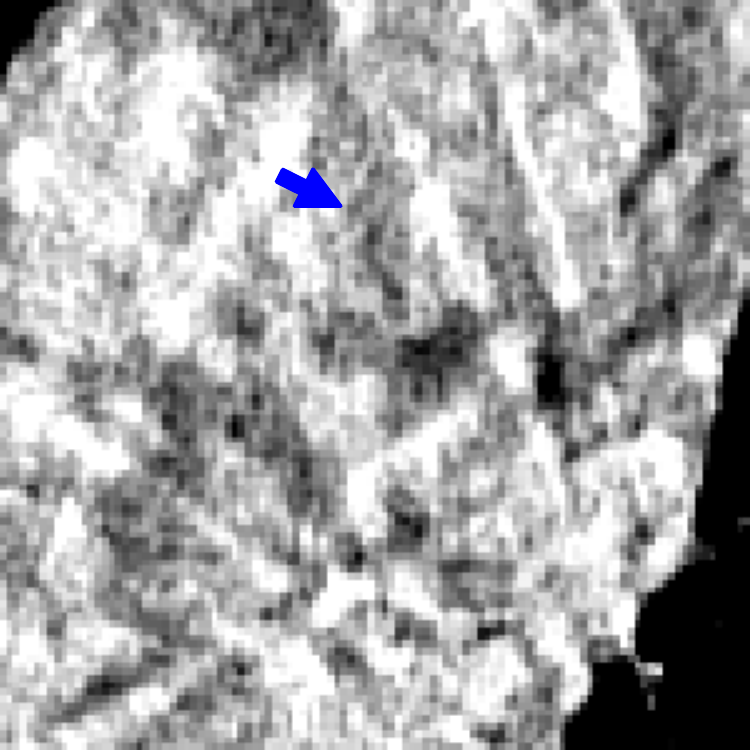} &
    \subfigimg[width=\linewidth]{}{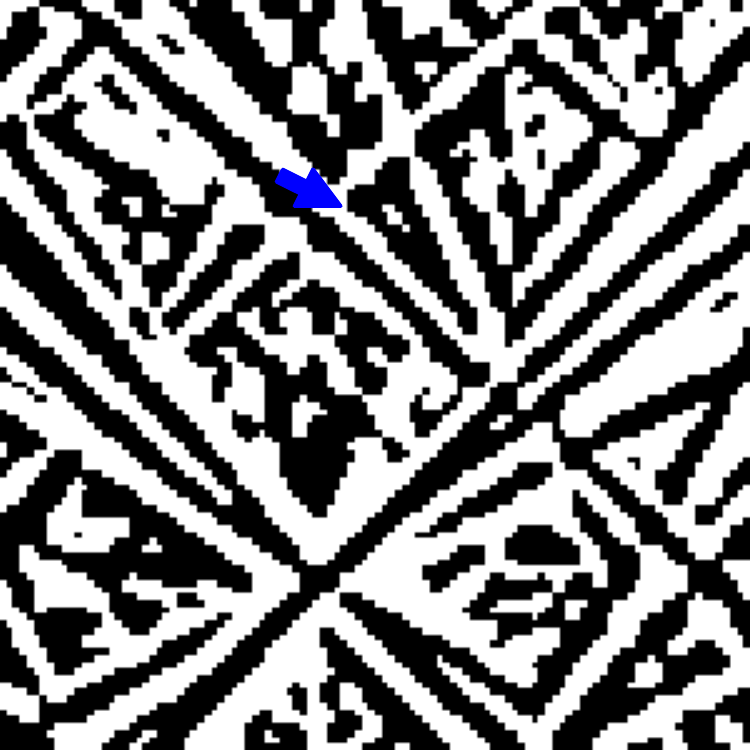} &
    \subfigimg[width=\linewidth]{}{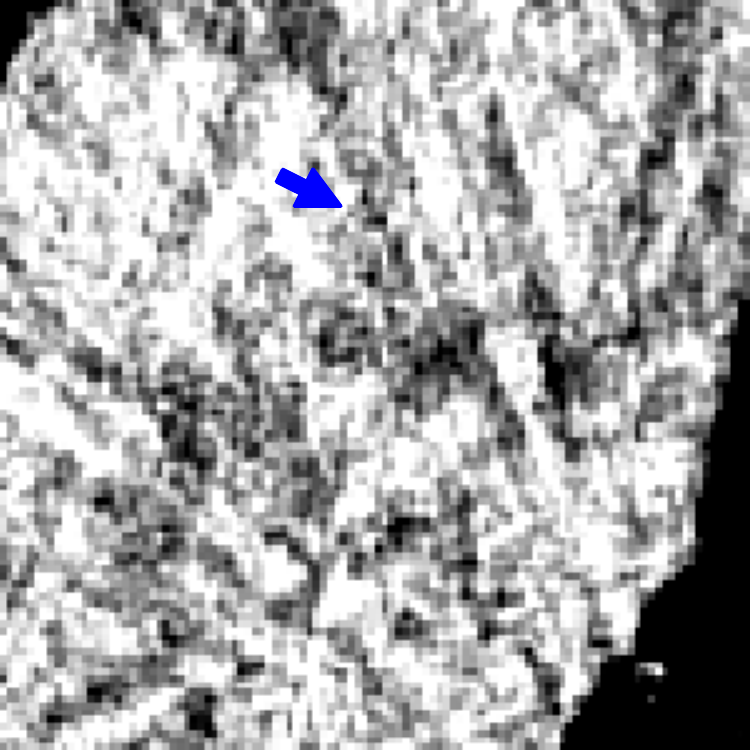} &
    \subfigimg[width=\linewidth]{}{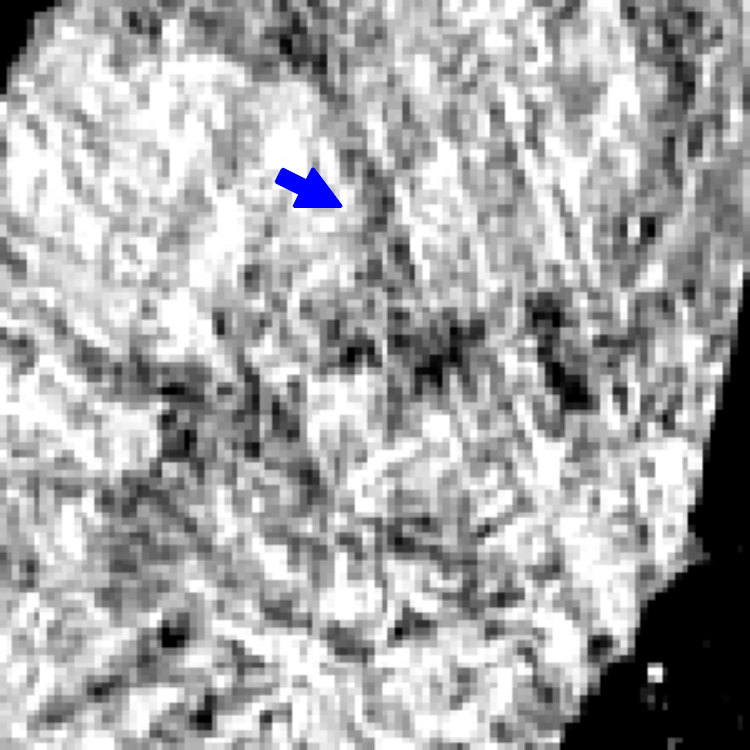} &
    \subfigimg[width=\linewidth]{}{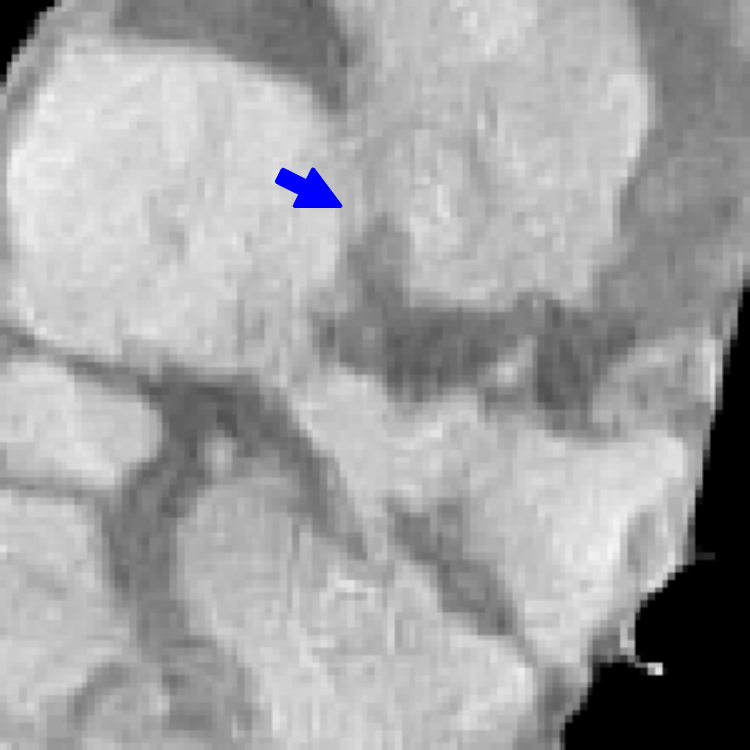} \\

    \subfigimg[width=\linewidth]{}{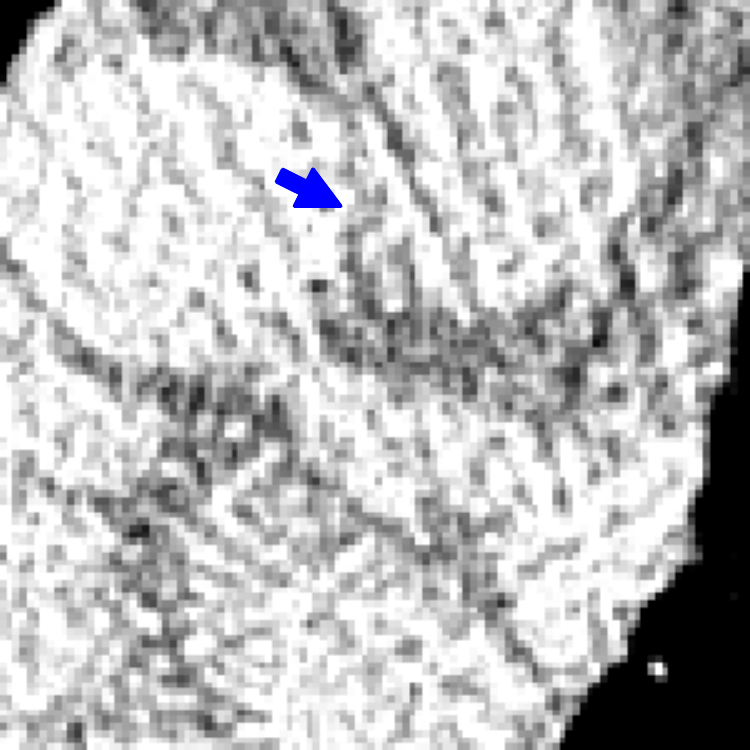} &
    \subfigimg[width=\linewidth]{}{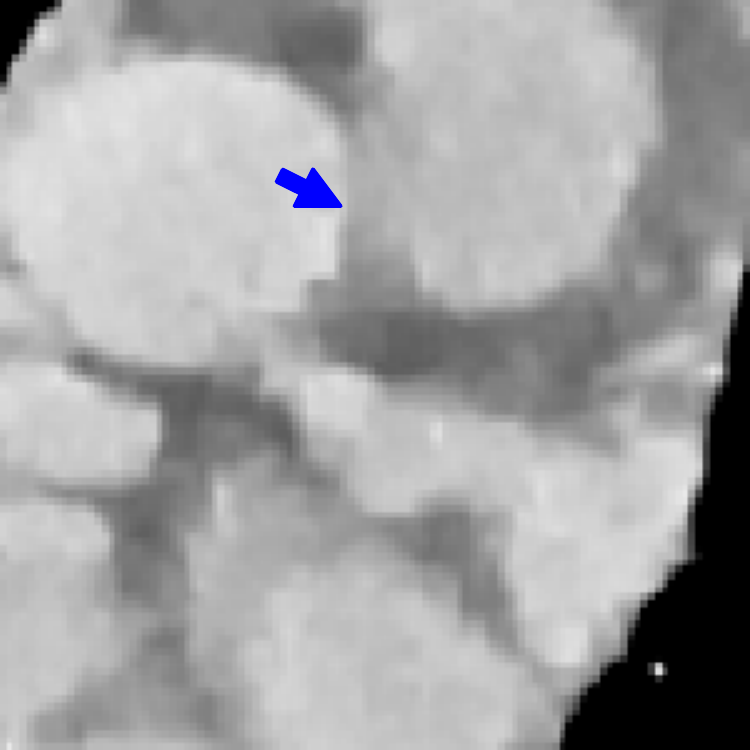} &
    \subfigimg[width=\linewidth]{}{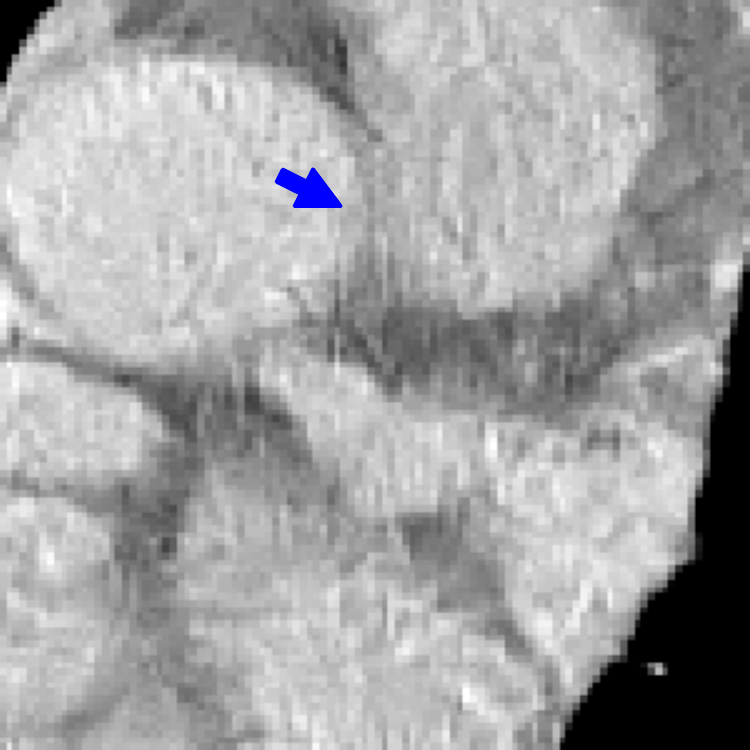} &
    \subfigimg[width=\linewidth]{}{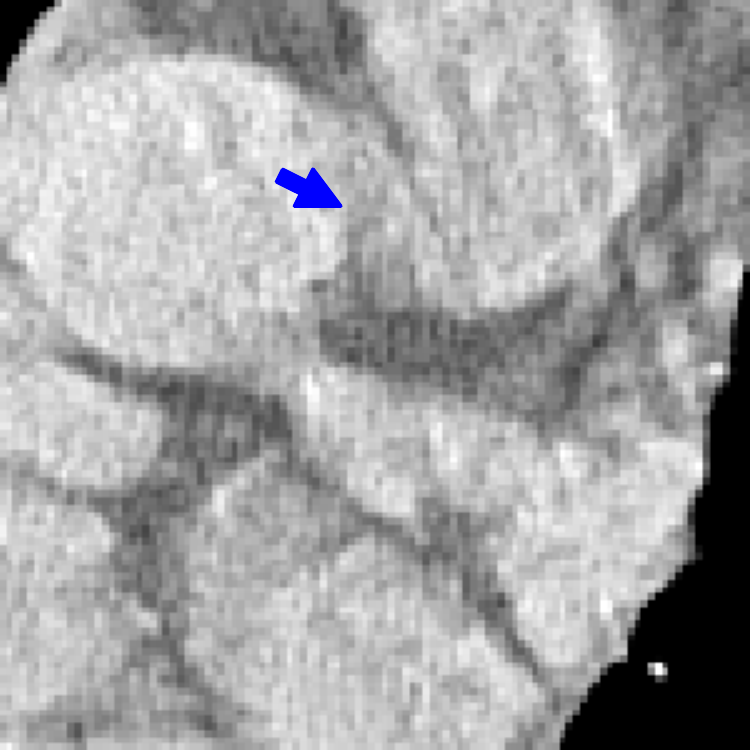} &
    \subfigimg[width=\linewidth]{}{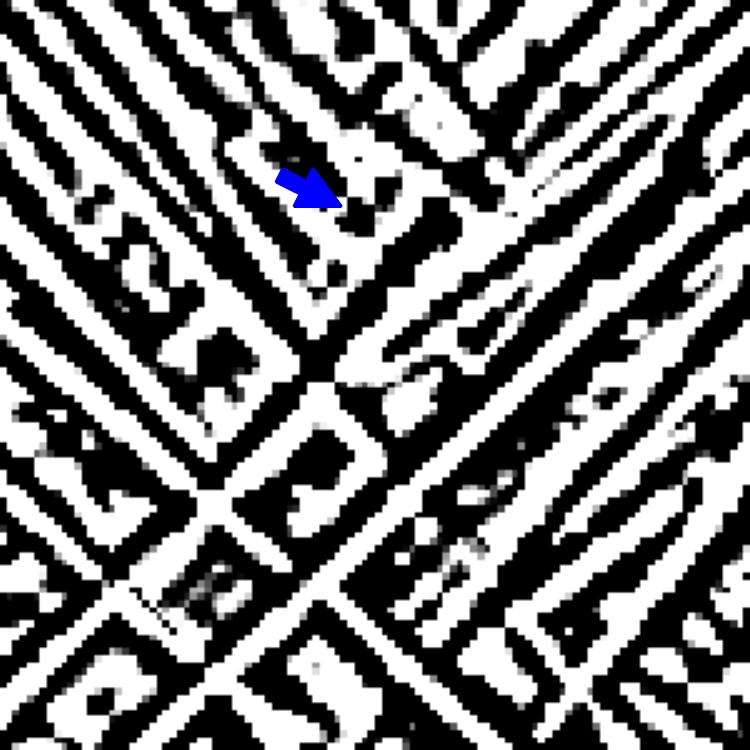} &
    \subfigimg[width=\linewidth]{}{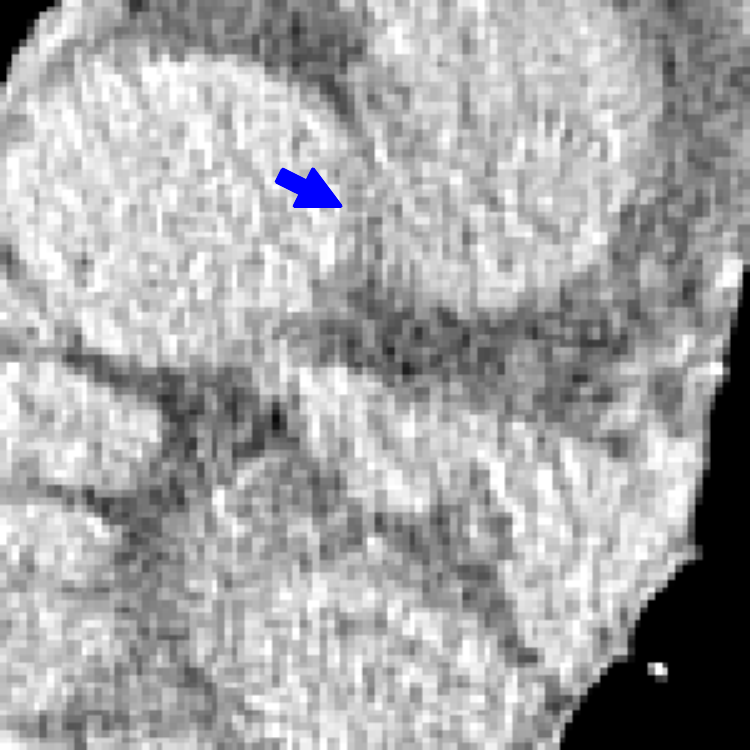} &
    \subfigimg[width=\linewidth]{}{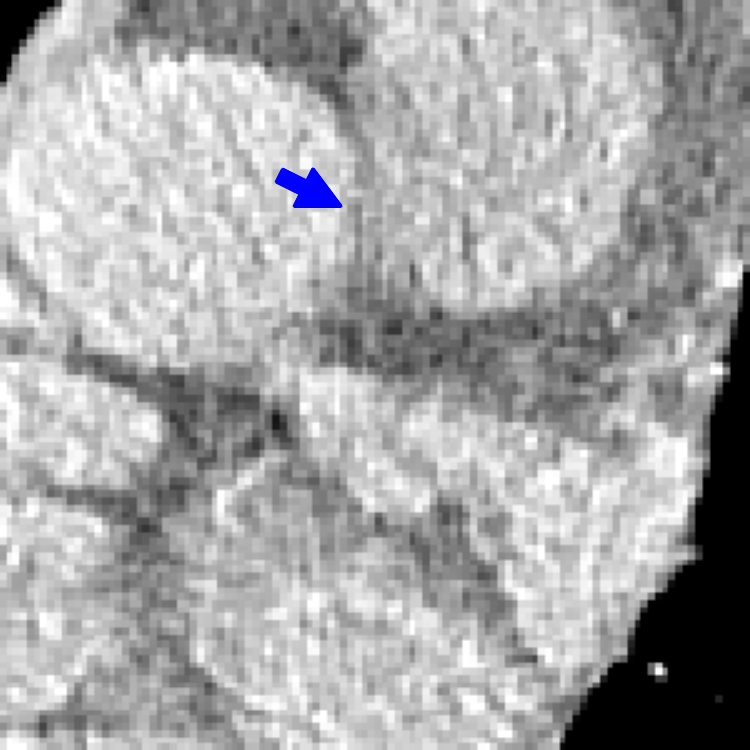} &
    \subfigimg[width=\linewidth]{}{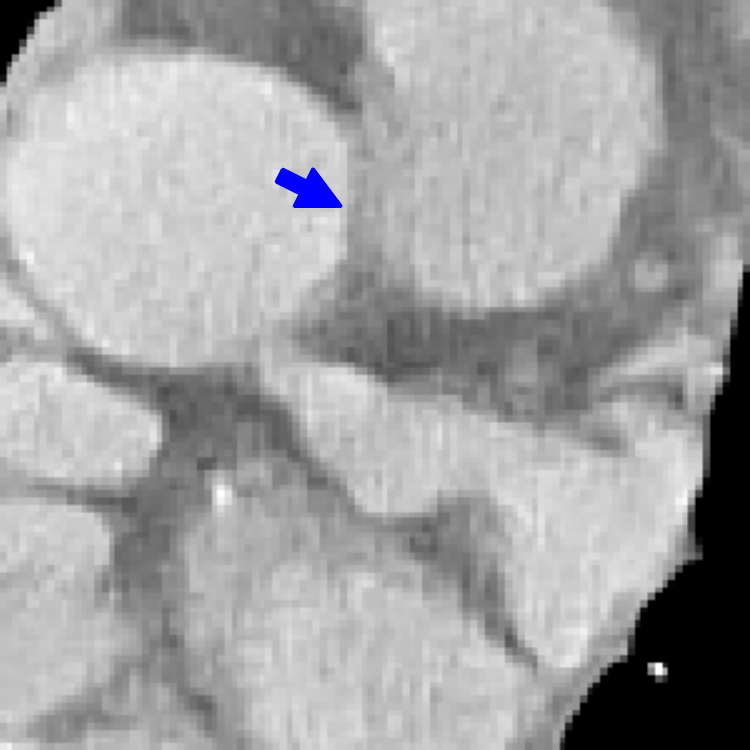} \\

    \subfigimg[width=\linewidth]{}{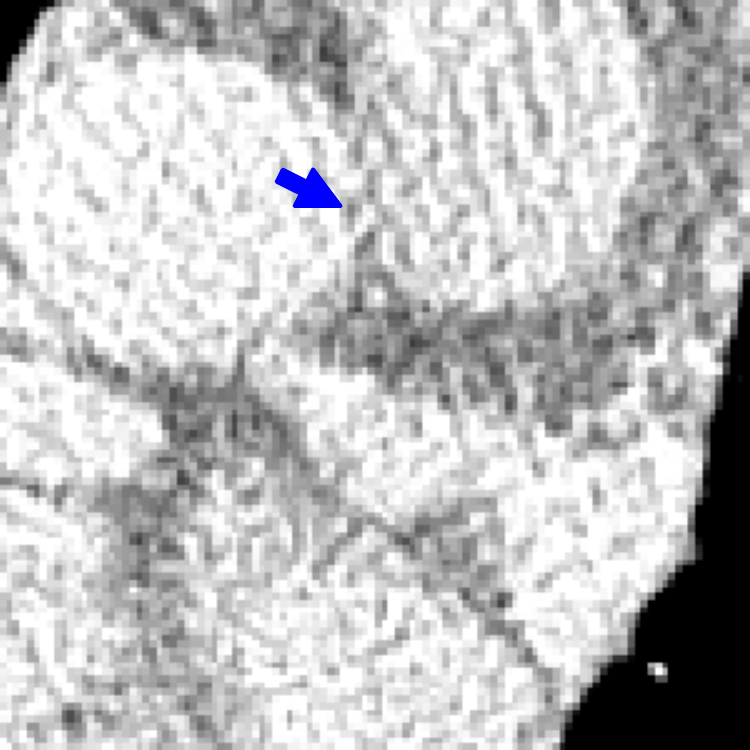}&
    \subfigimg[width=\linewidth]{}{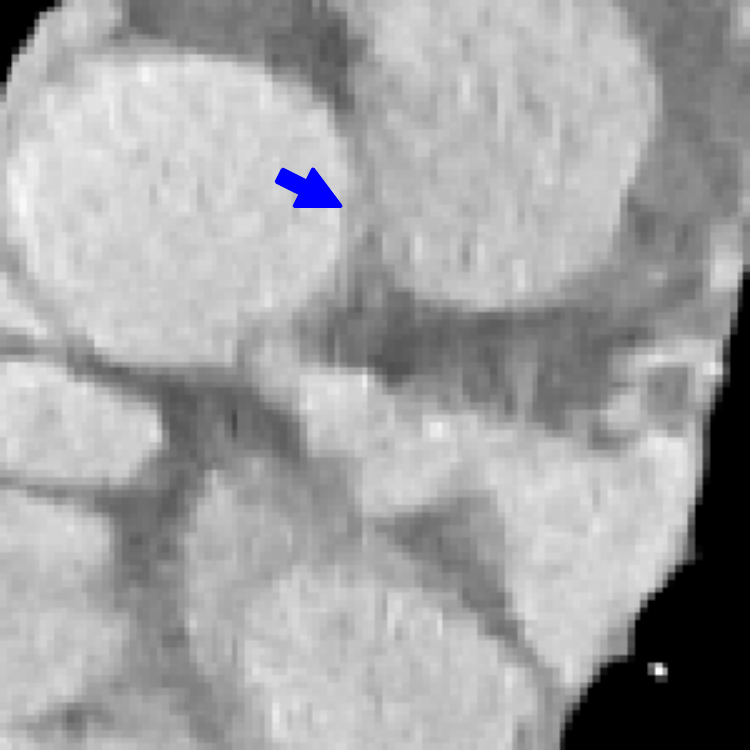} &
    \subfigimg[width=\linewidth]{}{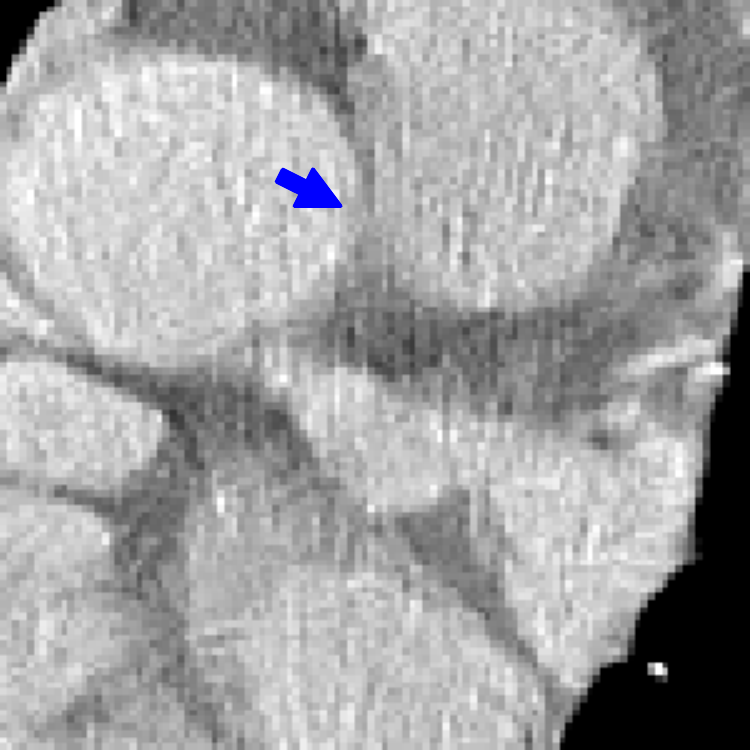} &
    \subfigimg[width=\linewidth]{}{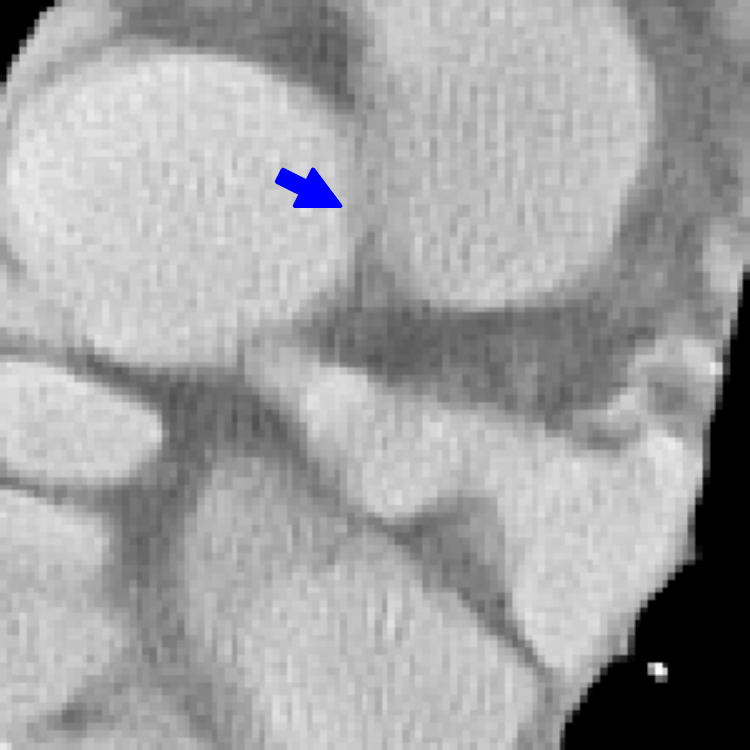}&
    \subfigimg[width=\linewidth]{}{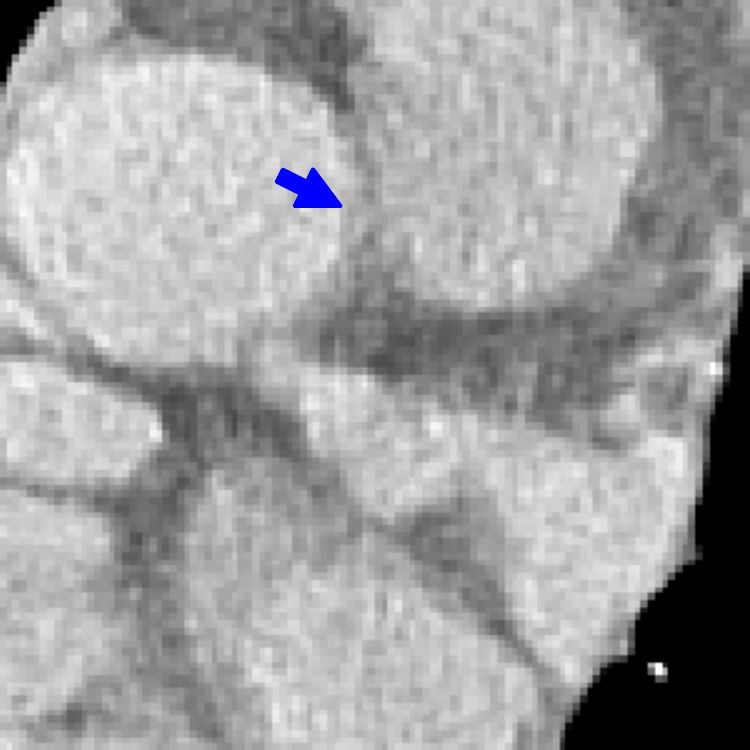}&
    \subfigimg[width=\linewidth]{}{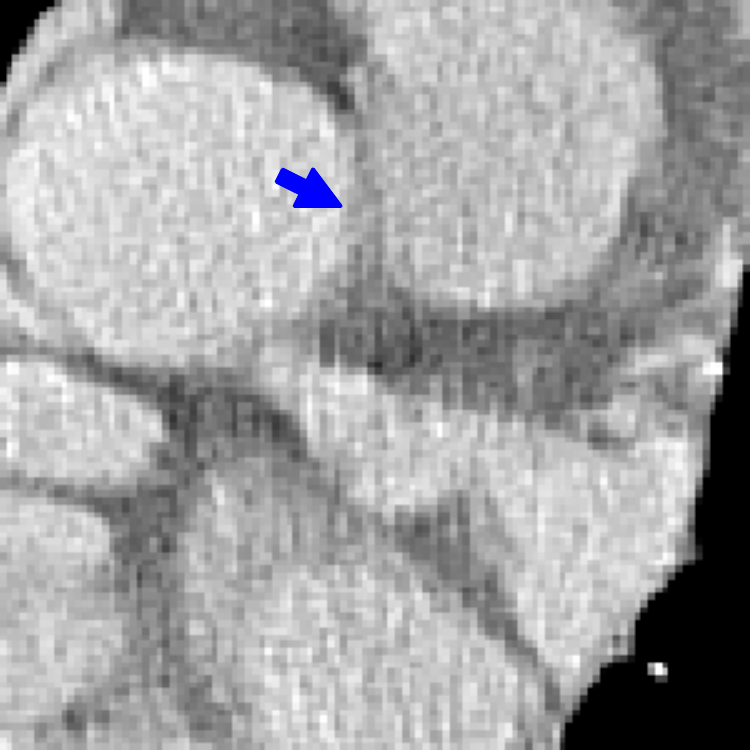} &
    \subfigimg[width=\linewidth]{}{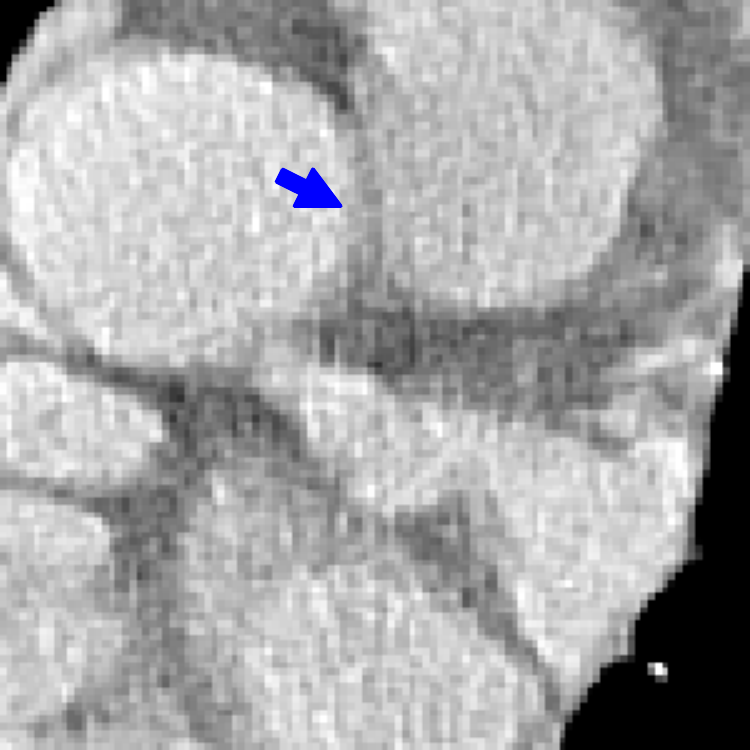} &
    \subfigimg[width=\linewidth]{}{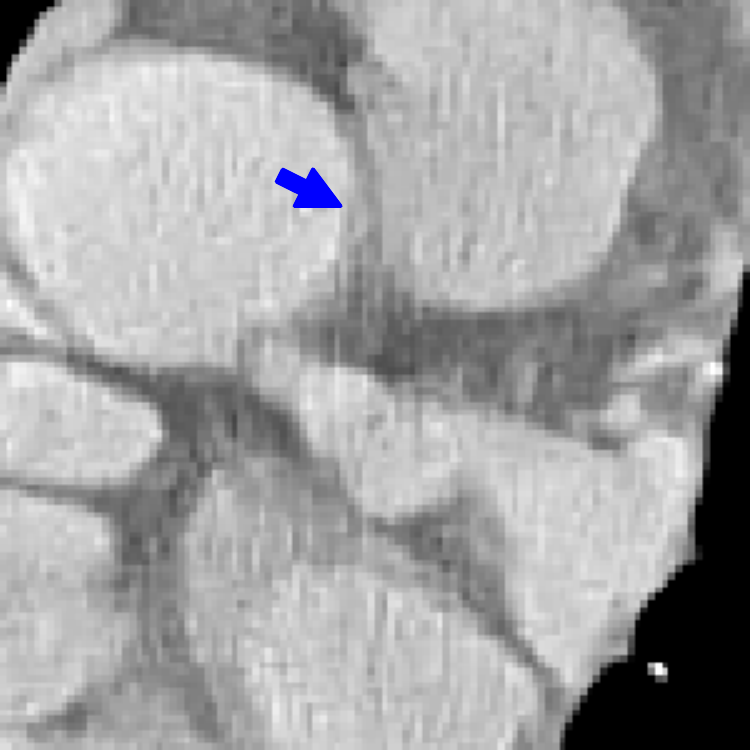} \\
  \end{tabular}
  \caption{Sparse-view CT image (GT1) reconstruction results. Sinogram is contaminated by Noise-2. The $1^{\mbox{st}}$ to $8^{\mbox{th}}$ columns correspond to FBP, HQS-CG, PWLS-CSCGR, FPBConv-Net, PFBS-AIR, PD-Net, JSR-Net, and MetaInv-Net (H). The $1^{\mbox{st}}$ to $5^{\mbox{th}}$ rows correspond to numbers of sparse views $15, 30, 60, 120,$ and $180$. The display window is $[-240,300]$HU. 
   }
\label{fig:rec-img-AAPM-Noise-2}
\end{figure*}

\begin{figure*}[ht]
  \centering
  \begin{tabular}{@{}p{0.12\linewidth}@{}p{0.12\linewidth}@{}p{0.12\linewidth}@{}p{0.12\linewidth}@{}p{0.12\linewidth}@{}p{0.12\linewidth}@{}p{0.12\linewidth}@{}p{0.12\linewidth}}
    \subfigimg[width=\linewidth]{}{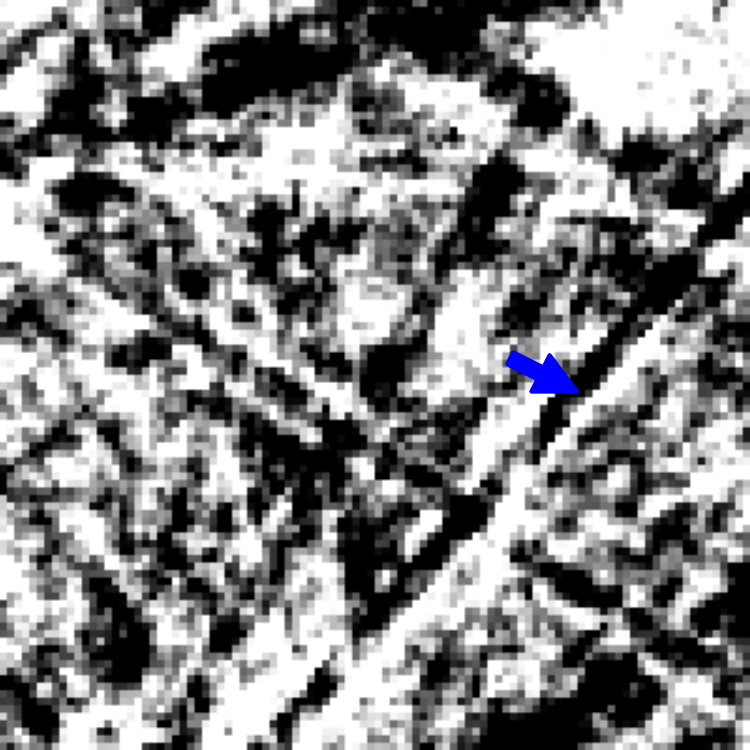} &
    \subfigimg[width=\linewidth]{}{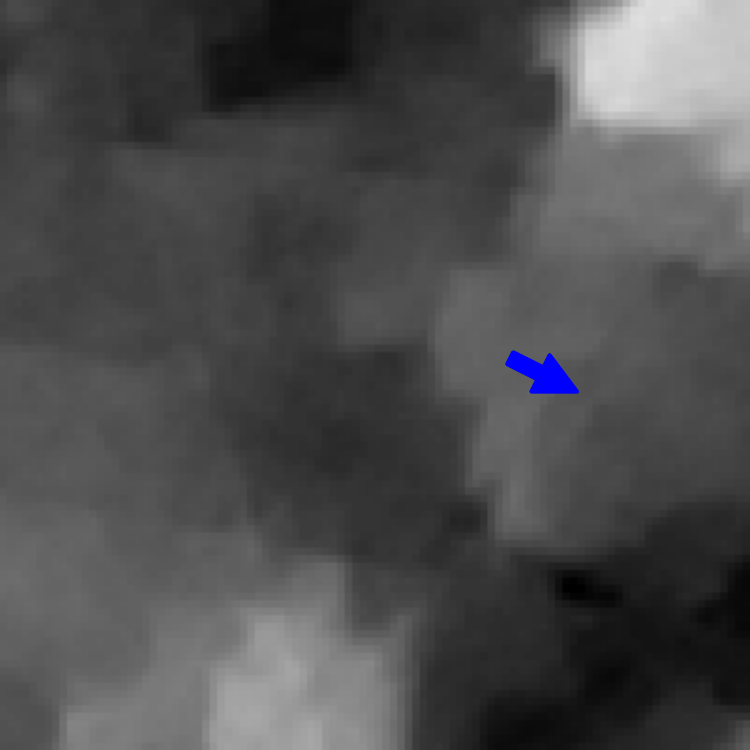} &
    \subfigimg[width=\linewidth]{}{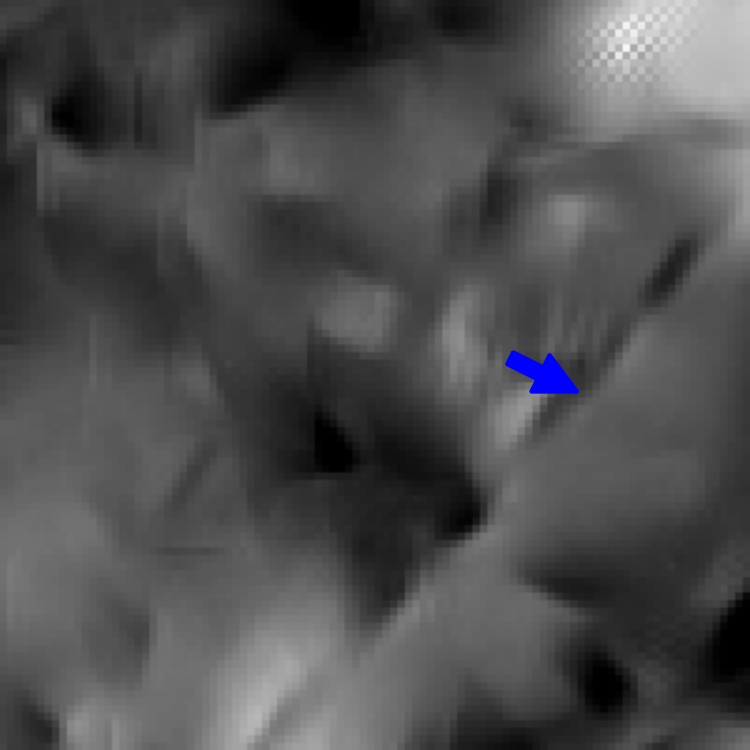} &
    \subfigimg[width=\linewidth]{}{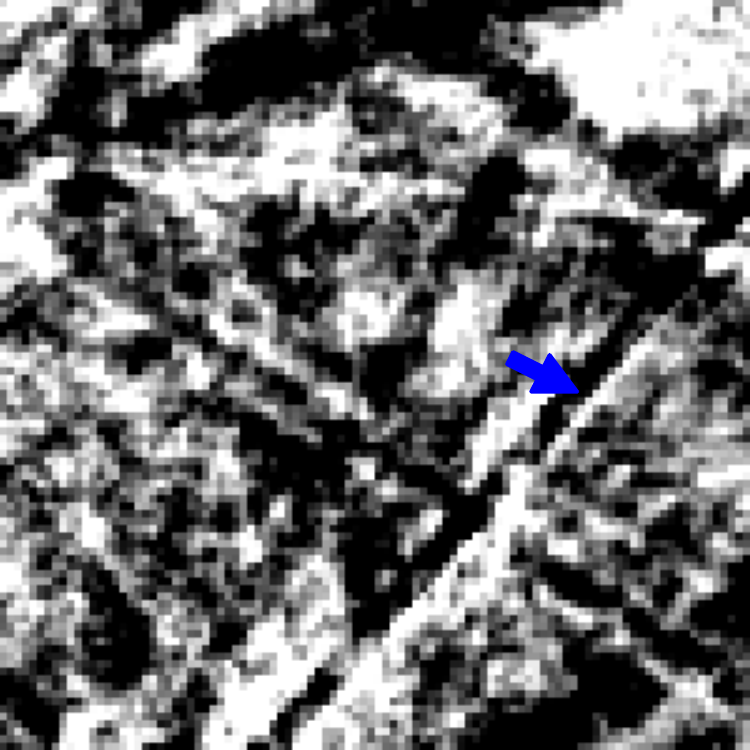} &
    \subfigimg[width=\linewidth]{}{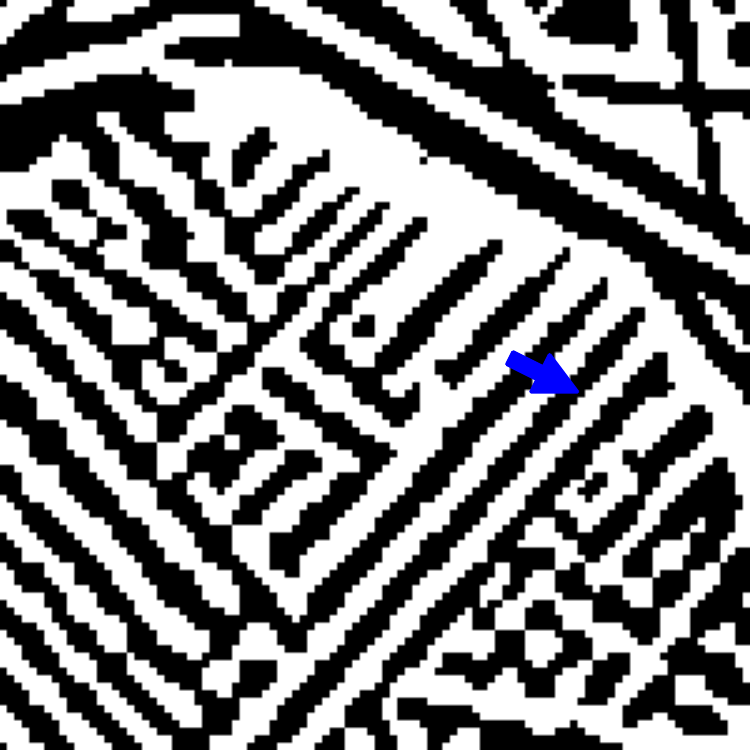} &
    \subfigimg[width=\linewidth]{}{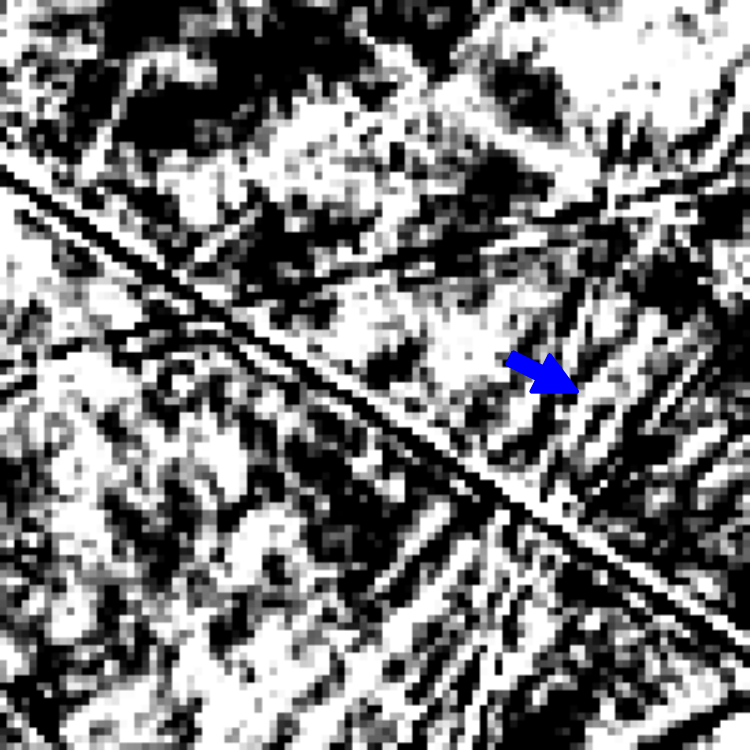} &
    \subfigimg[width=\linewidth]{}{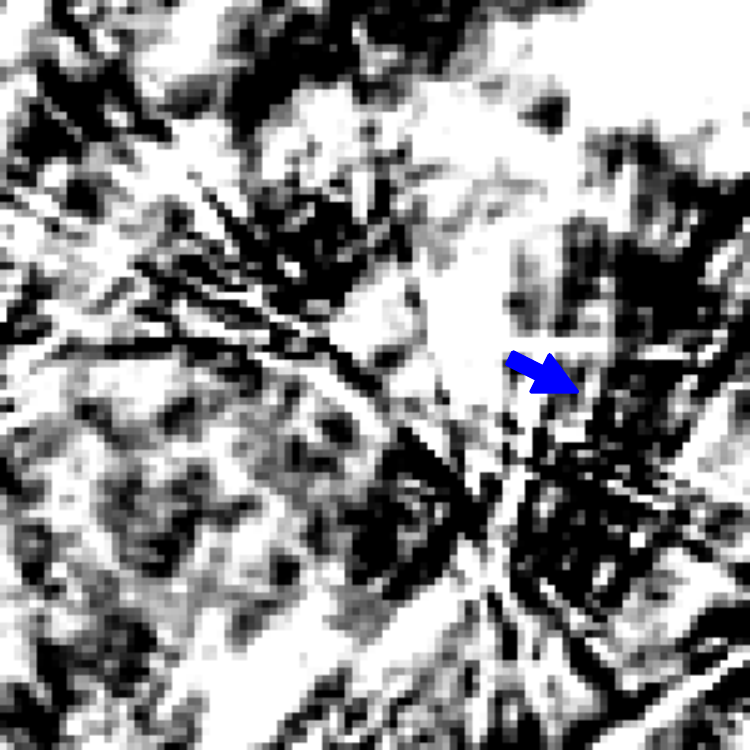} &
    \subfigimg[width=\linewidth]{}{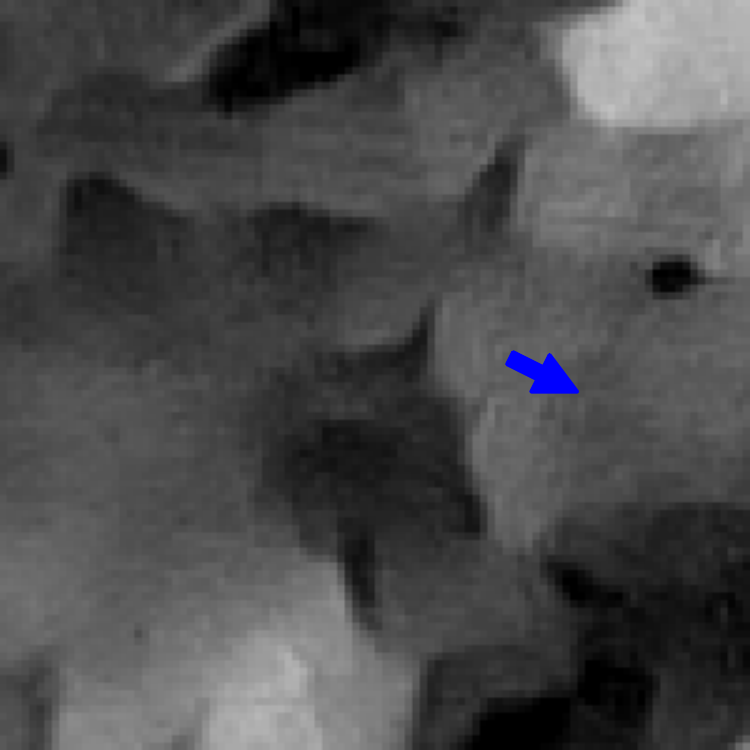} \\

    \subfigimg[width=\linewidth]{}{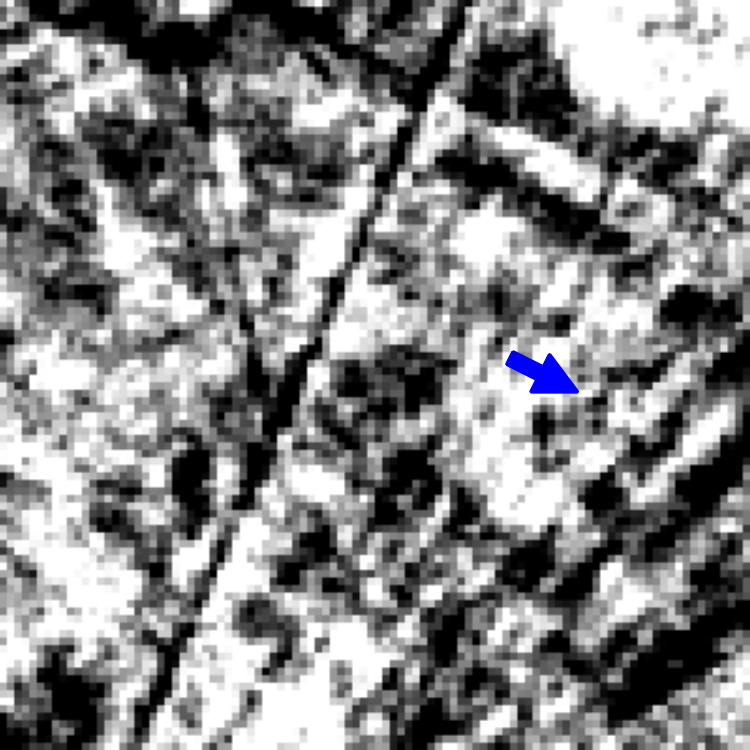} &
    \subfigimg[width=\linewidth]{}{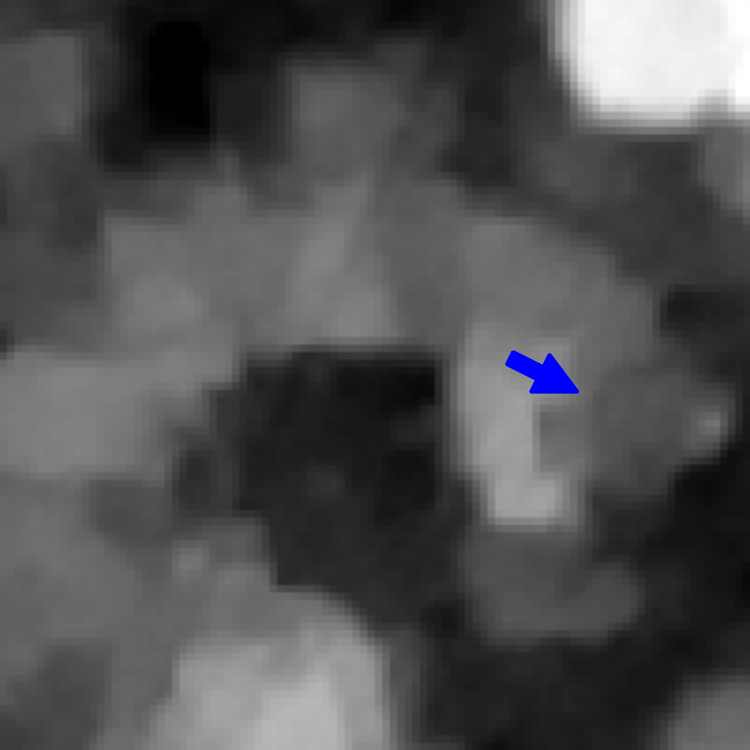} &
    \subfigimg[width=\linewidth]{}{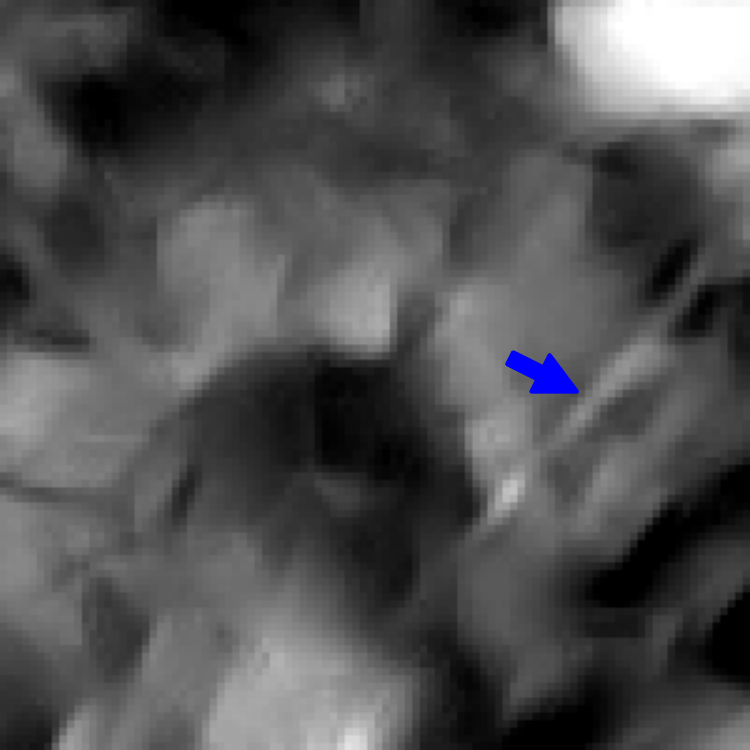} &
    \subfigimg[width=\linewidth]{}{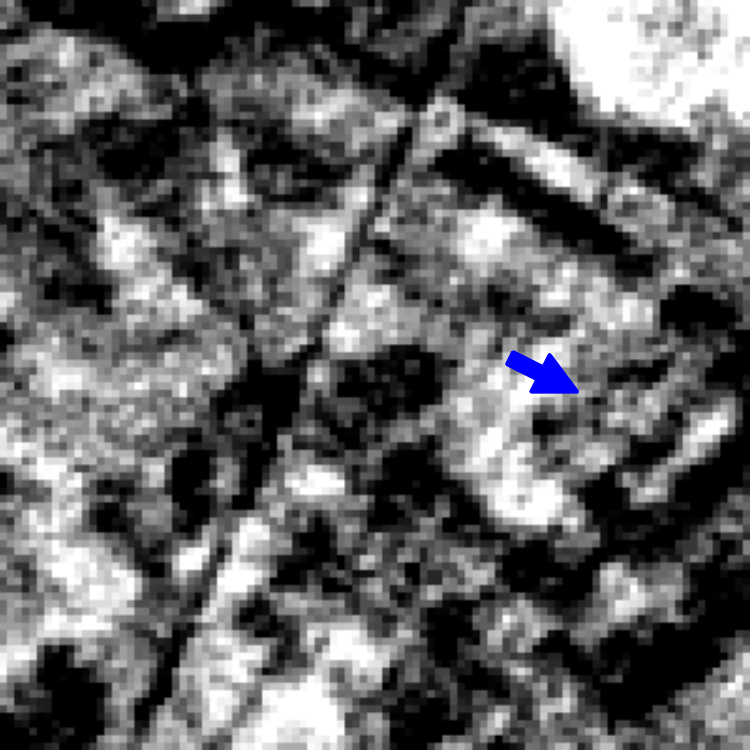} &
    \subfigimg[width=\linewidth]{}{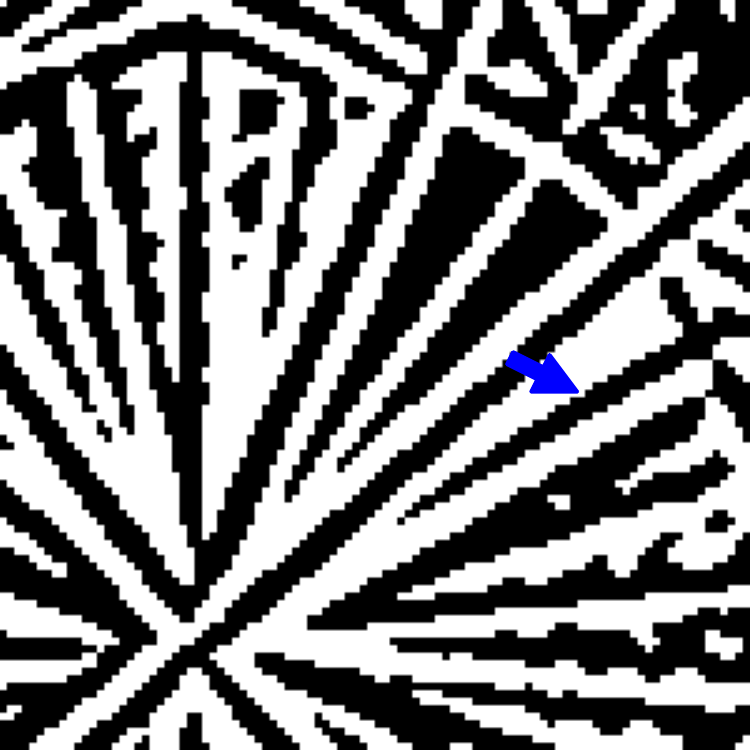} &
    \subfigimg[width=\linewidth]{}{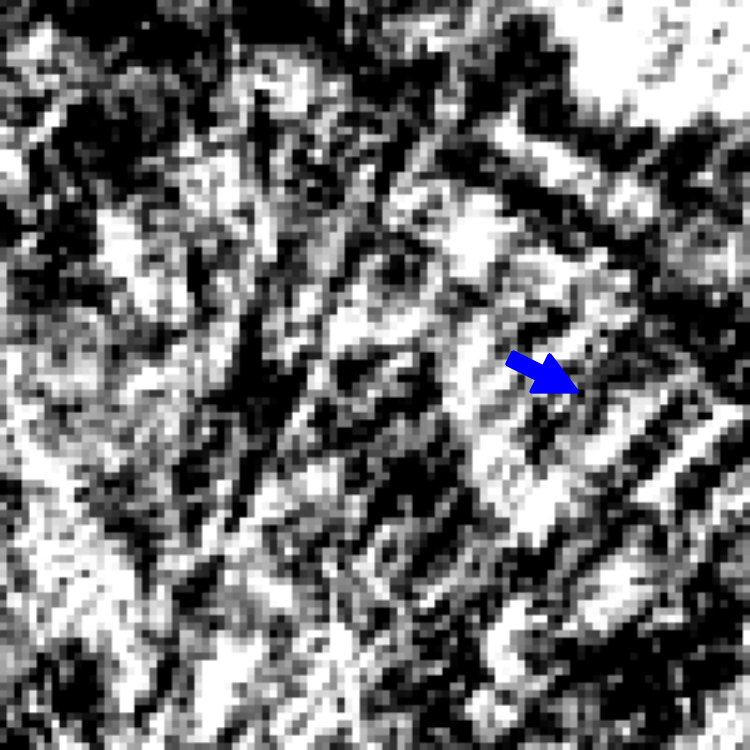} &
    \subfigimg[width=\linewidth]{}{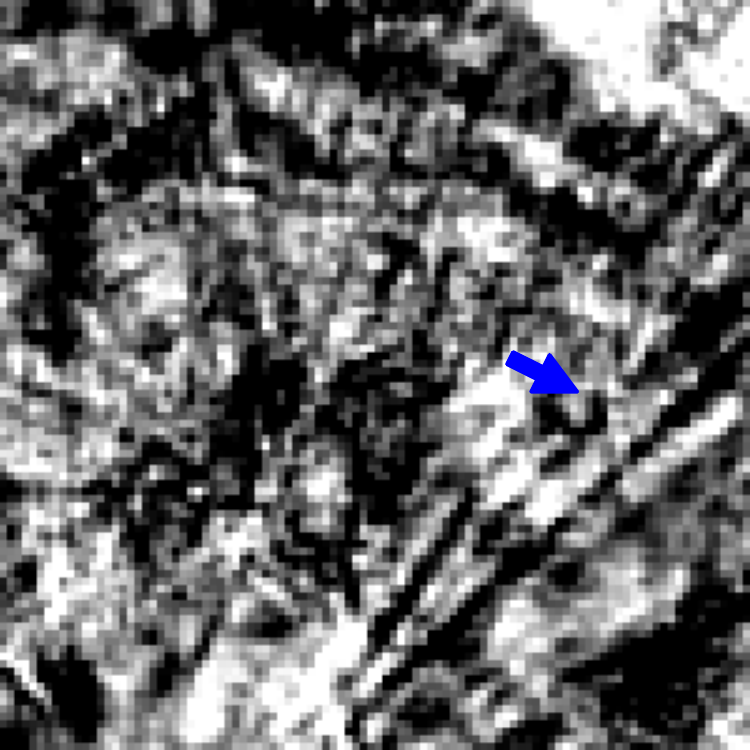} &
    \subfigimg[width=\linewidth]{}{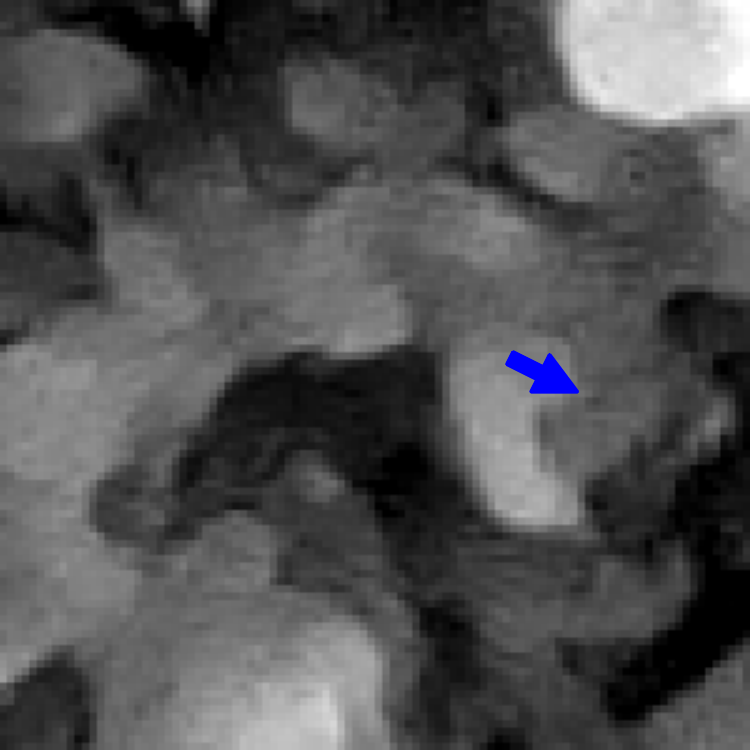} \\
    
    \subfigimg[width=\linewidth]{}{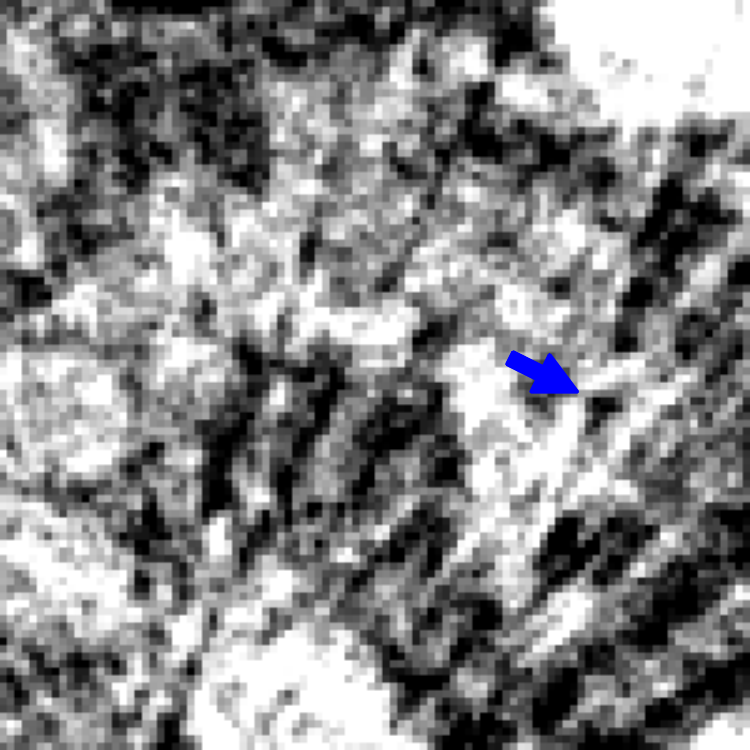} &
    \subfigimg[width=\linewidth]{}{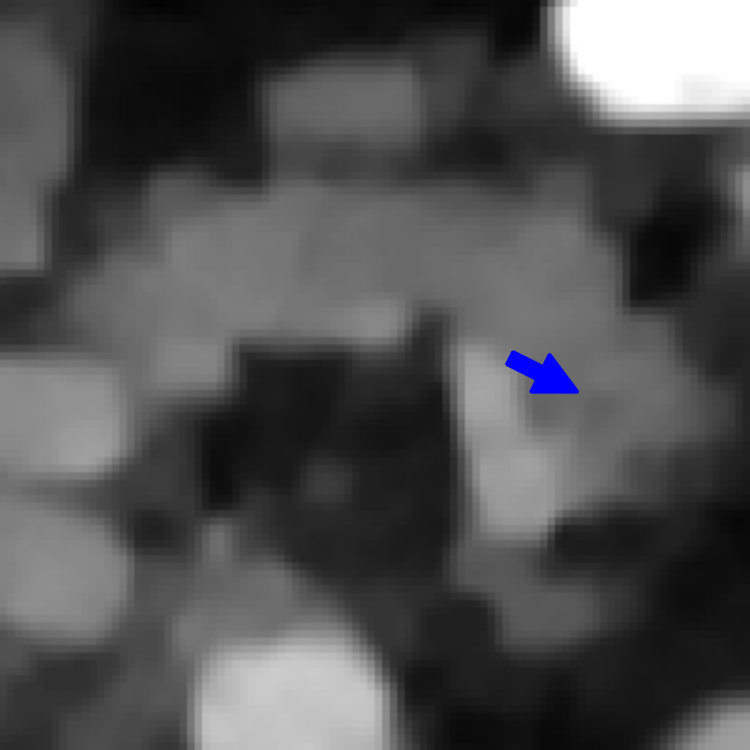} &
    \subfigimg[width=\linewidth]{}{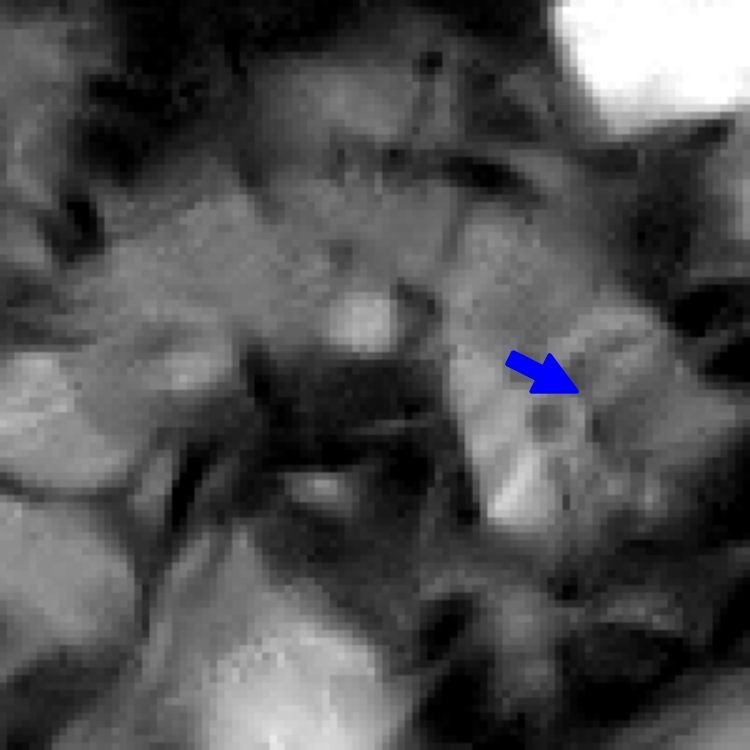} &
    \subfigimg[width=\linewidth]{}{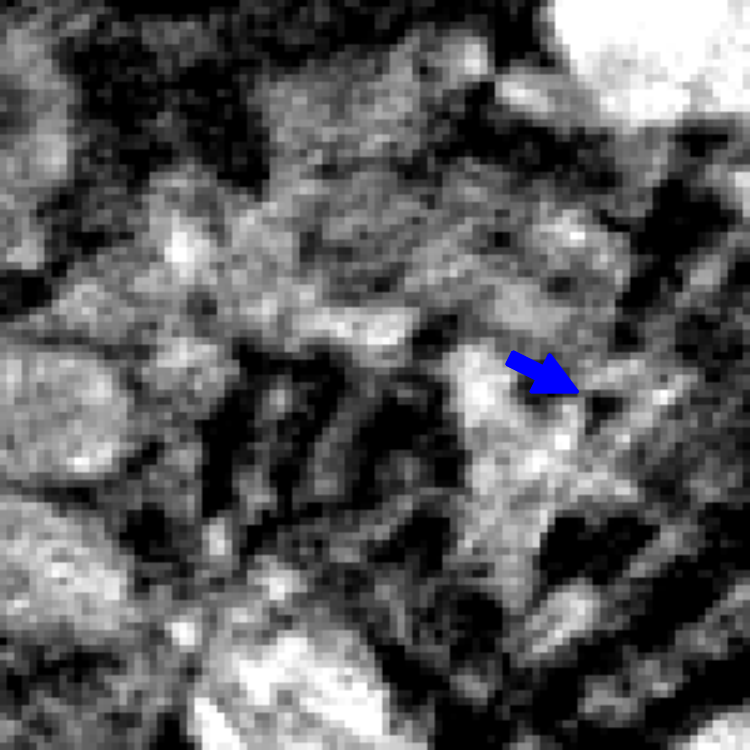} &
    \subfigimg[width=\linewidth]{}{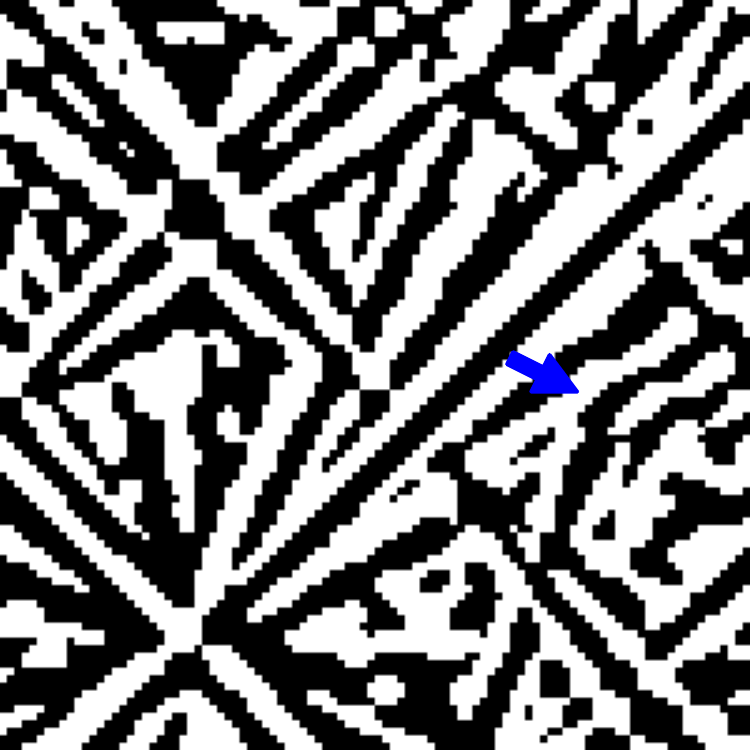} &
    \subfigimg[width=\linewidth]{}{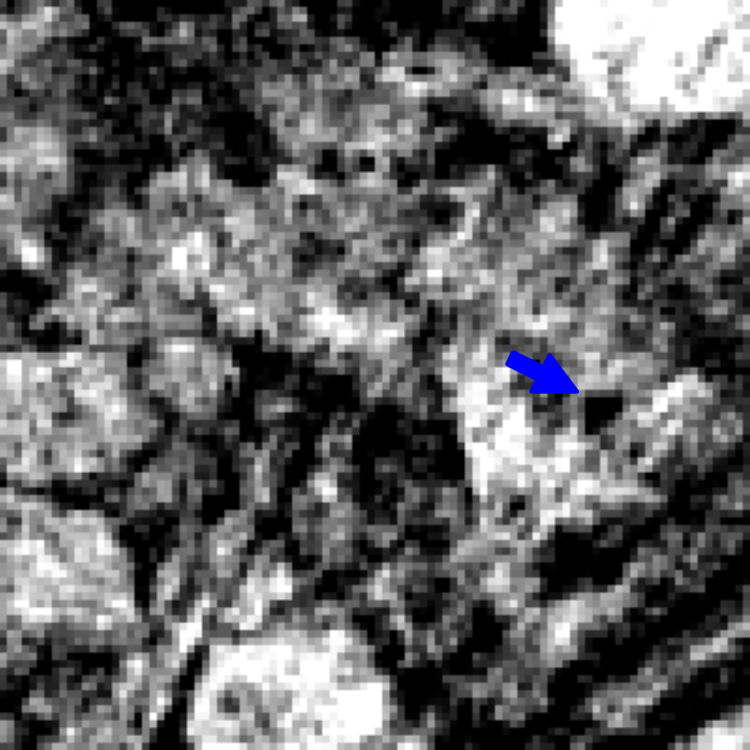} &
    \subfigimg[width=\linewidth]{}{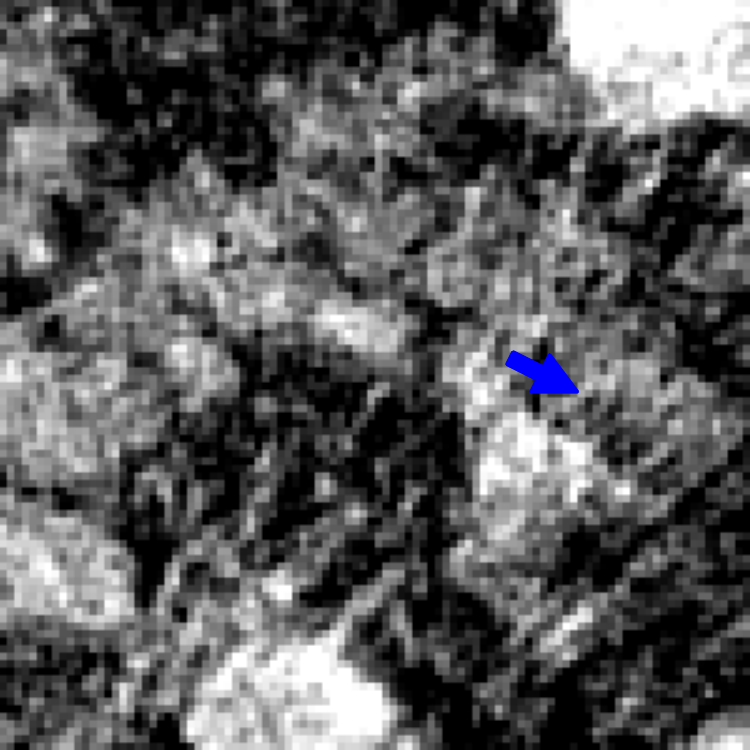} &
    \subfigimg[width=\linewidth]{}{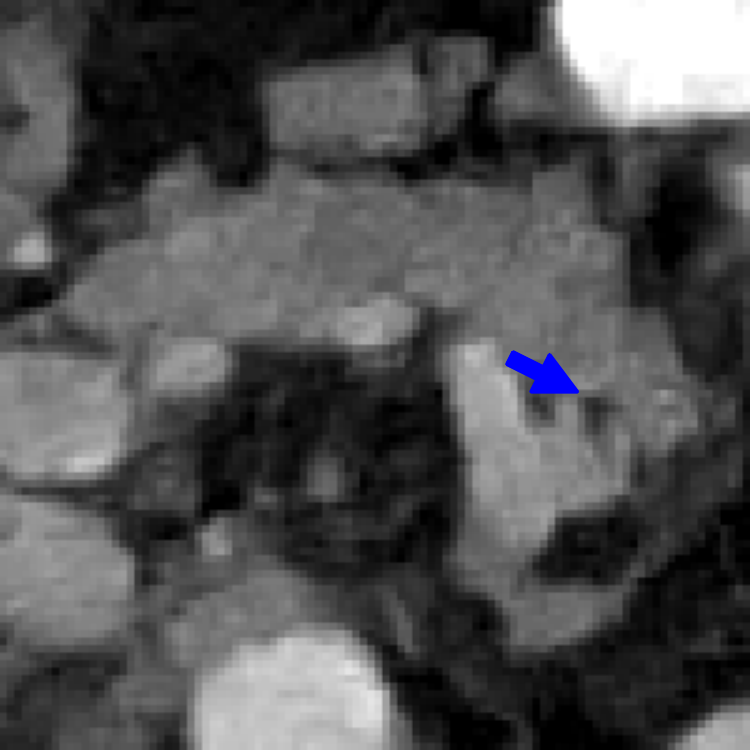} \\

    \subfigimg[width=\linewidth]{}{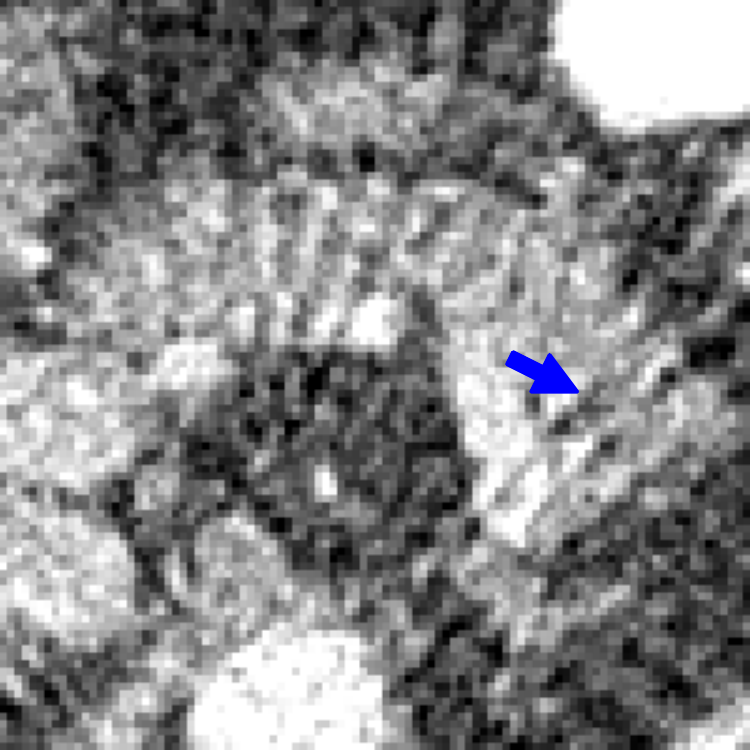} &
    \subfigimg[width=\linewidth]{}{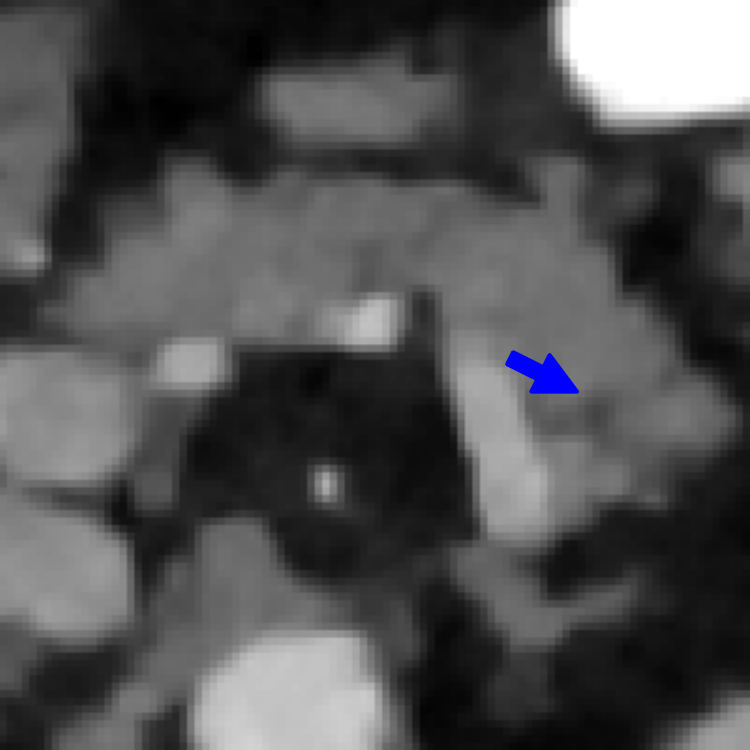} &
    \subfigimg[width=\linewidth]{}{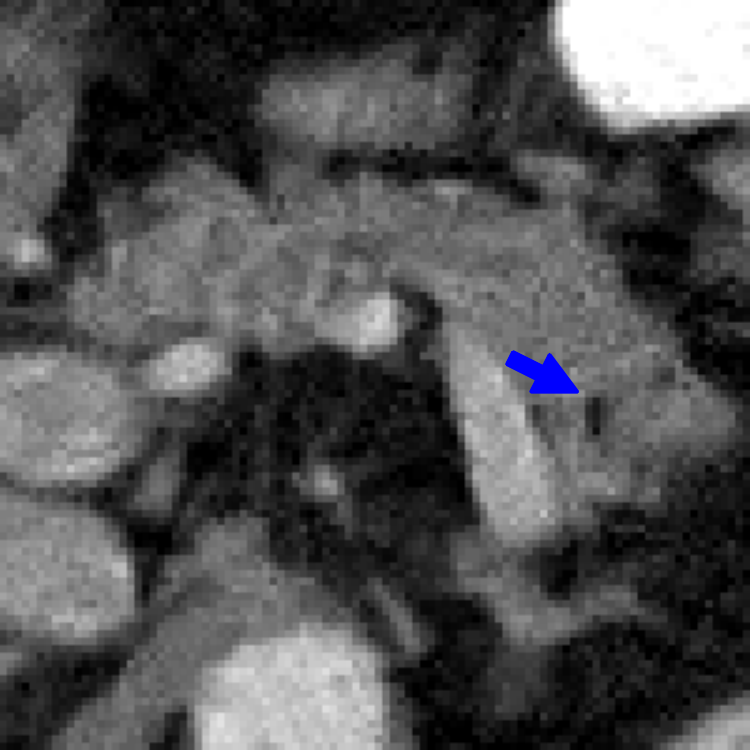} &
    \subfigimg[width=\linewidth]{}{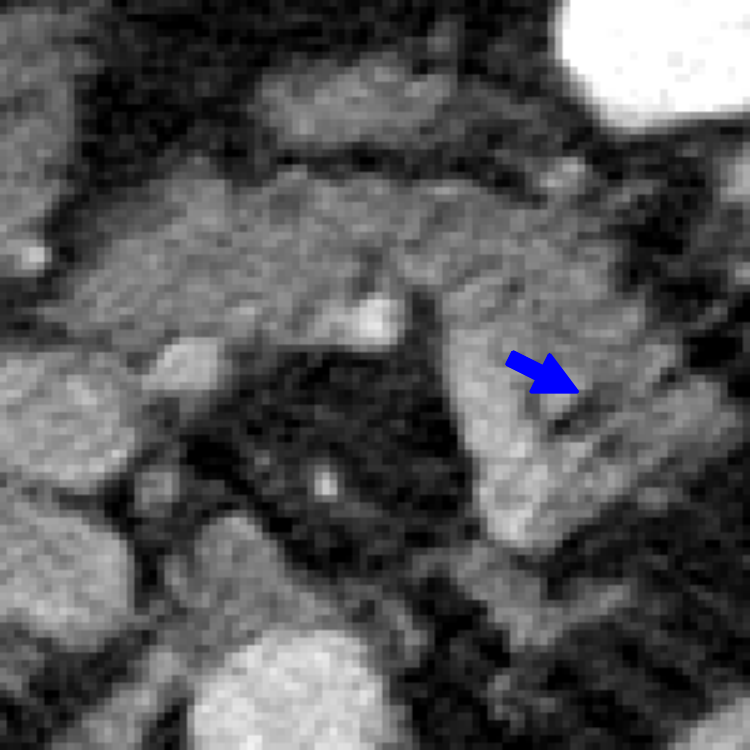} &
    \subfigimg[width=\linewidth]{}{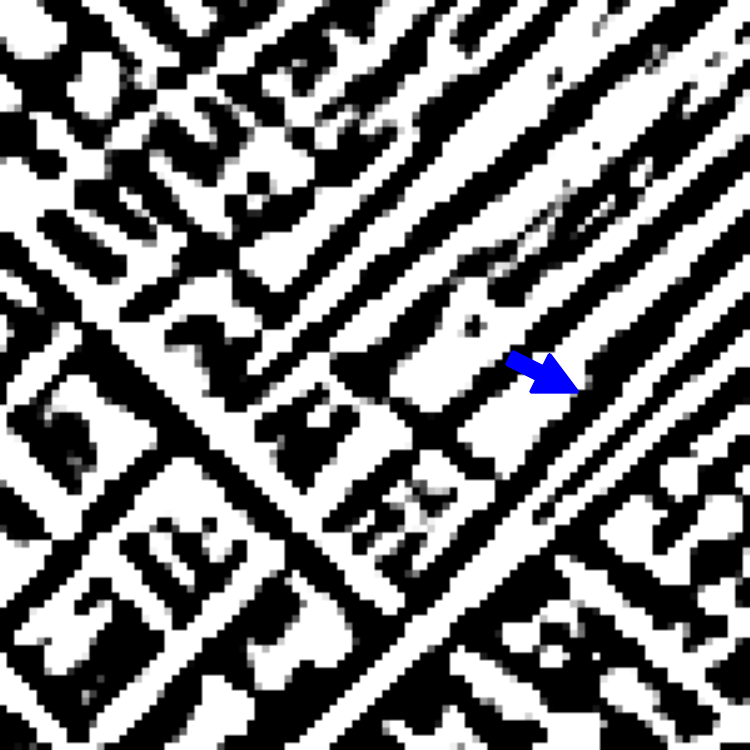} &
    \subfigimg[width=\linewidth]{}{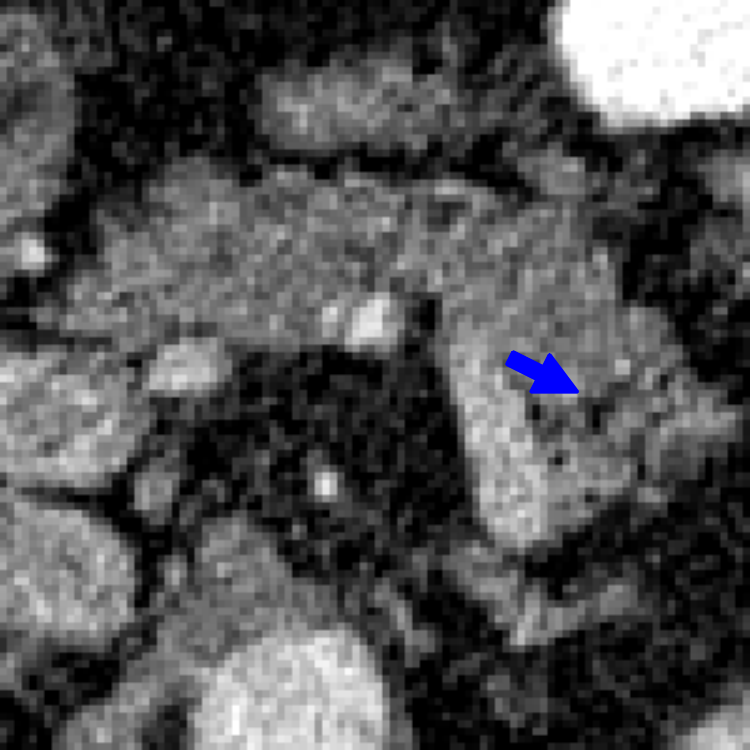} &
    \subfigimg[width=\linewidth]{}{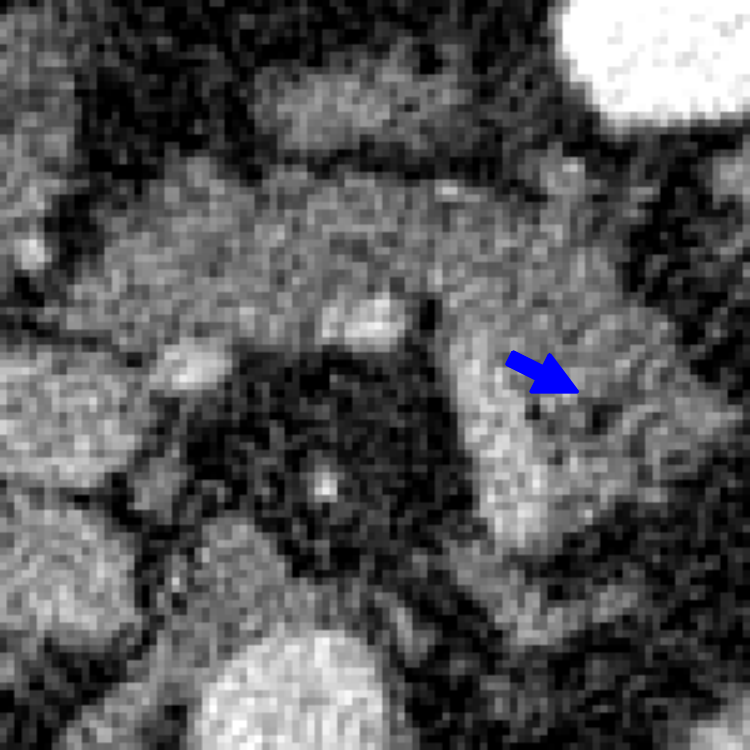} &
    \subfigimg[width=\linewidth]{}{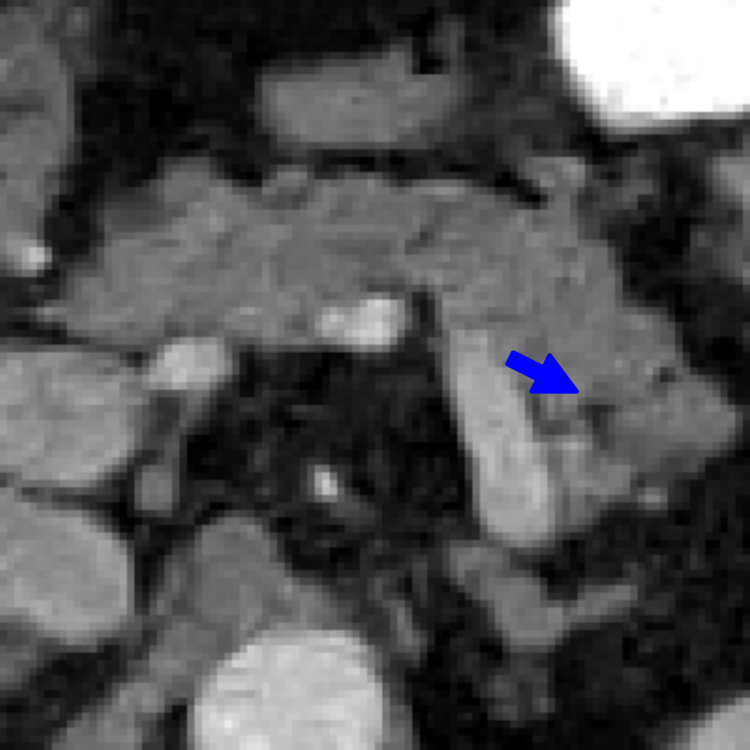} \\

    \subfigimg[width=\linewidth]{}{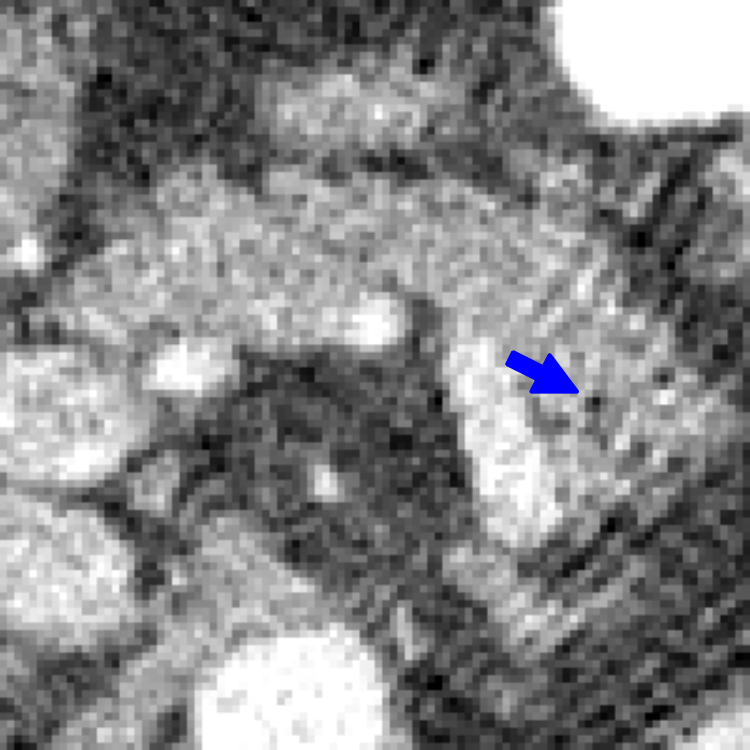}&
    \subfigimg[width=\linewidth]{}{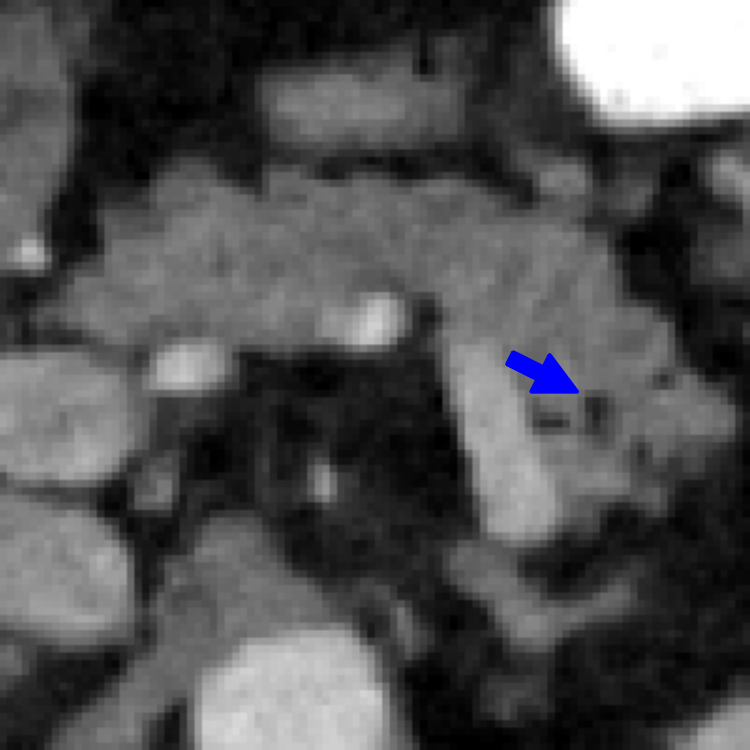} &
    \subfigimg[width=\linewidth]{}{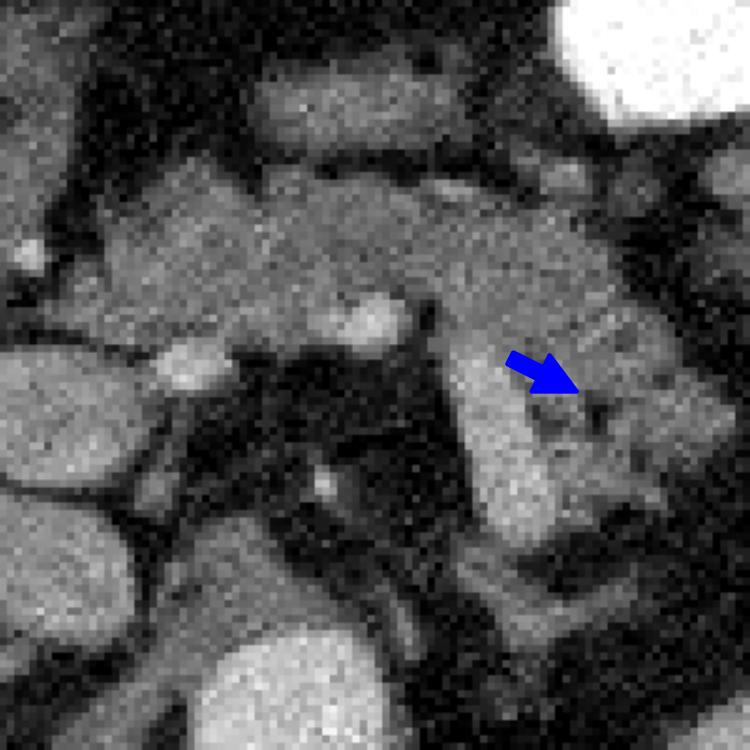} &
    \subfigimg[width=\linewidth]{}{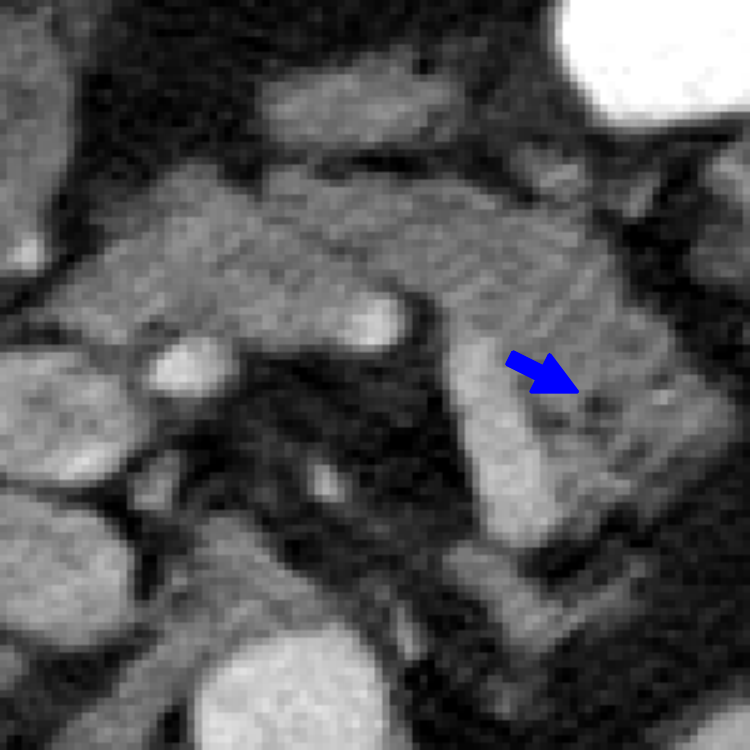}&
    \subfigimg[width=\linewidth]{}{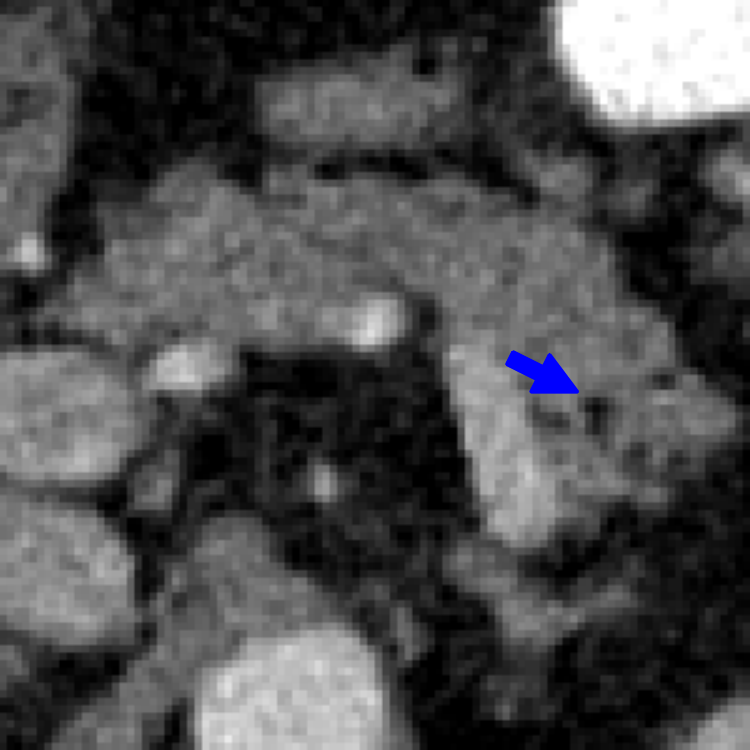}&
    \subfigimg[width=\linewidth]{}{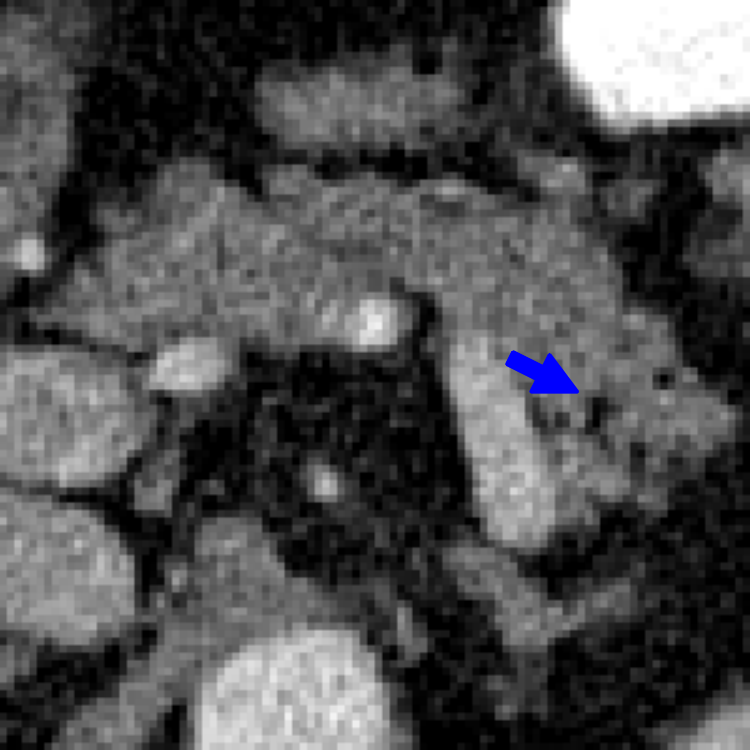} &
    \subfigimg[width=\linewidth]{}{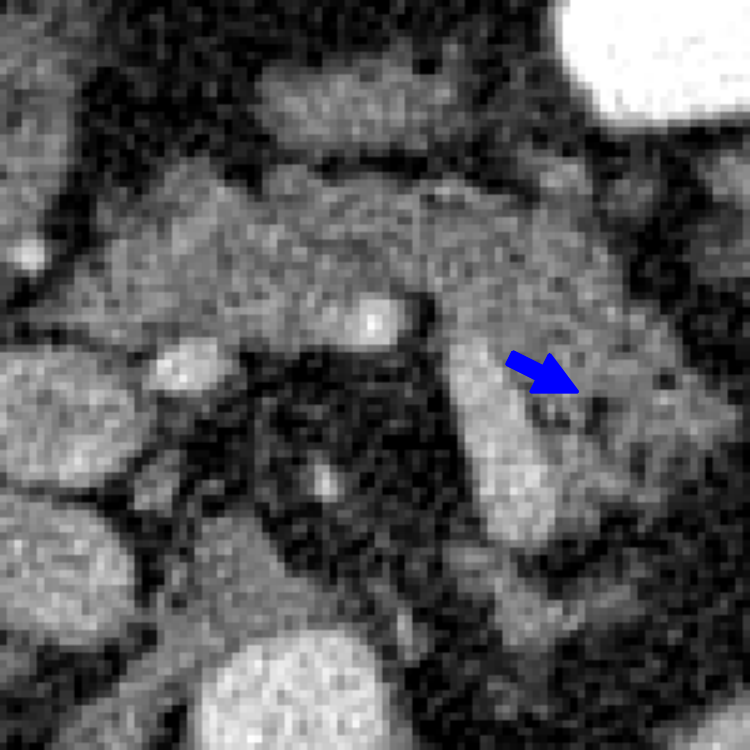} &
    \subfigimg[width=\linewidth]{}{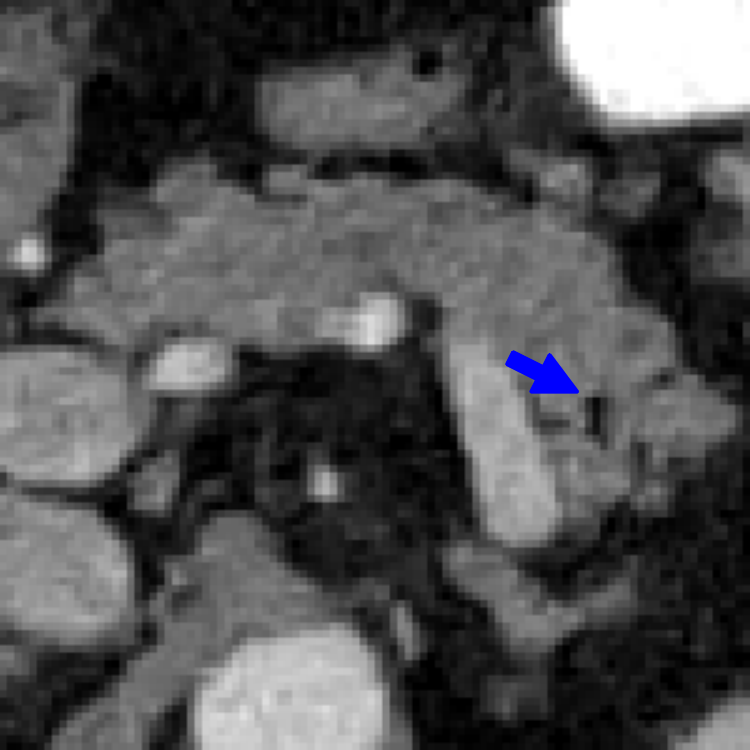} \\
  \end{tabular}
  \caption{Sparse-view CT image (GT2) reconstruction results. Sinogram is contaminated by Noise-2. The $1^{\mbox{st}}$ to $8^{\mbox{th}}$ columns correspond to FBP, HQS-CG, PWLS-CSCGR, FPBConv-Net, PFBS-AIR, PD-Net, JSR-Net, and MetaInv-Net (H). The $1^{\mbox{st}}$ to $5^{\mbox{th}}$ rows correspond to numbers of sparse views $15, 30, 60, 120,$ and $180$. The display window is $[-180,420]$HU.
   }
\label{fig:rec-img-Panc-Noise-2}
\end{figure*}

\begin{figure*}[ht]
  \centering
  \begin{tabular}{@{}p{0.2\linewidth}@{}p{0.2\linewidth}@{}@{}p{0.2\linewidth}@{}p{0.2\linewidth}@{}p{0.2\linewidth}}
    \subfigimg[width=1.\linewidth]{}{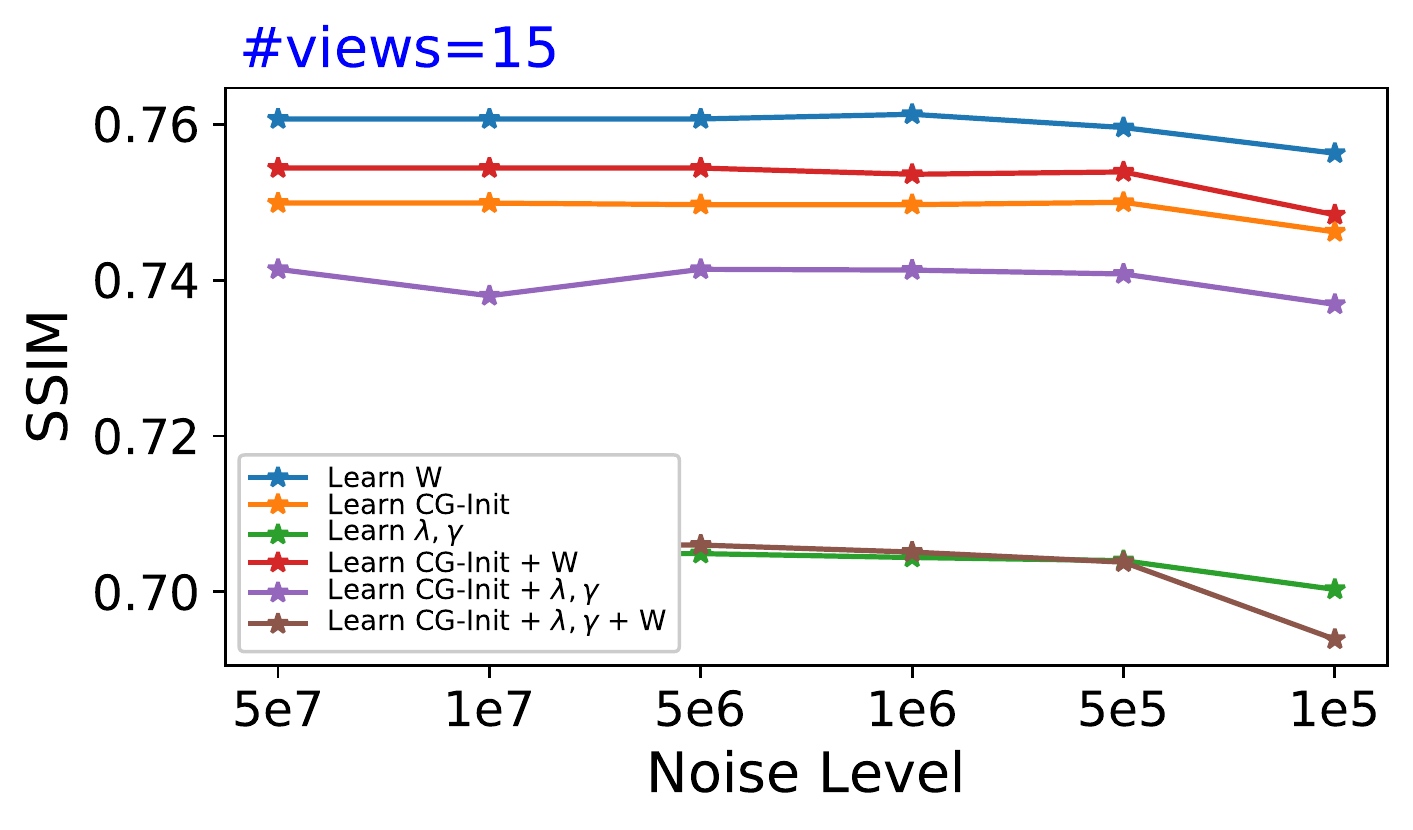} &
    \subfigimg[width=1.\linewidth]{}{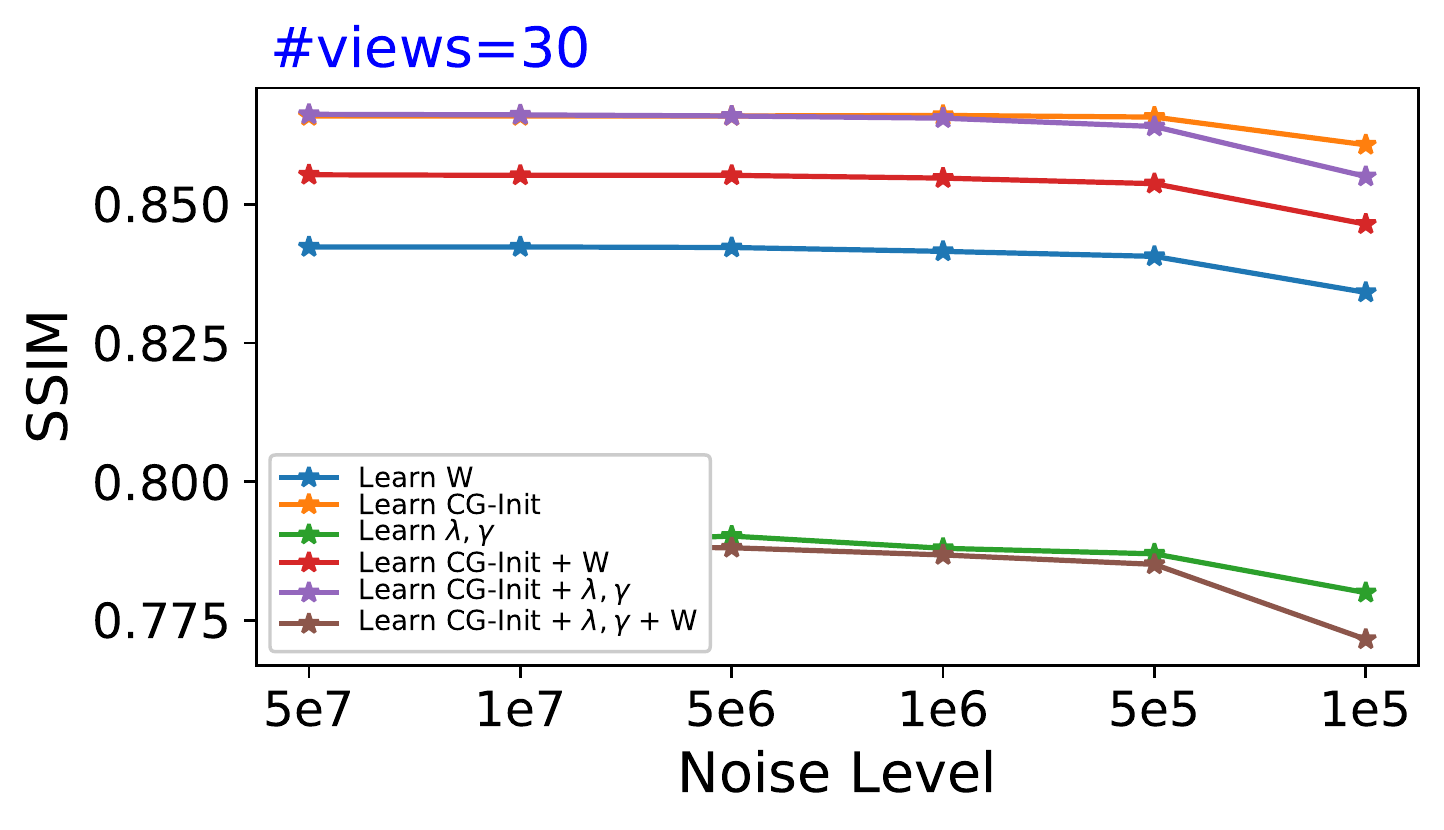} &
    \subfigimg[width=1.\linewidth]{}{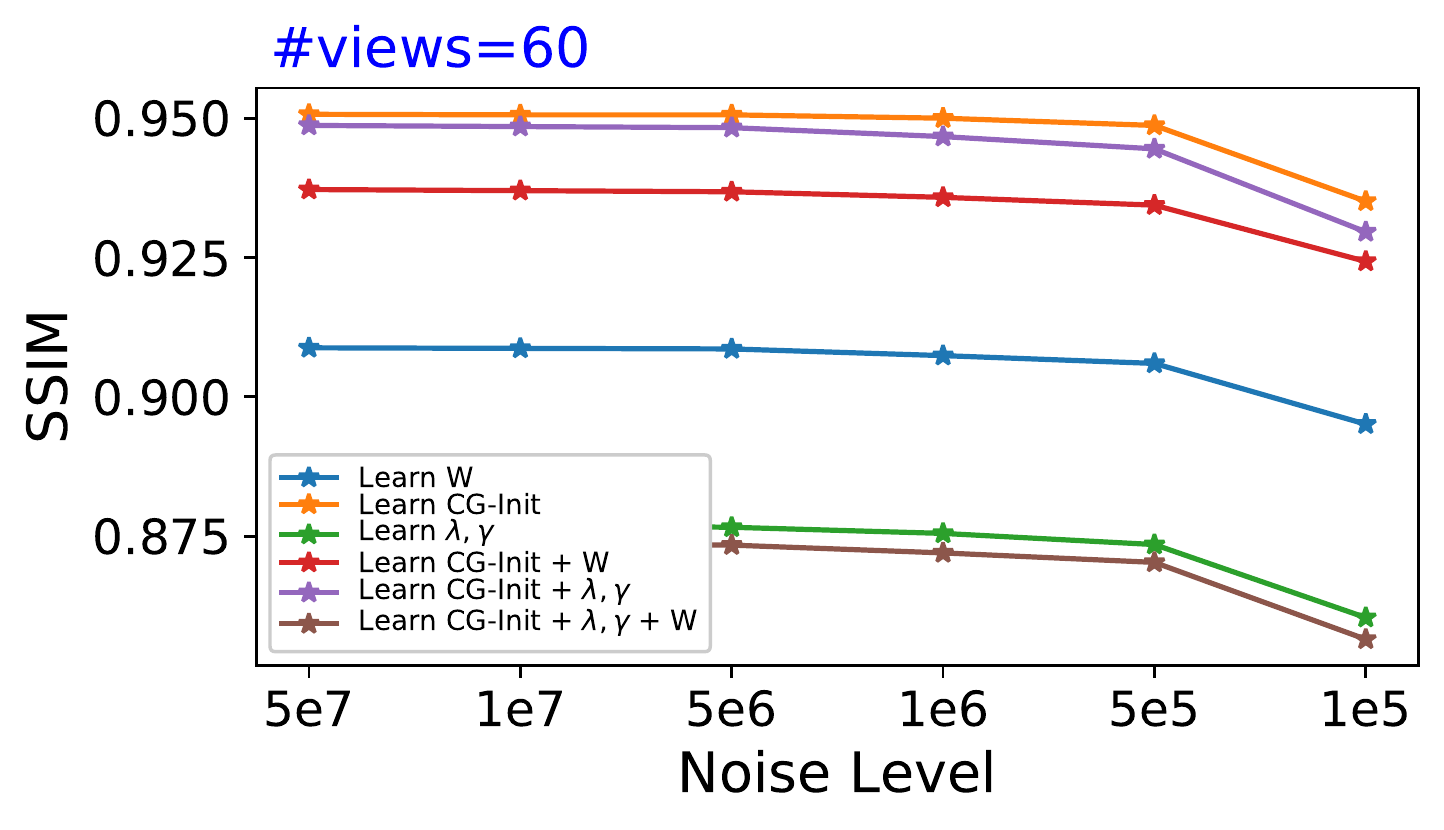} &
    \subfigimg[width=1.\linewidth]{}{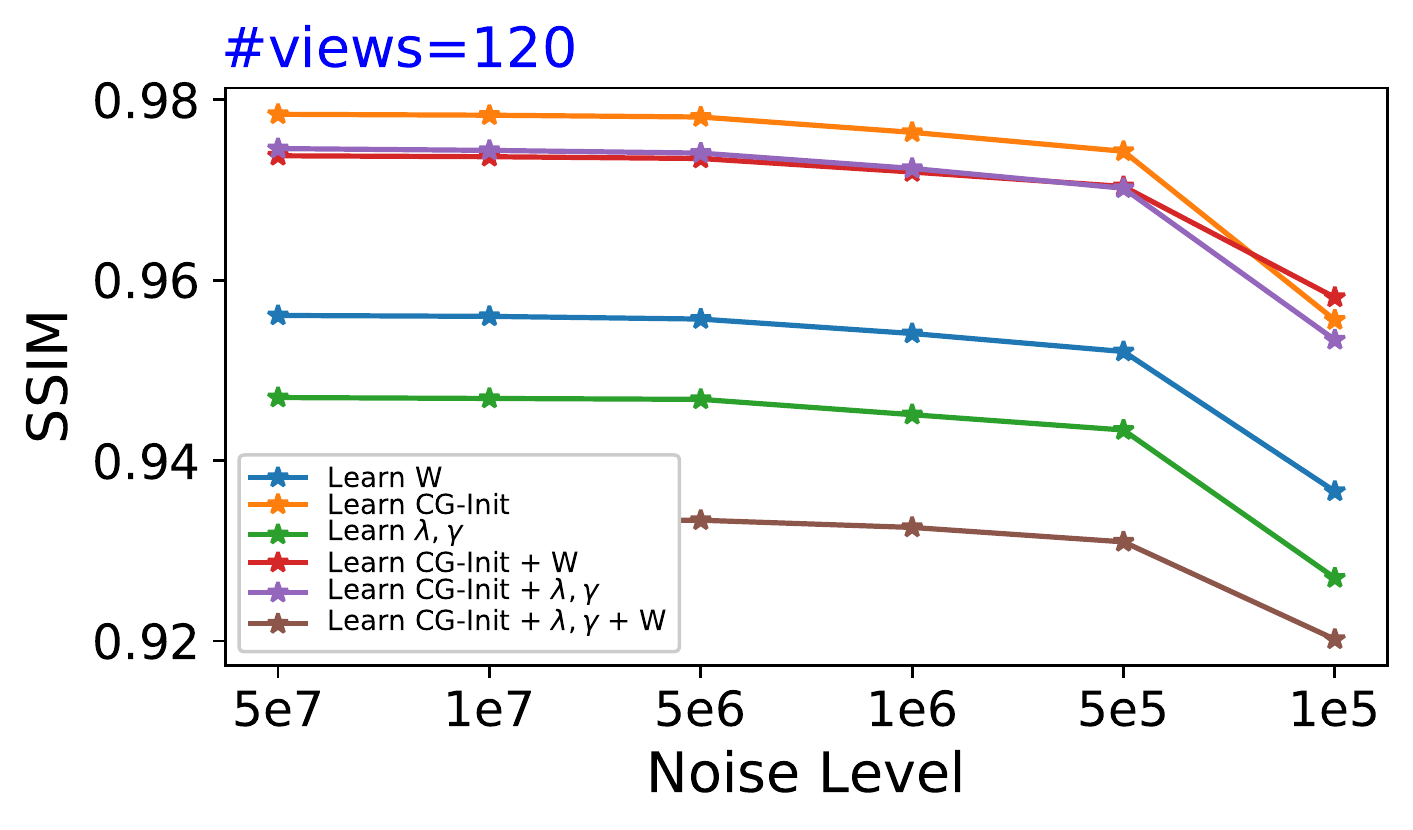} &
    \subfigimg[width=1.\linewidth]{}{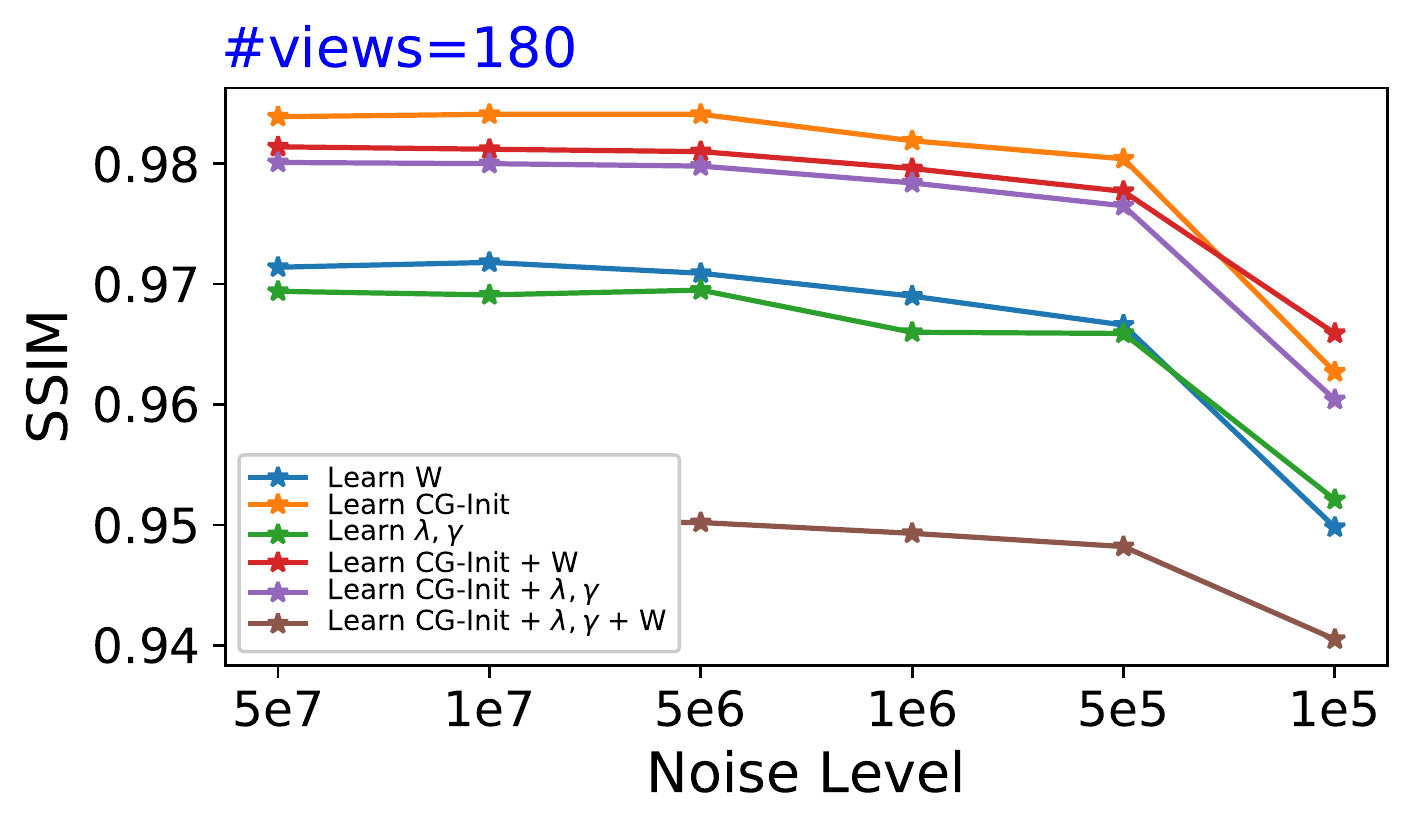} \\

    \subfigimg[width=1.\linewidth]{}{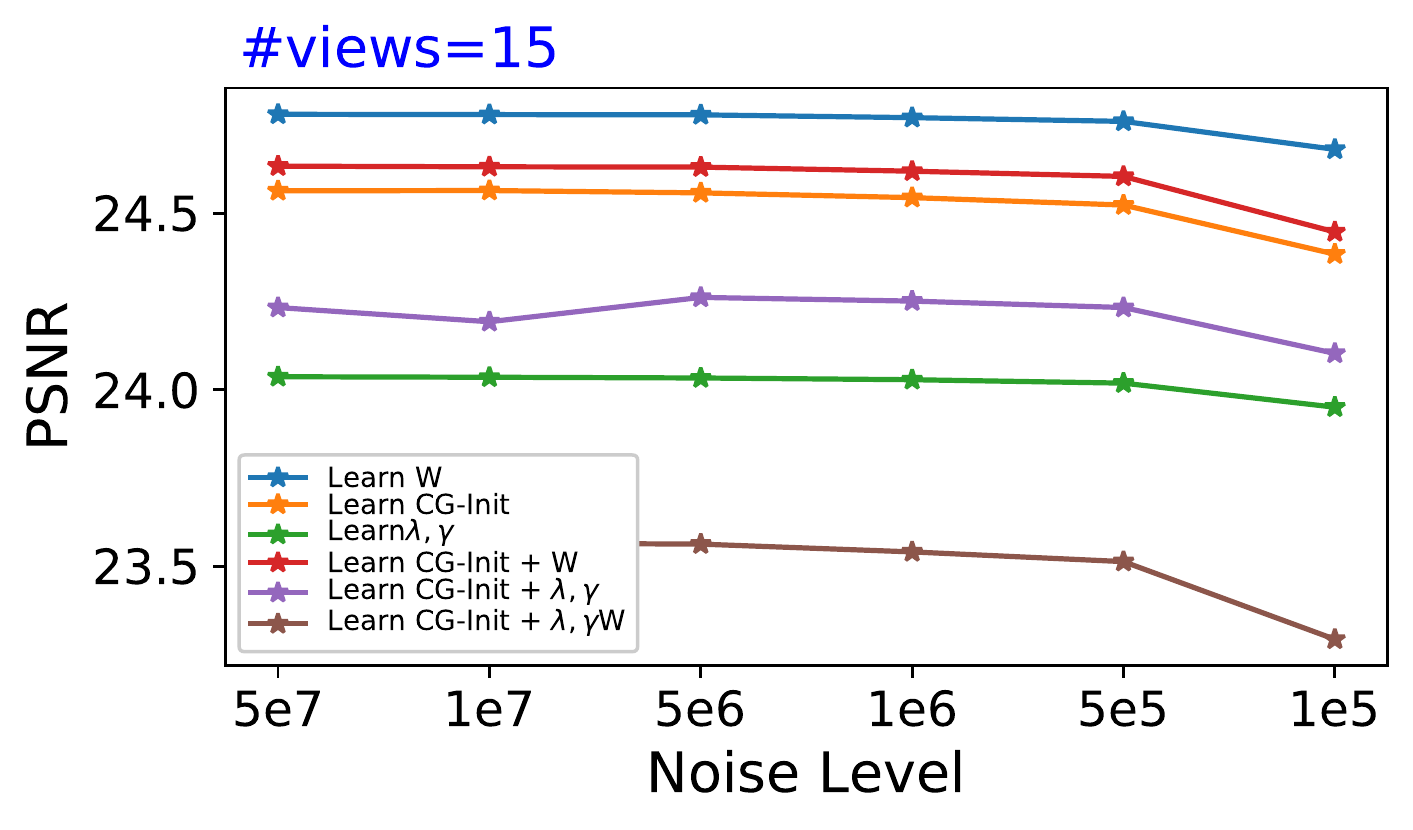} &
    \subfigimg[width=1.\linewidth]{}{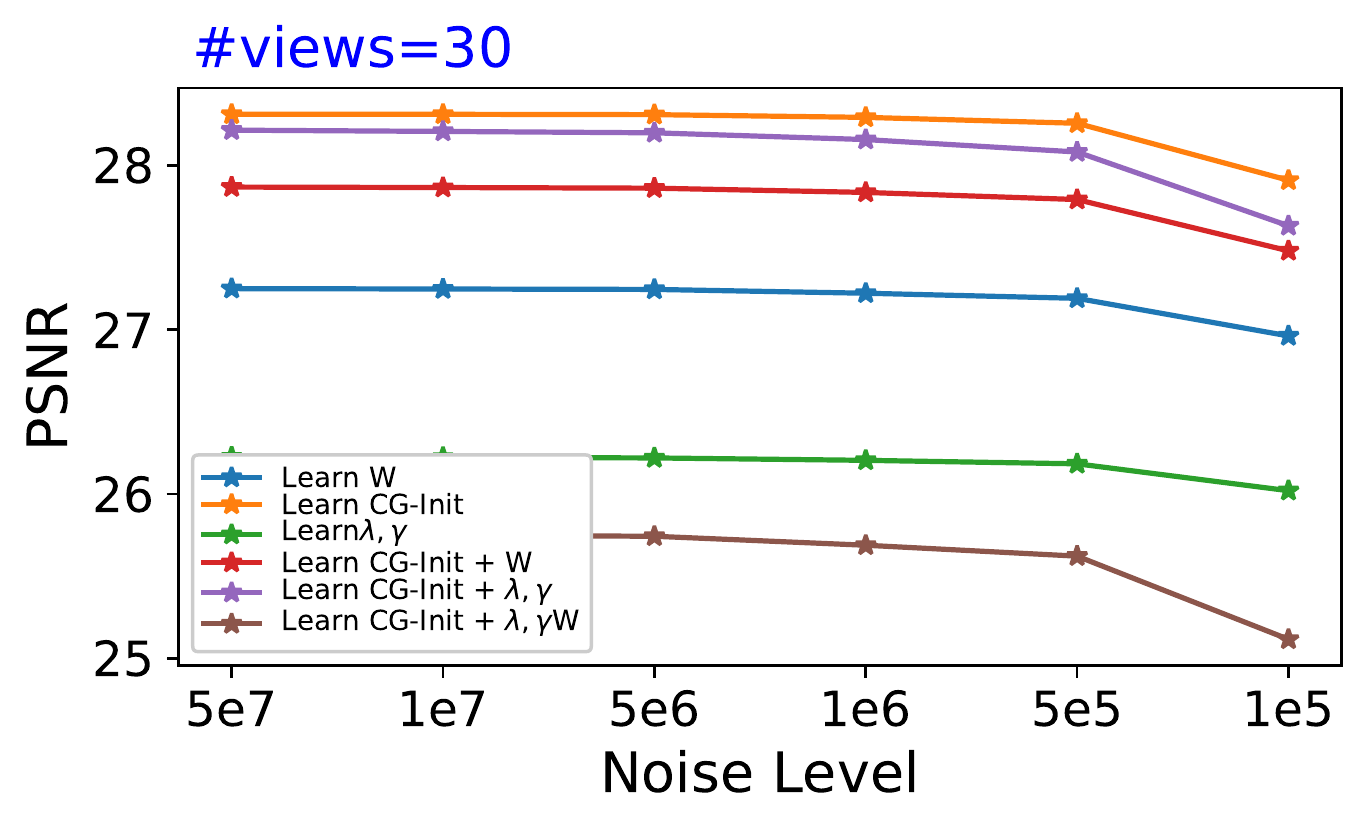} &
    \subfigimg[width=1.\linewidth]{}{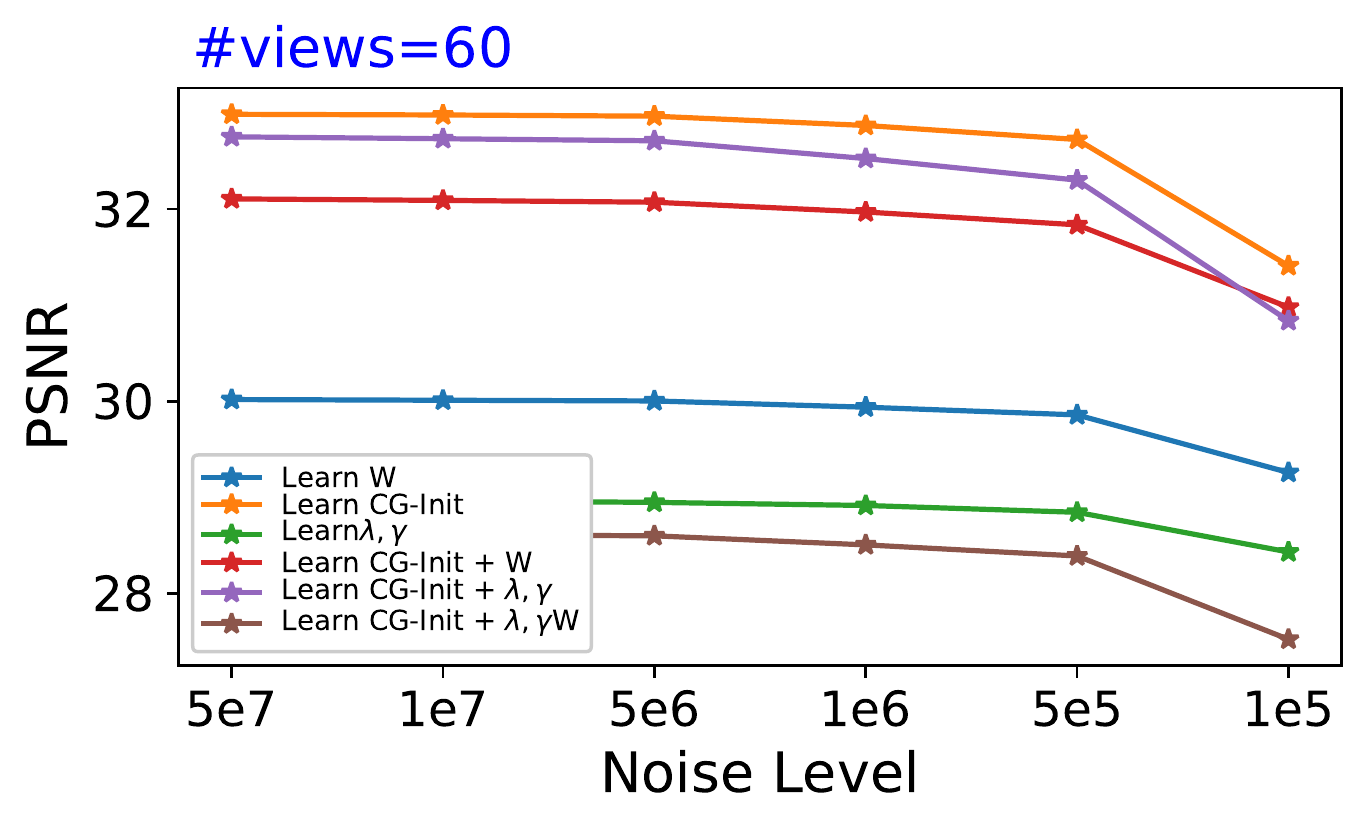} &
    \subfigimg[width=1.\linewidth]{}{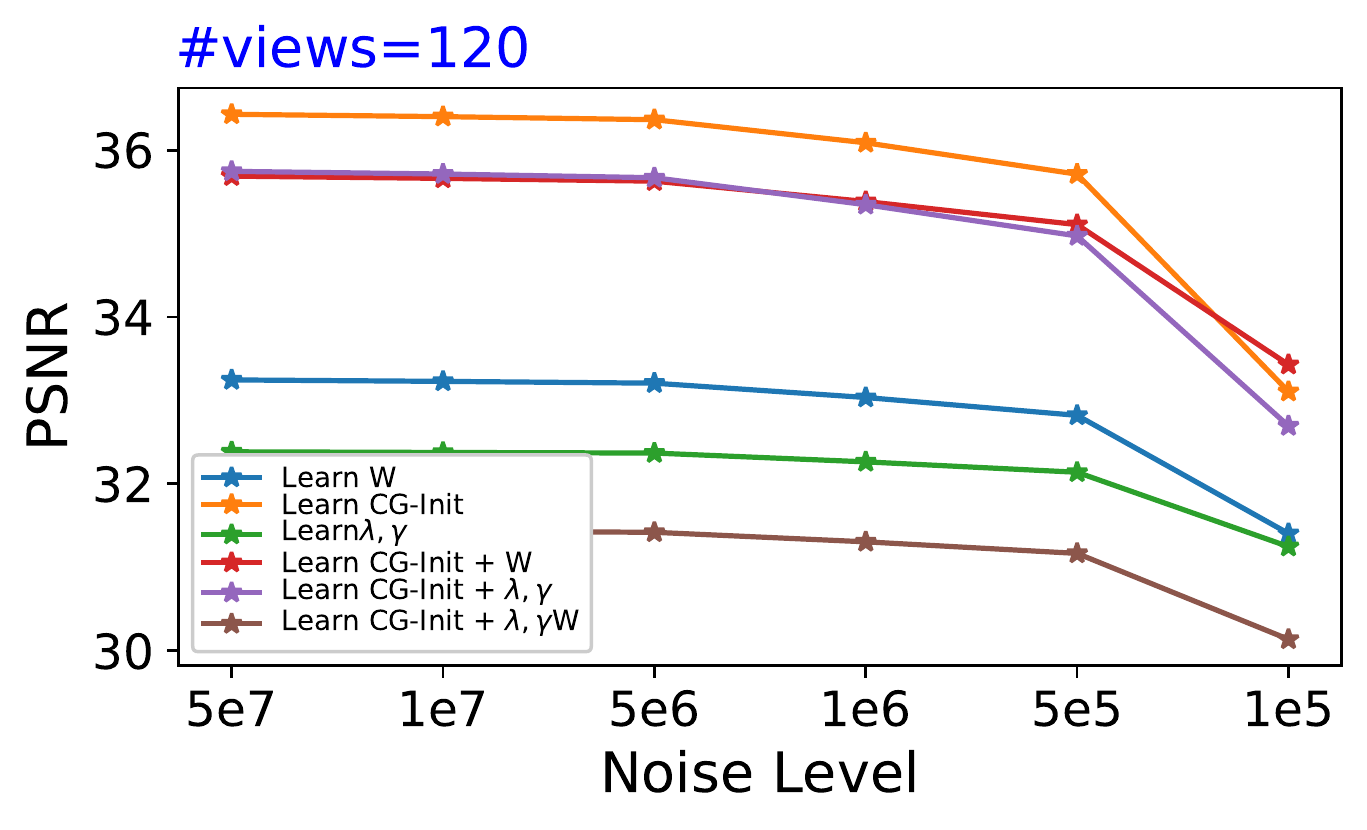} &
    \subfigimg[width=1.\linewidth]{}{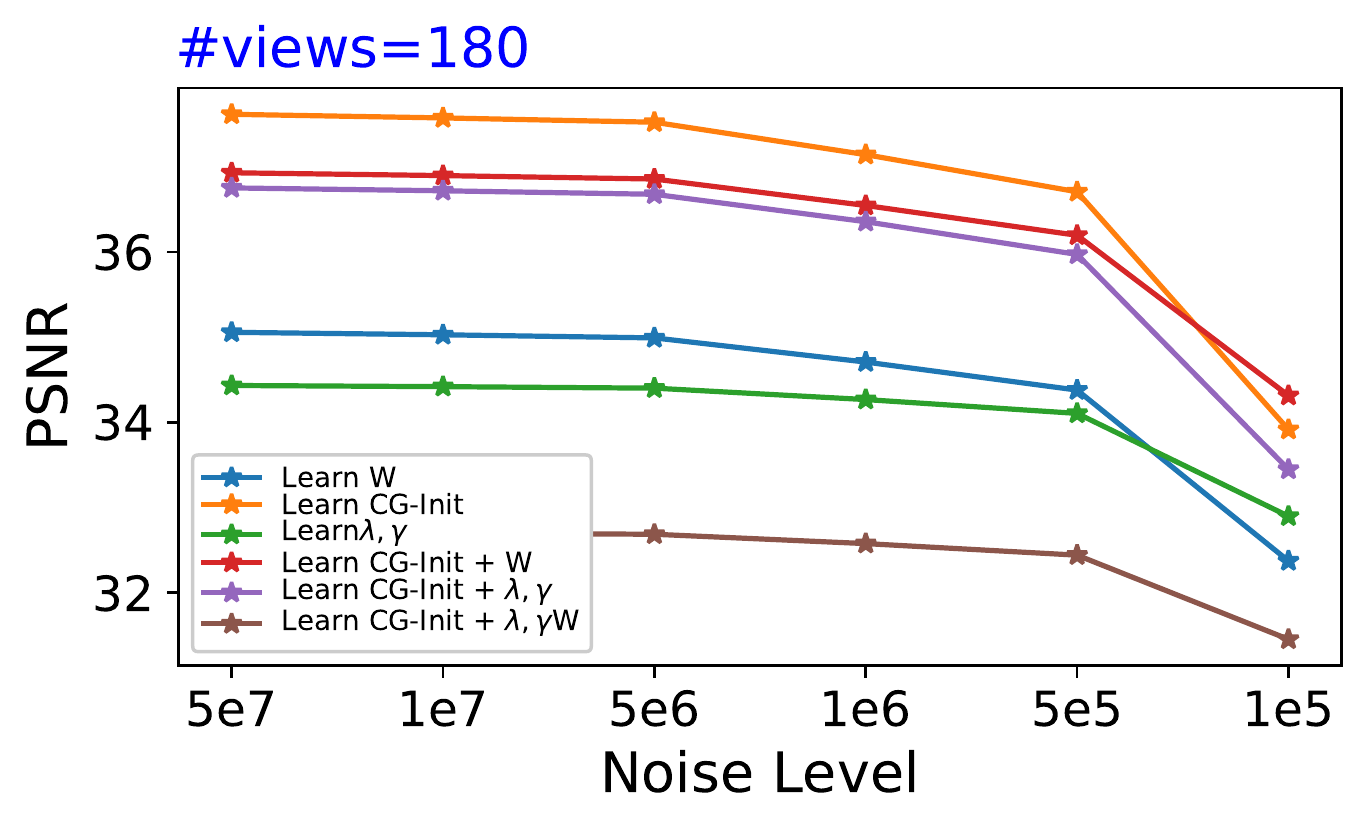} \\
  \end{tabular}
\caption{Learned hyperparameters in MetaInv-Net. The SSIM/PSNR versus different noise level are evaluated on AAPM-Test set. The first and second rows correspond to SSIM and PSNR curve, respectively. The $1^{\mbox{st}}$-$5^{\mbox{th}}$ columns show the results from different scanning settings, i.e., the number of views (\#views) are 15, 30, 60, 120, and 180, respectively. }
\label{fig:learn-diff-hyper-parameters-metainv-net}
\end{figure*}

\begin{figure}[ht]
  \centering
  \begin{tabular}{@{}p{0.28\linewidth}@{}p{0.28\linewidth}}
    \subfigimg[width=\linewidth]{}{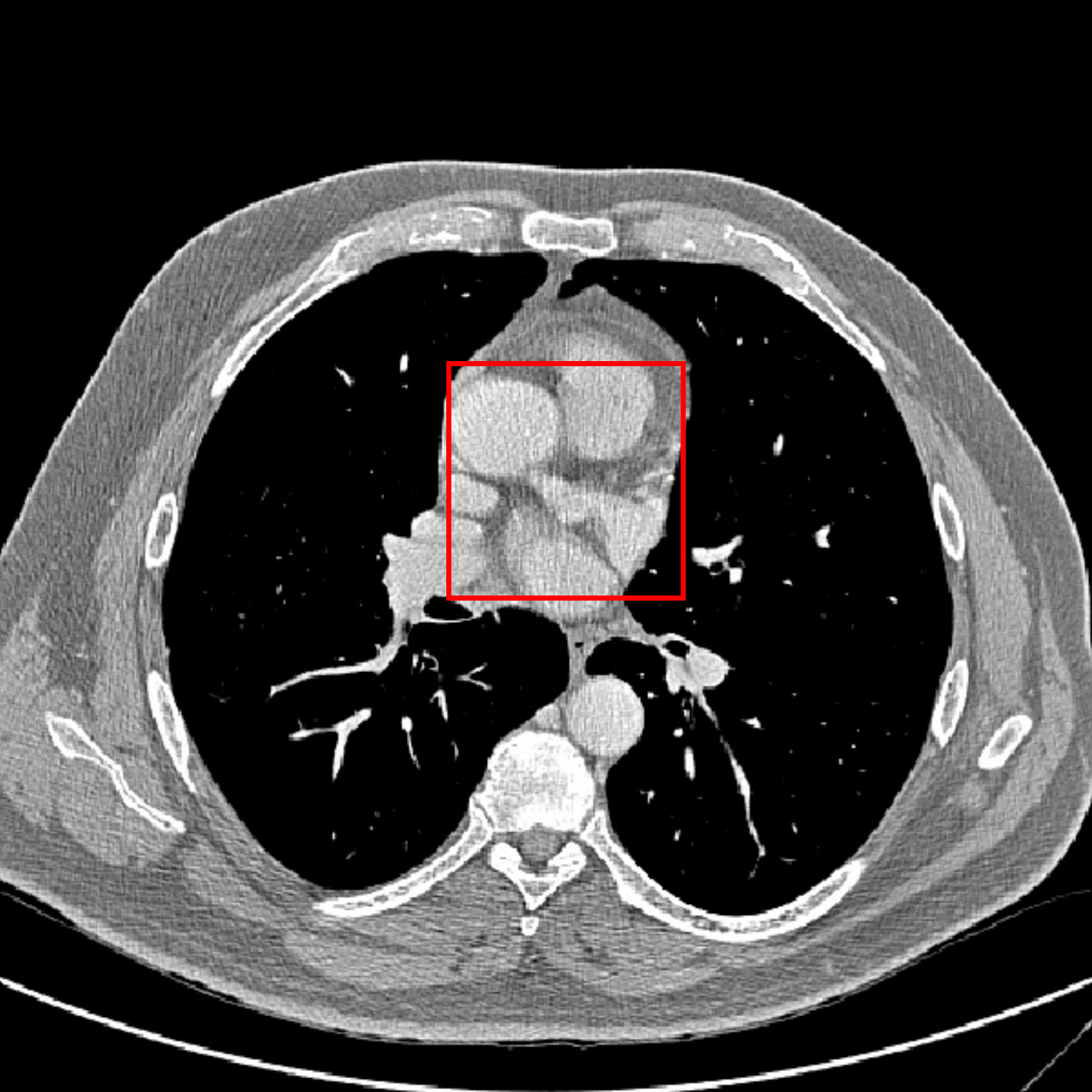} &
    \subfigimg[width=\linewidth]{}{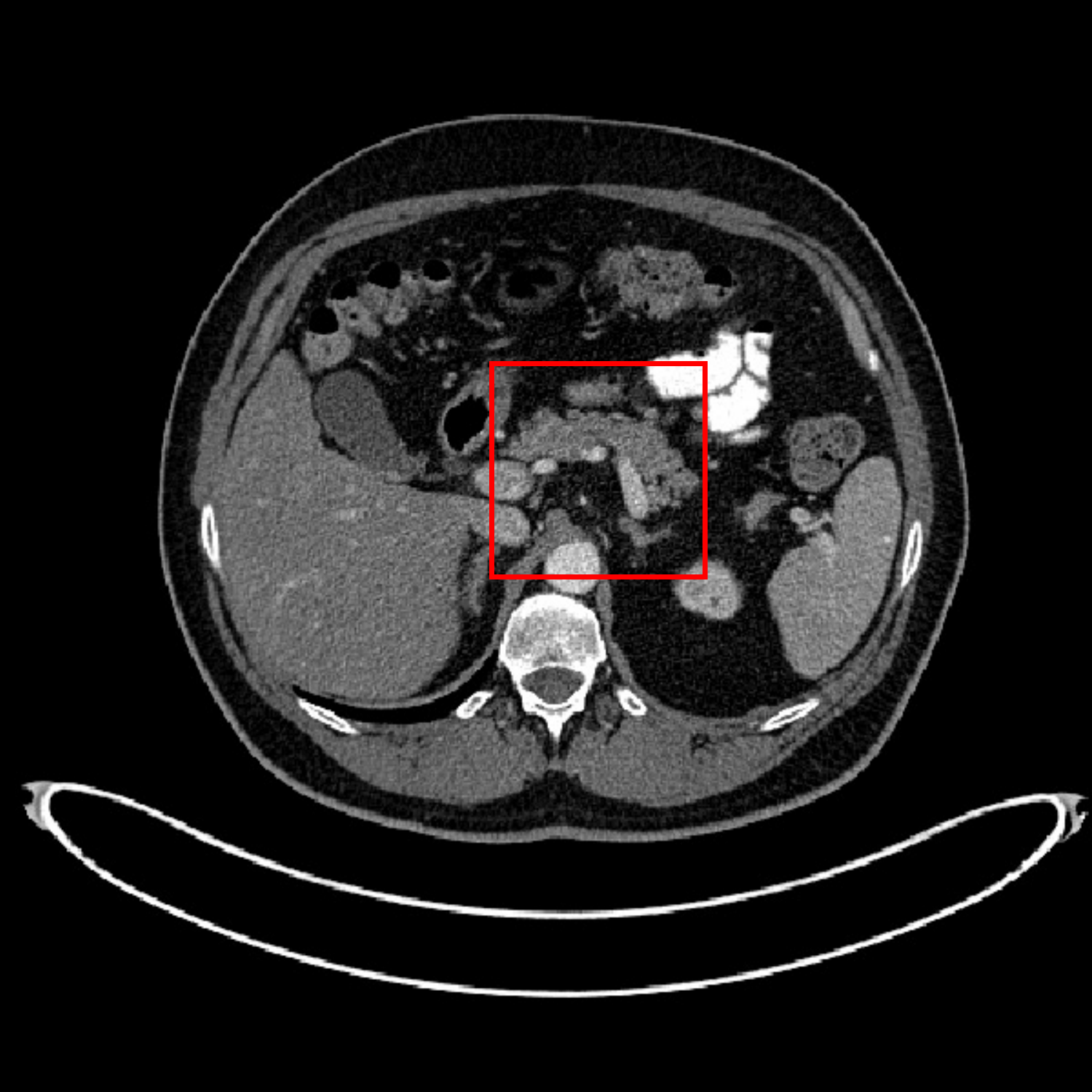}
  \end{tabular}
  \caption{Groundtruth images. The left is GT1 and the right is GT2. The display window for GT1 is $[-240,300]$HU and the display window for GT2 is  $[-180,420]$HU.}
\label{fig:GT-aapm-pancreas}
\end{figure}

\begin{figure}[ht]
  \centering
  \begin{tabular}{@{}p{0.5\linewidth}@{}p{0.5\linewidth}}
    \subfigimg[width=\linewidth]{}{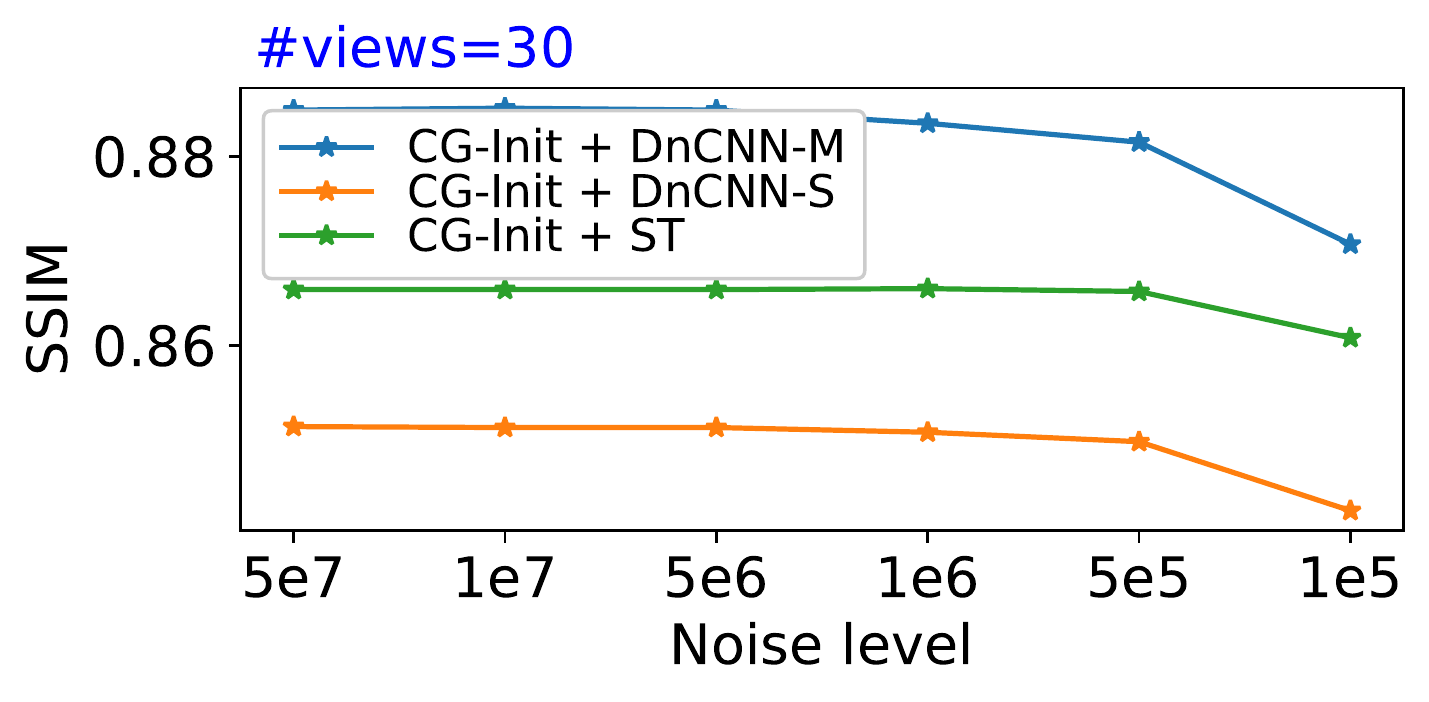} &
    \subfigimg[width=\linewidth]{}{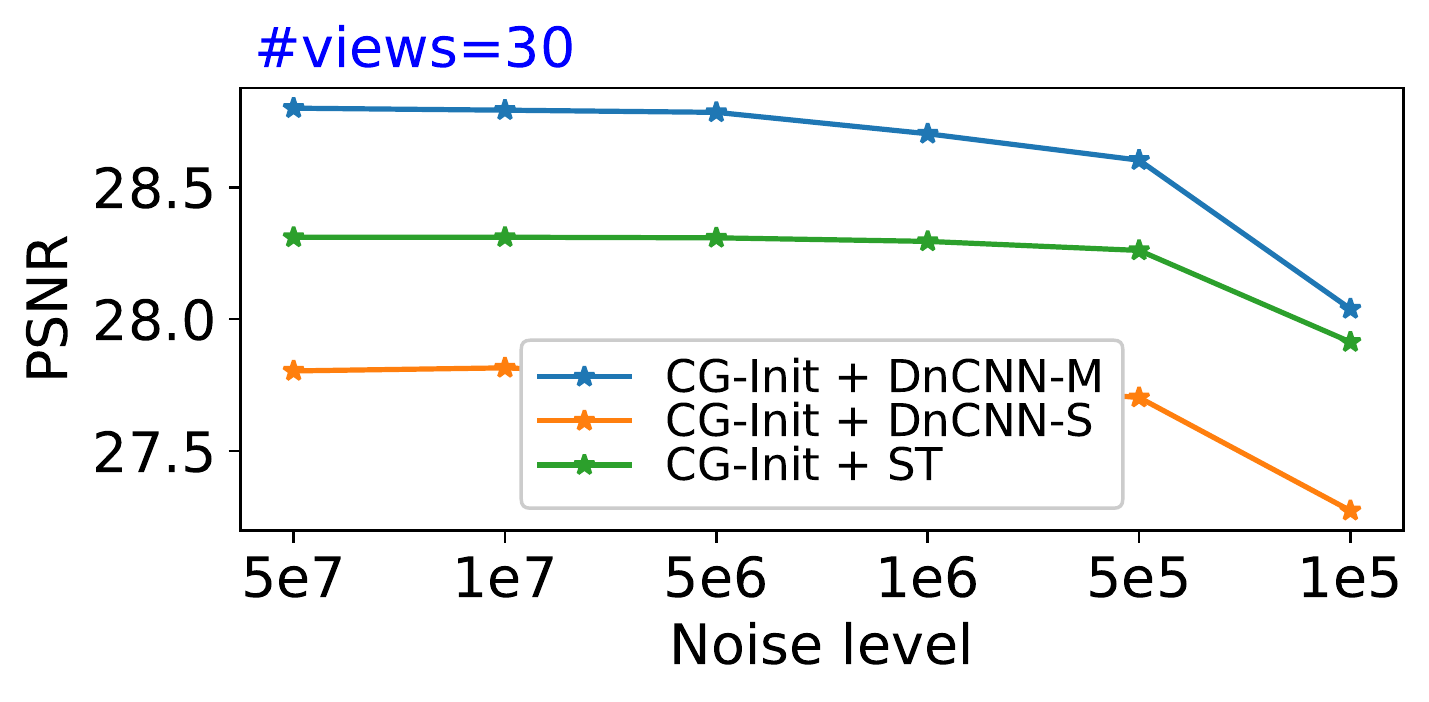} \\

    \subfigimg[width=\linewidth]{}{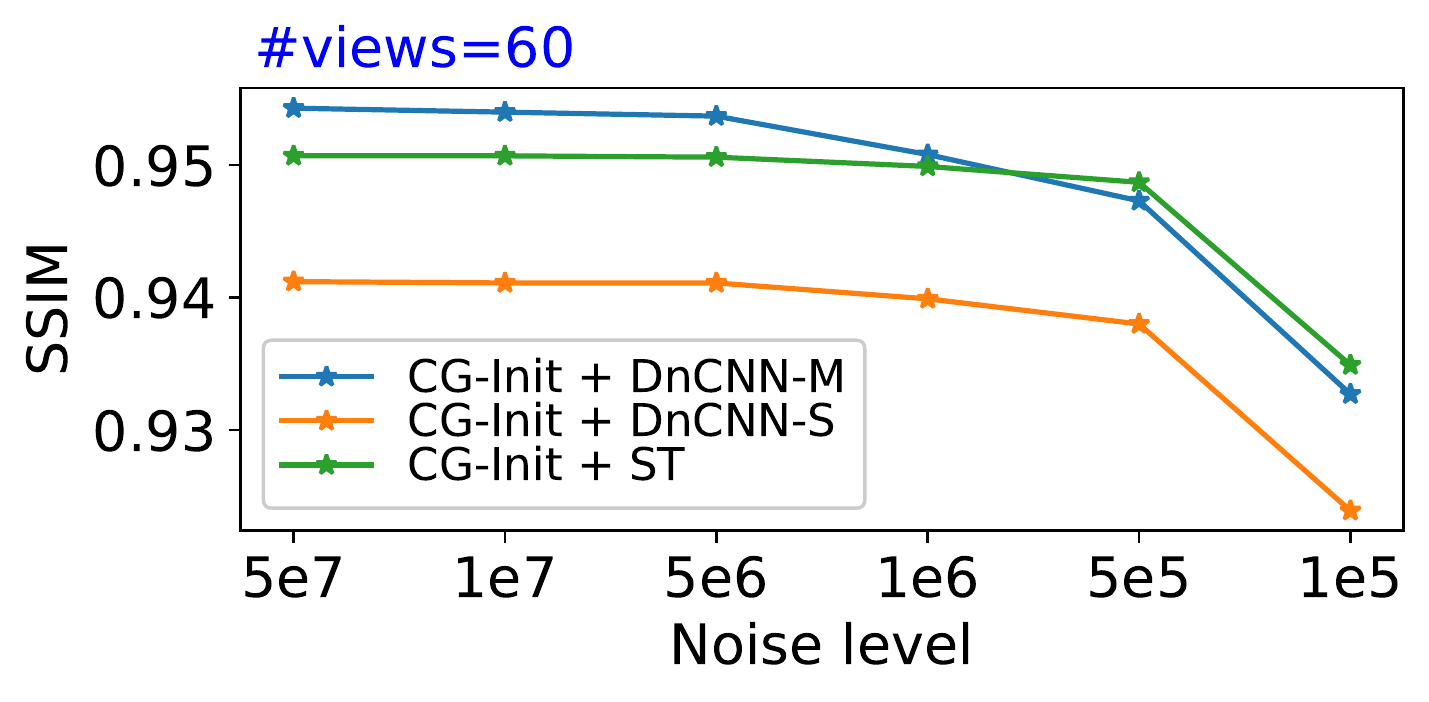} &
    \subfigimg[width=\linewidth]{}{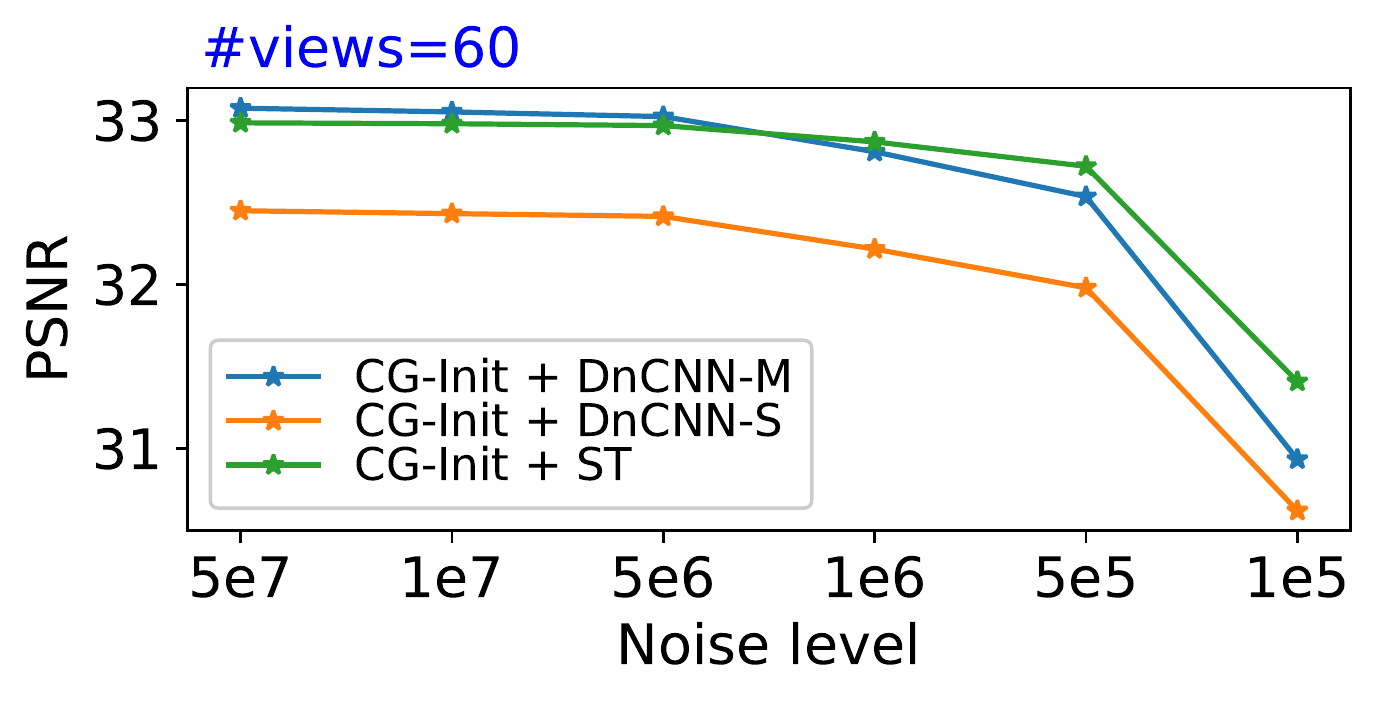} \\

    \subfigimg[width=\linewidth]{}{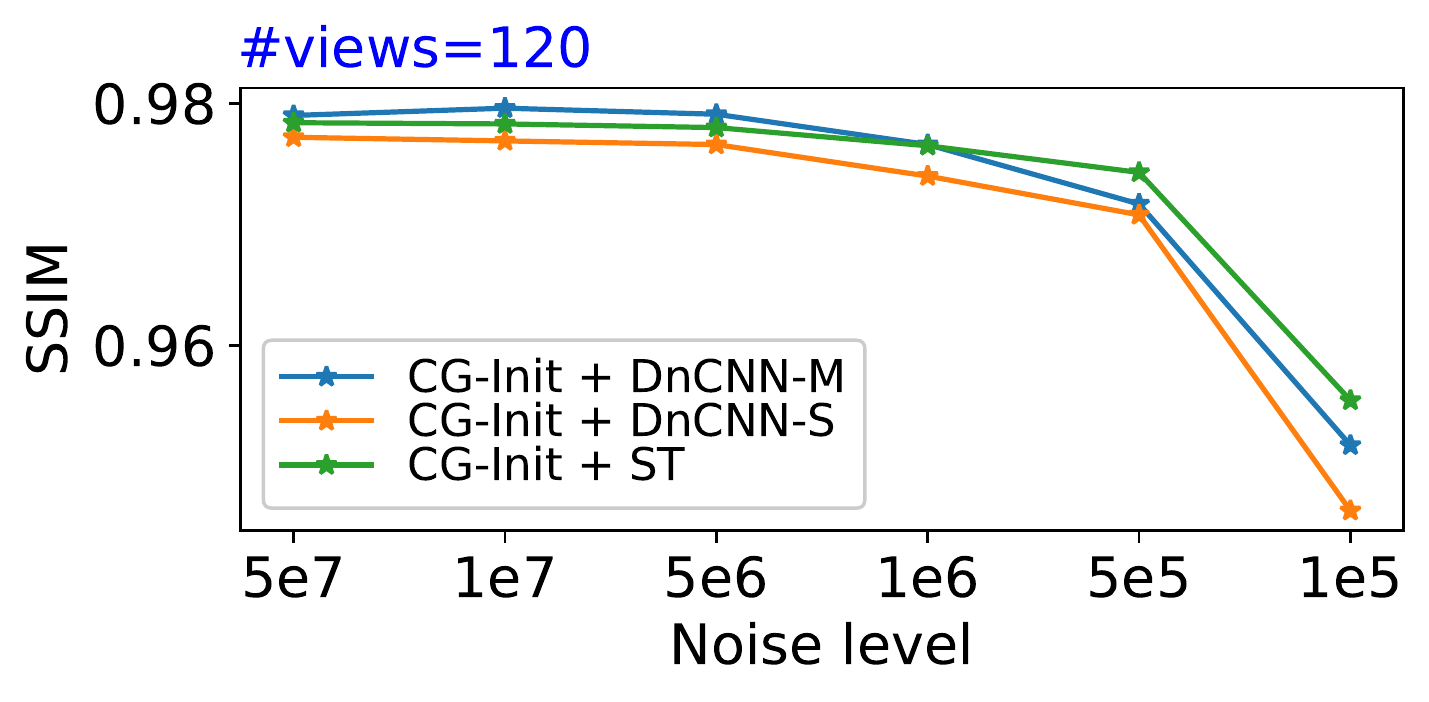} &
    \subfigimg[width=\linewidth]{}{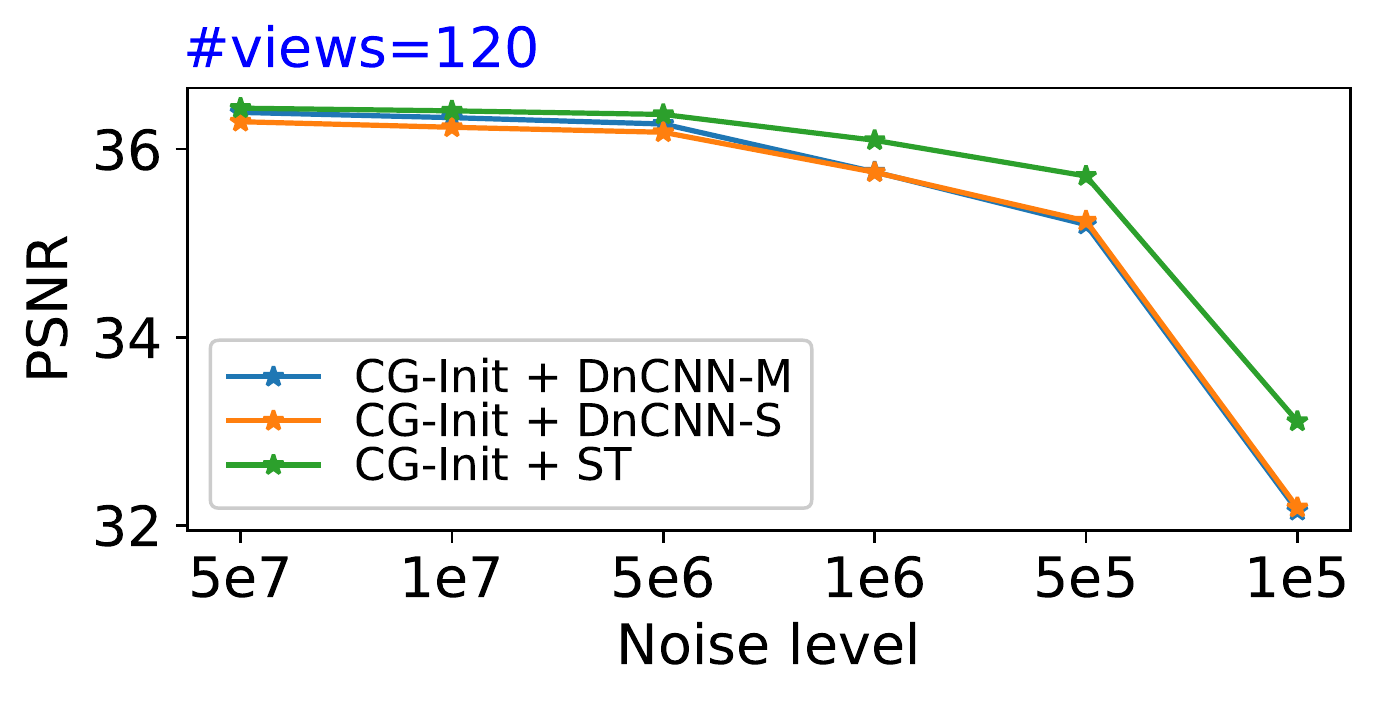} \\
  \end{tabular}
  \caption{MetaInv-Net with different denoiser modules, i.e., CG-Init+ST, CG-Init+DnCNN-M, and CG-Init+DnCNN-S, evaluated at various Poisson noise degraded sinogram data. From the $1^{st}$ to $3^{rd}$ rows show the quantitative results from different number of scanning views, i.e., \#views= 30, 60, 120, respectively. The left column and right column establish the SSIM and PSNR evaluated on AAPM-Test set, respectively.}
\label{fig:diff-denoiser-metainv-net}
\end{figure}

\begin{figure}[ht]
  \centering
  \begin{tabular}{@{}p{0.5\linewidth}@{}p{0.5\linewidth}}
    \subfigimg[width=\linewidth]{}{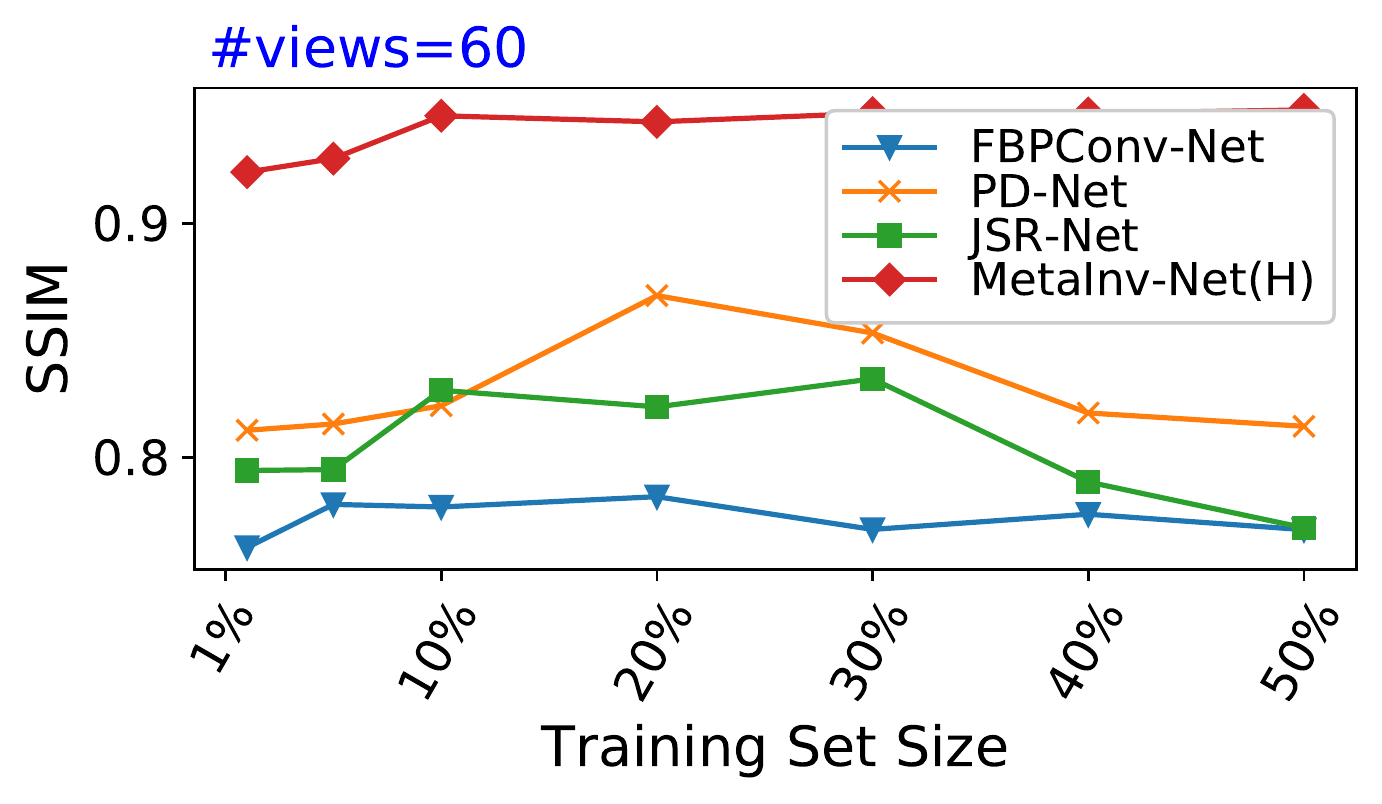} &
    \subfigimg[width=\linewidth]{}{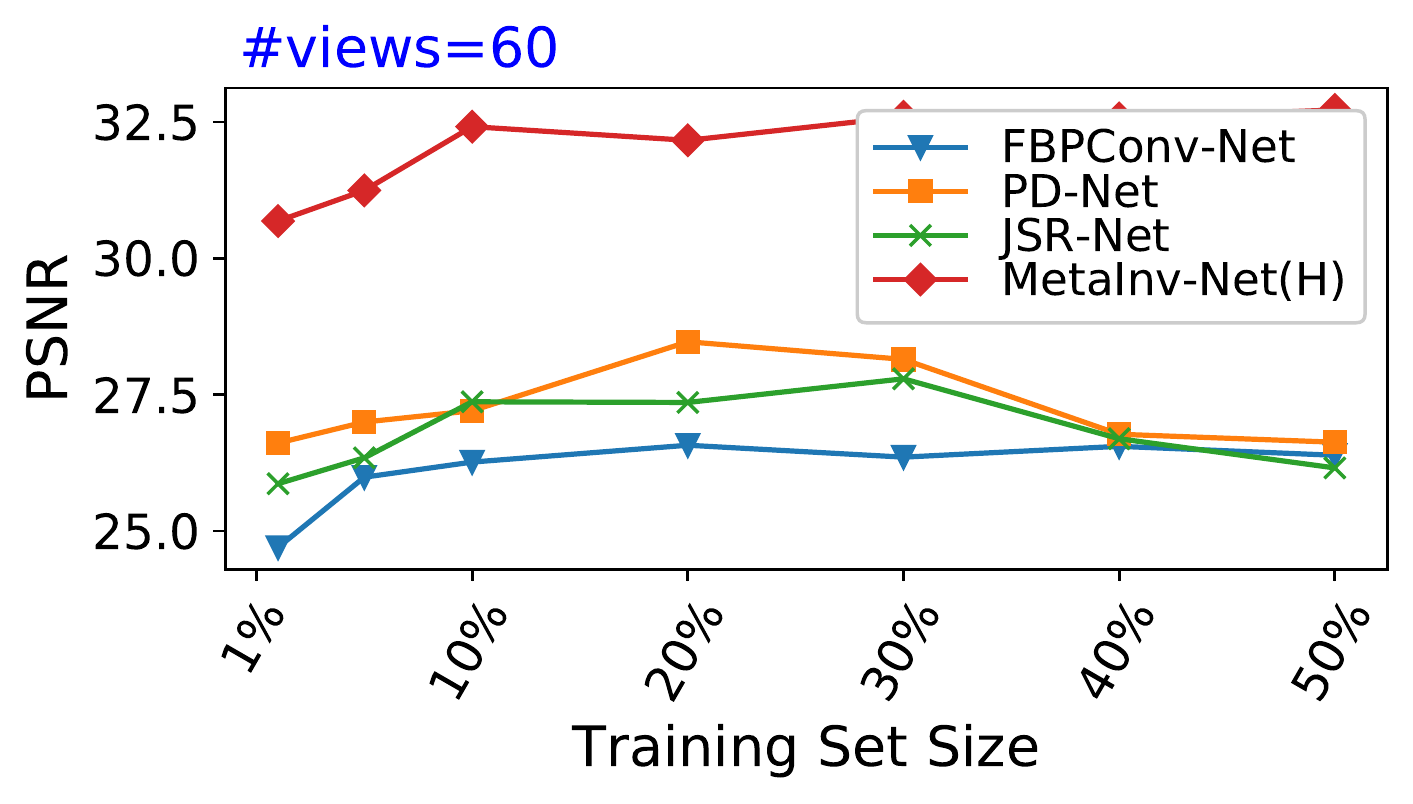} \\

    \subfigimg[width=\linewidth]{}{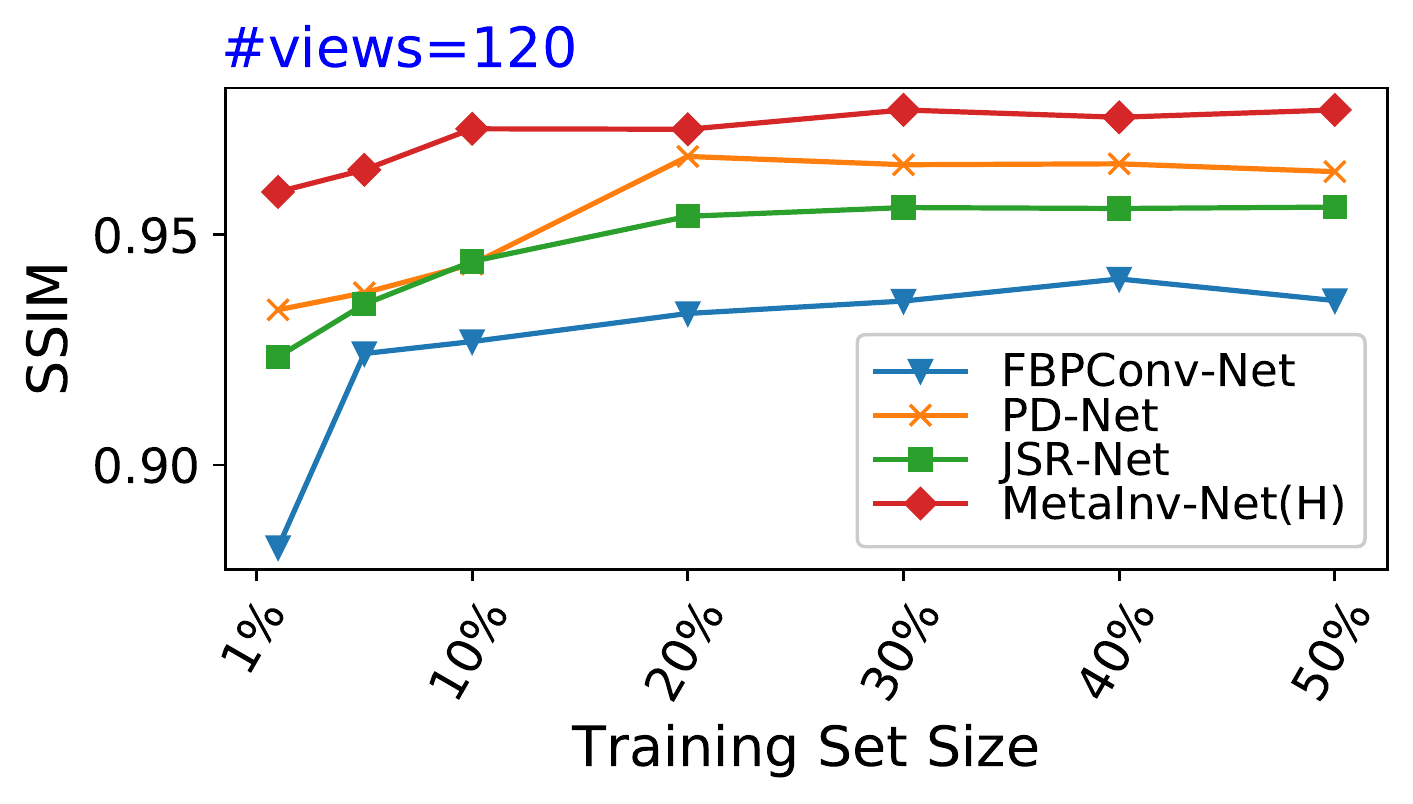} &
    \subfigimg[width=\linewidth]{}{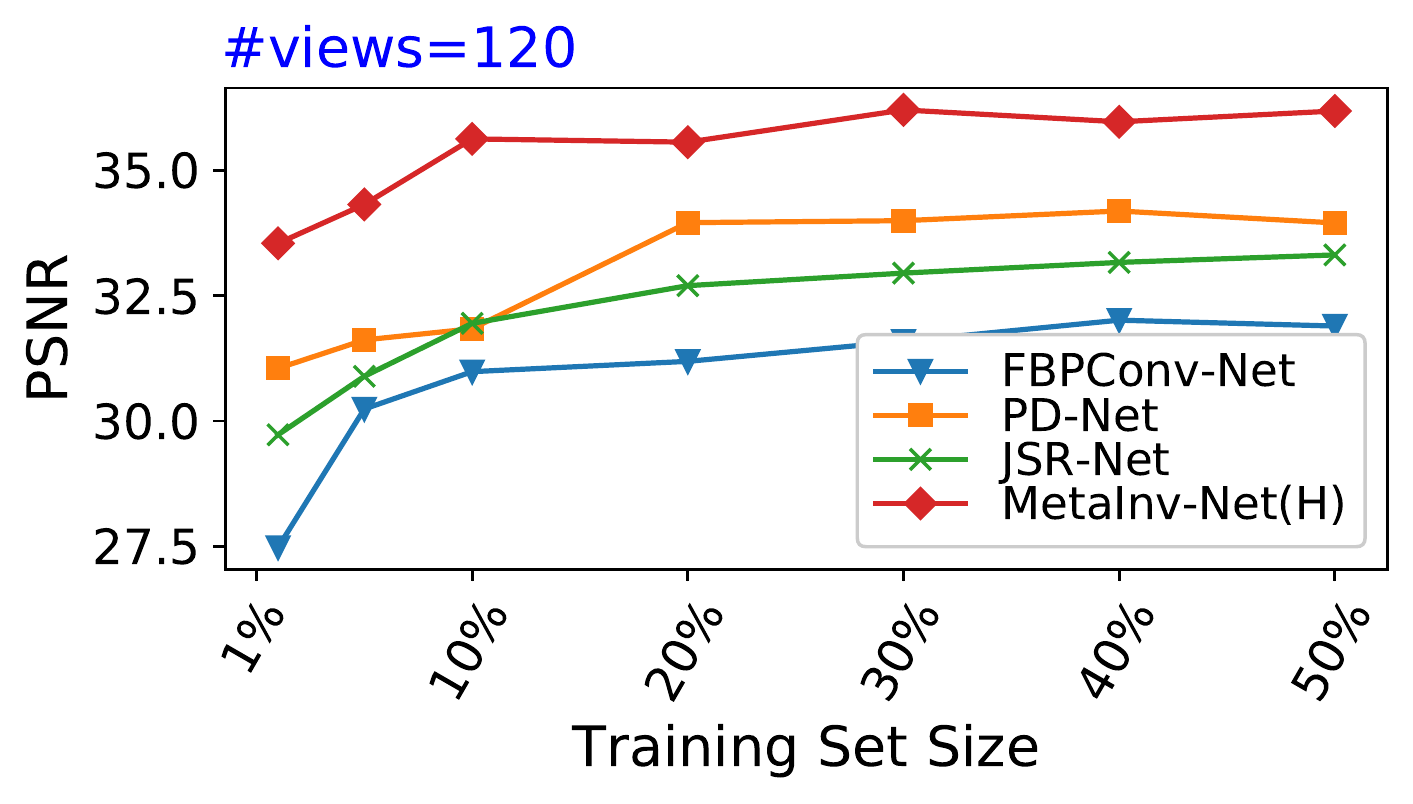} \\

    \subfigimg[width=\linewidth]{}{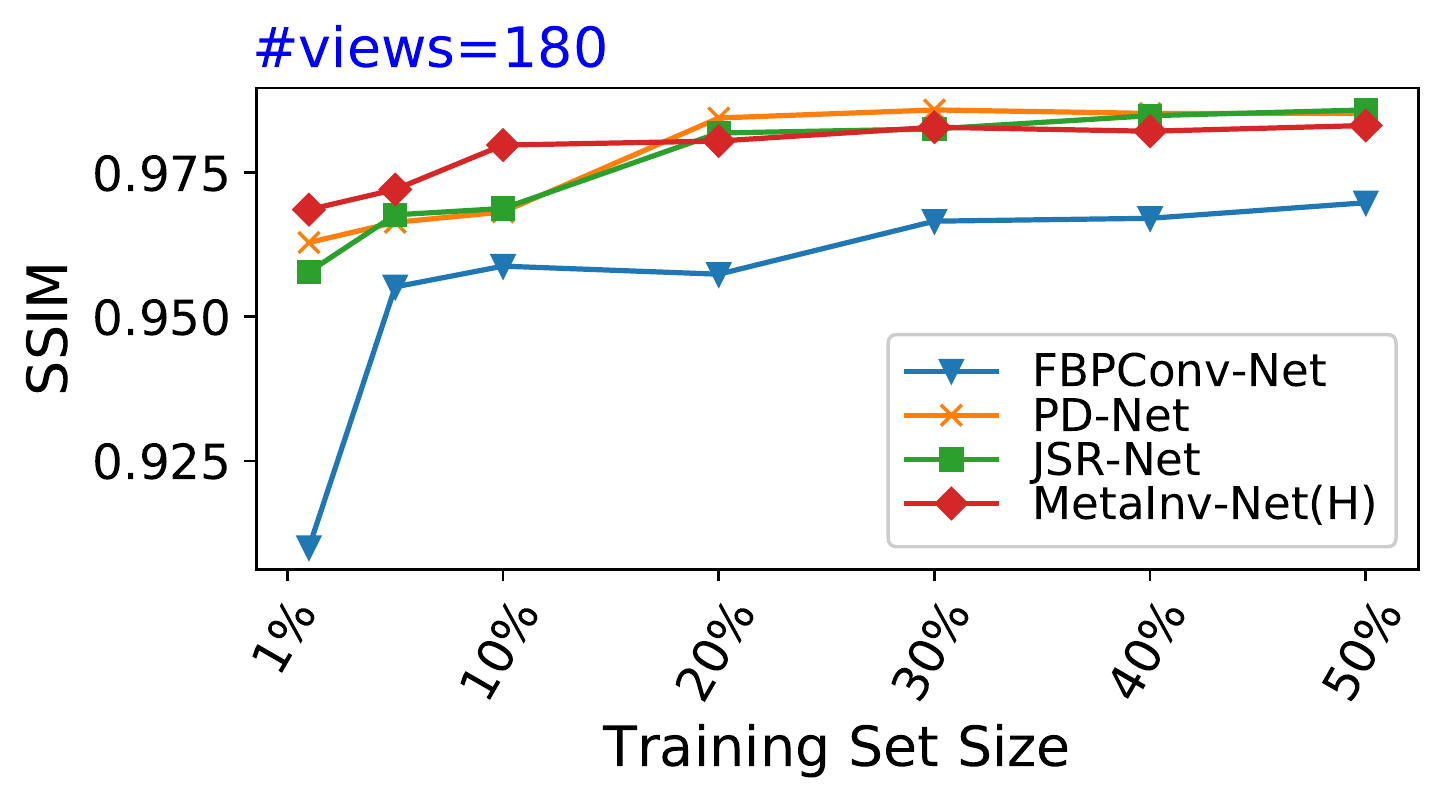} &
    \subfigimg[width=\linewidth]{}{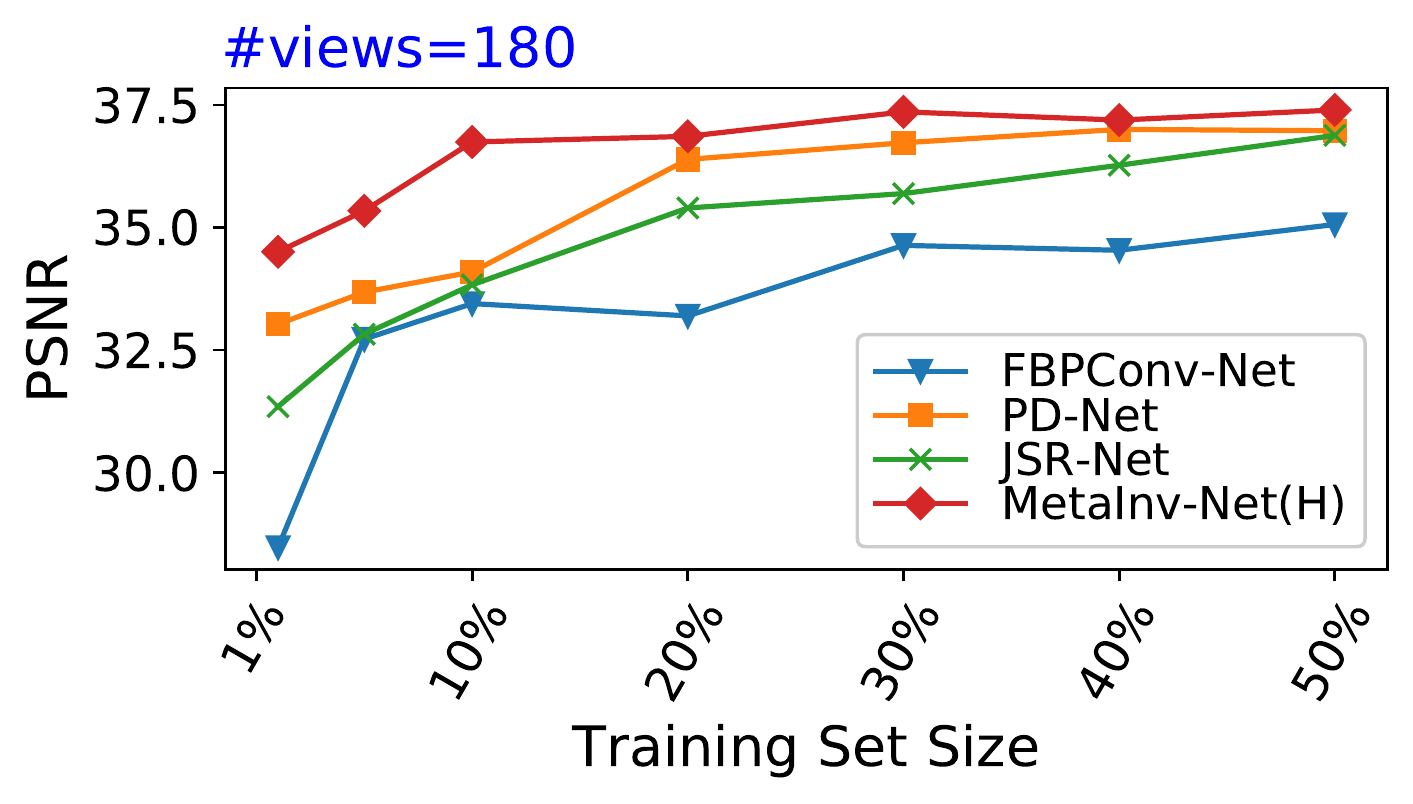} \\
  \end{tabular}
  \caption{Deep models (i.e., FBPConv-Net, PD-Net, JSR-Net, and MetaInv-Net (H)) trained with different training set size, i.e., $1\%, 5\%, 10\%, 20\%, 30\%, 40\%, \mbox{and}\;50\%$. From the $1^{st}$ to $3^{rd}$ rows show the quantitative values at number of scanning views $60, 120,$ and $180$, respectively. The left and right columns show the SSIM and PSNR, respectively.}
\label{fig:diff-train-set-size-aapm-test-set}
\end{figure}

\begin{figure*}[t]
  \centering
  \begin{tabular}{@{}p{0.1314\linewidth}@{}p{0.1314\linewidth}@{}p{0.1314\linewidth}@{}p{0.1314\linewidth}@{}p{0.1314\linewidth}@{}p{0.1314\linewidth}@{}p{0.1314\linewidth}} 
    \subfigimg[width=\linewidth]{}{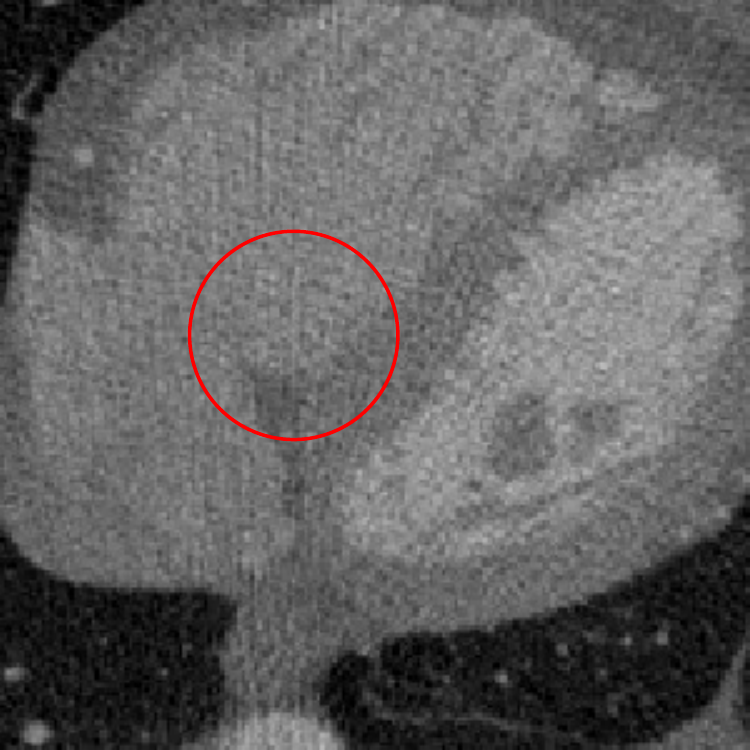}&
    \subfigimg[width=\linewidth]{}{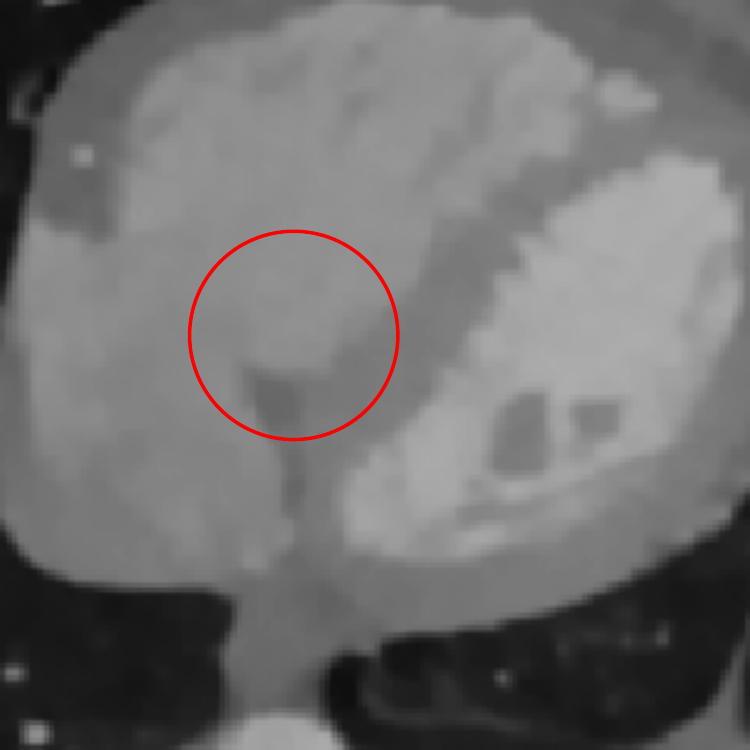} &
    \subfigimg[width=\linewidth]{}{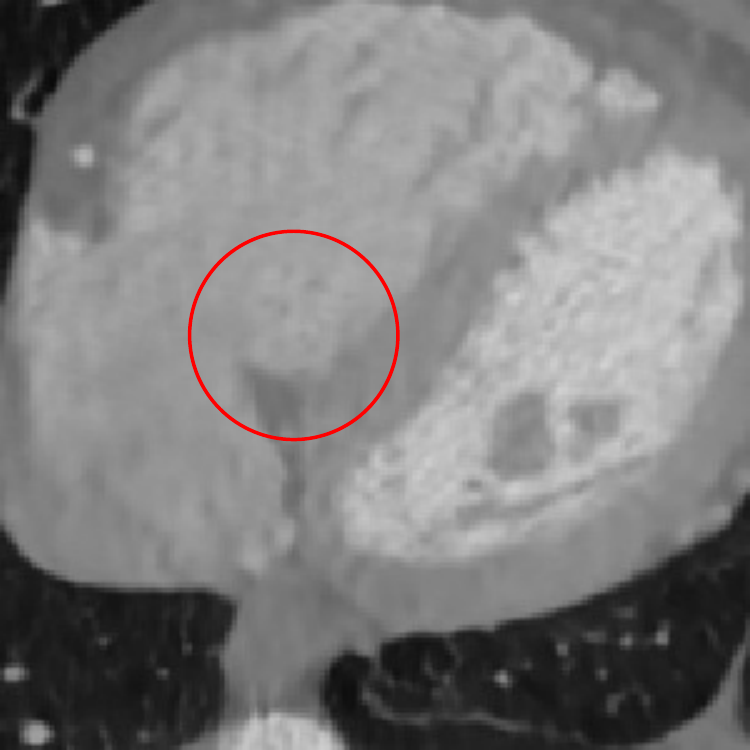} &
    \subfigimg[width=\linewidth]{}{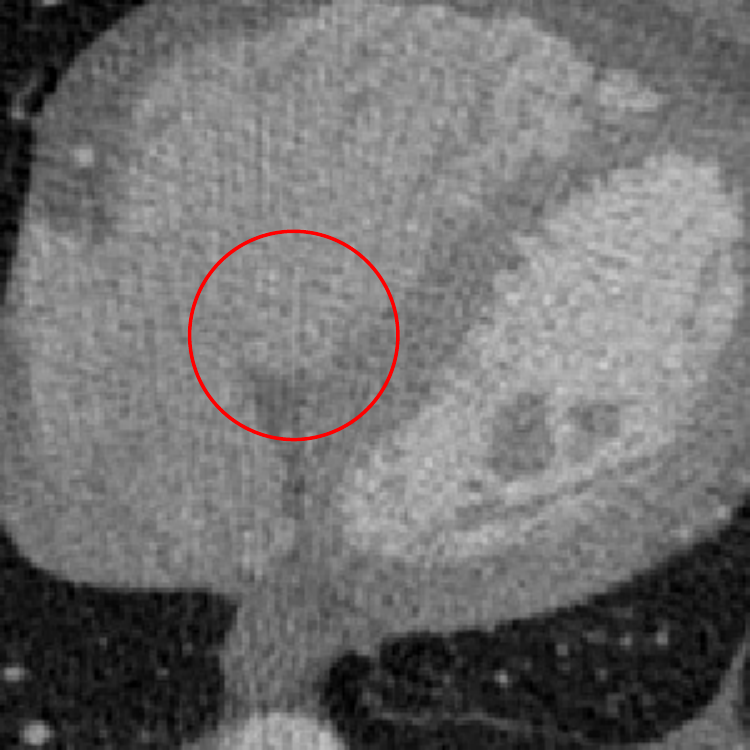} &
    \subfigimg[width=\linewidth]{}{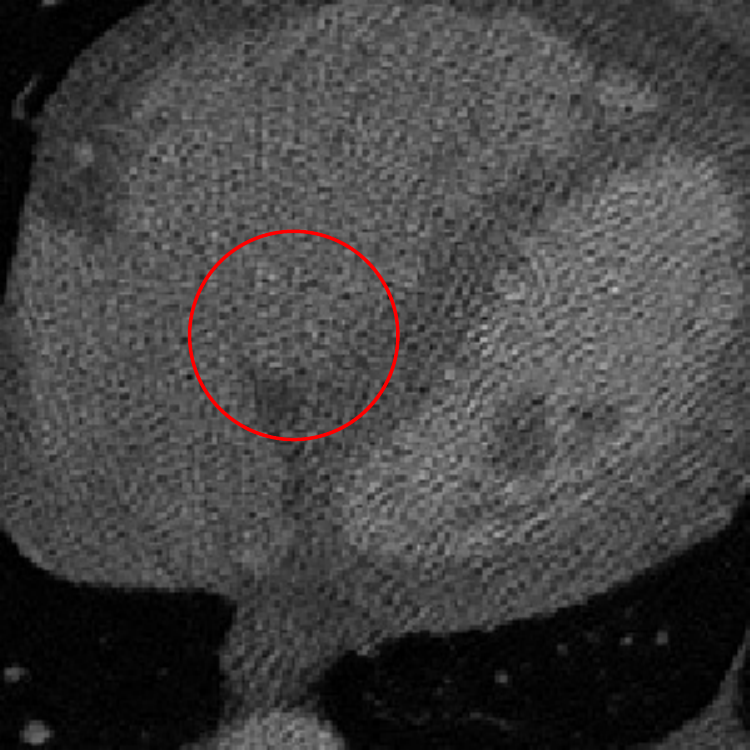} &
    \subfigimg[width=\linewidth]{}{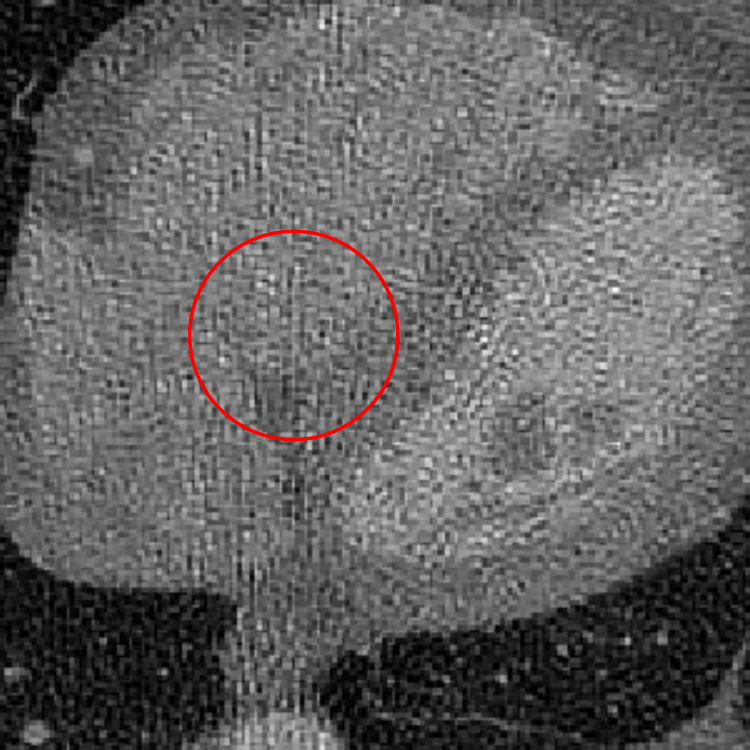} &
    \subfigimg[width=\linewidth]{}{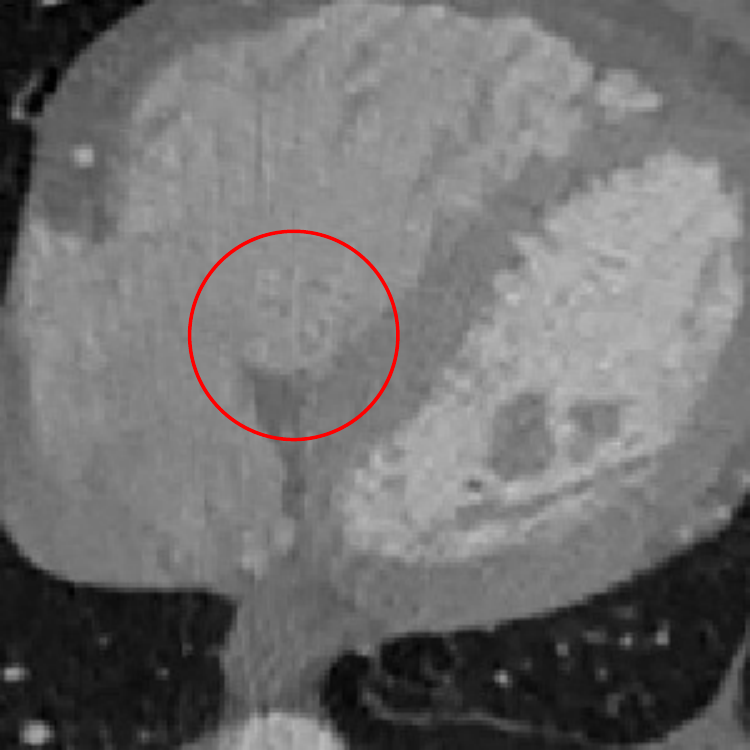} \\

    \subfigimg[width=\linewidth]{}{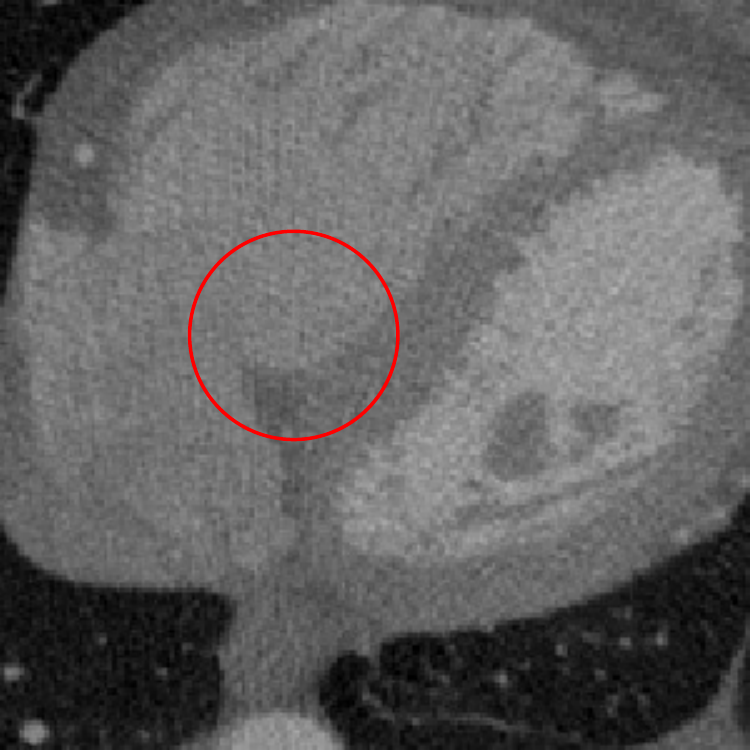}&
    \subfigimg[width=\linewidth]{}{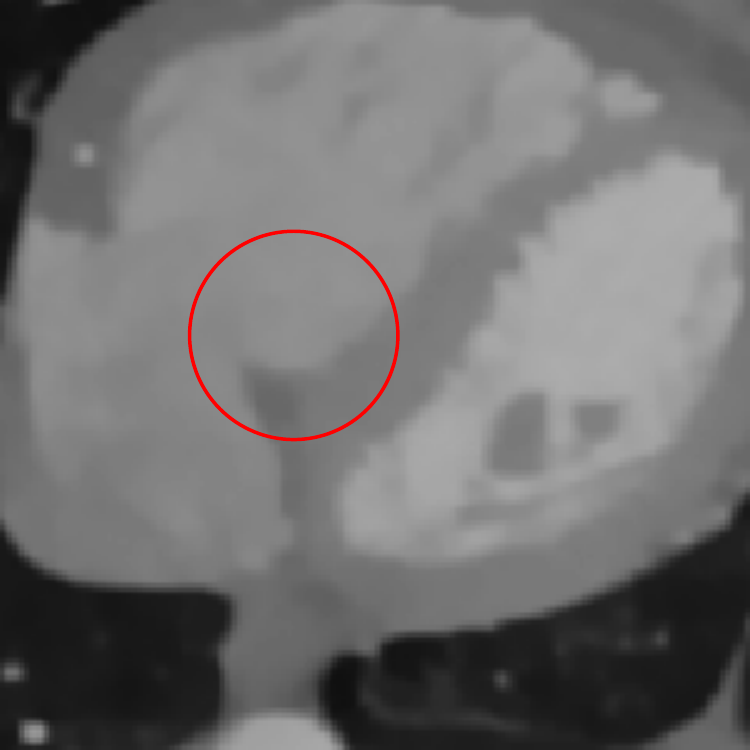} &
    \subfigimg[width=\linewidth]{}{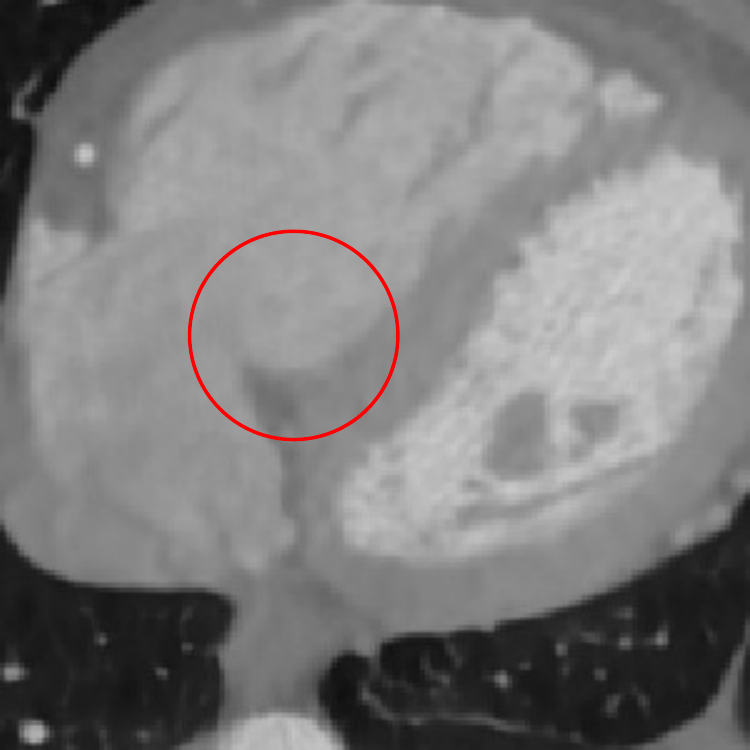} &
    \subfigimg[width=\linewidth]{}{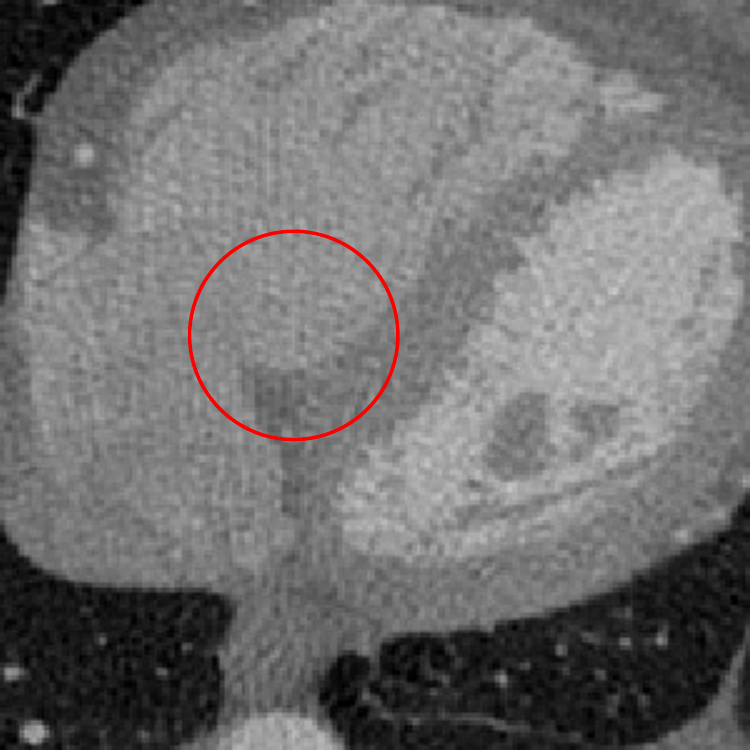} &
    \subfigimg[width=\linewidth]{}{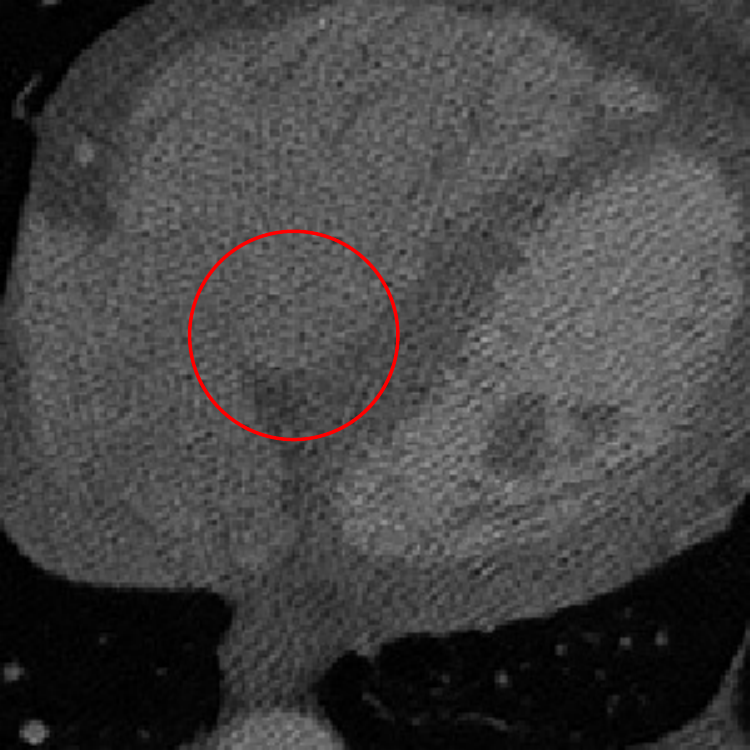} &
    \subfigimg[width=\linewidth]{}{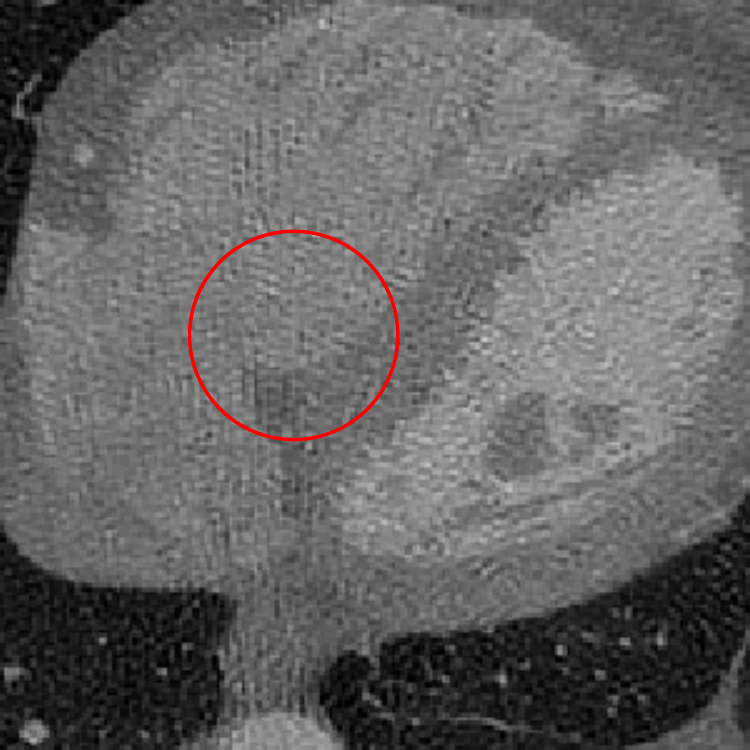} &
    \subfigimg[width=\linewidth]{}{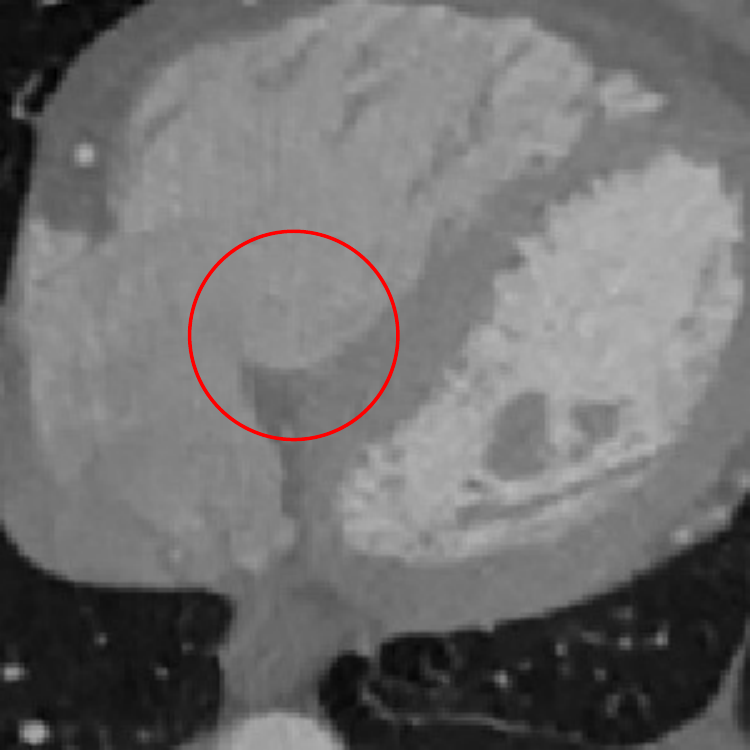} \\
    
    \subfigimg[width=\linewidth]{}{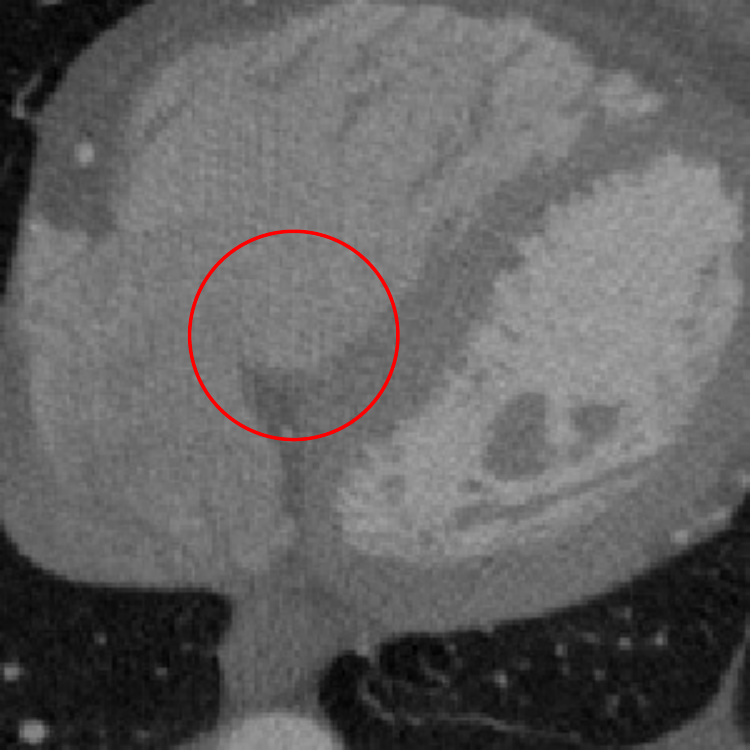}&
    \subfigimg[width=\linewidth]{}{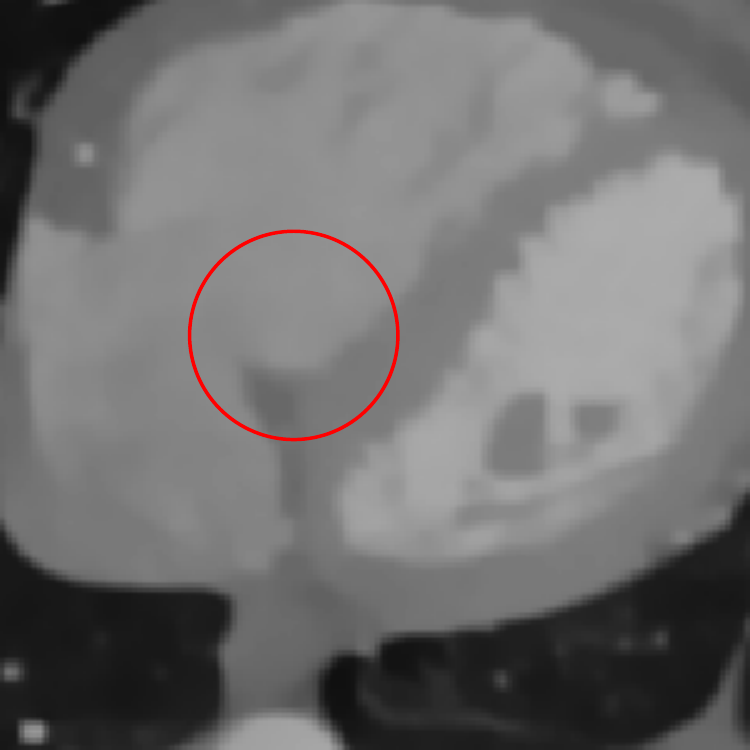} &
    \subfigimg[width=\linewidth]{}{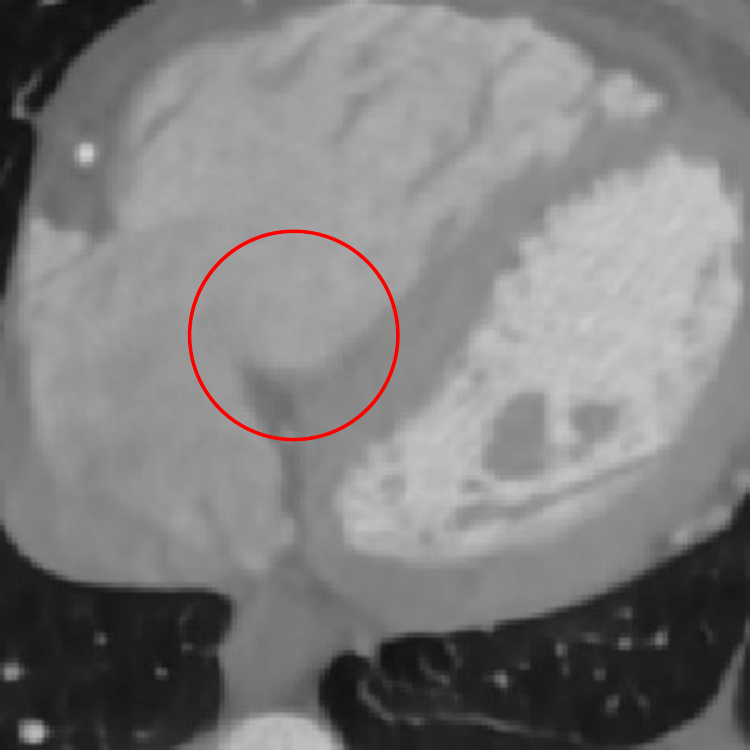} &
    \subfigimg[width=\linewidth]{}{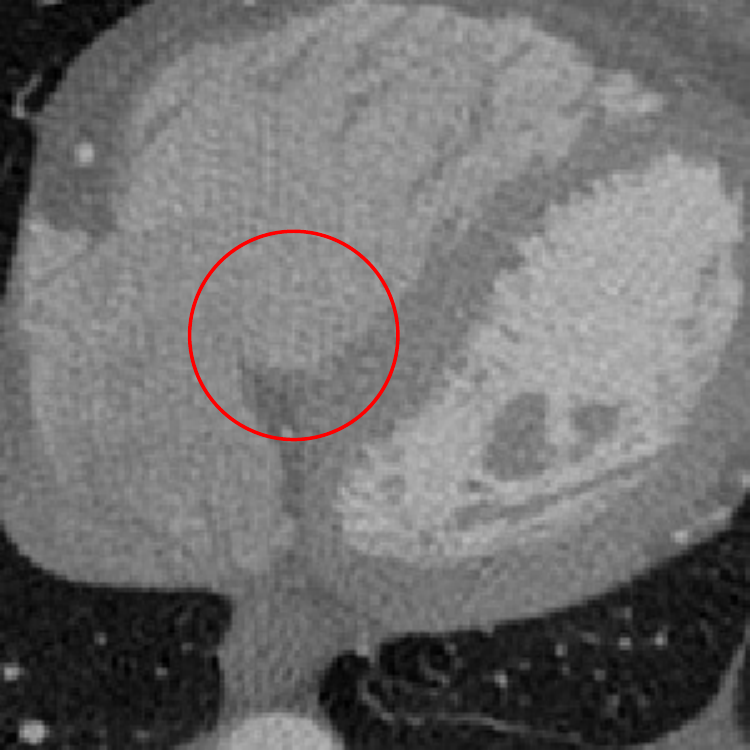} &
    \subfigimg[width=\linewidth]{}{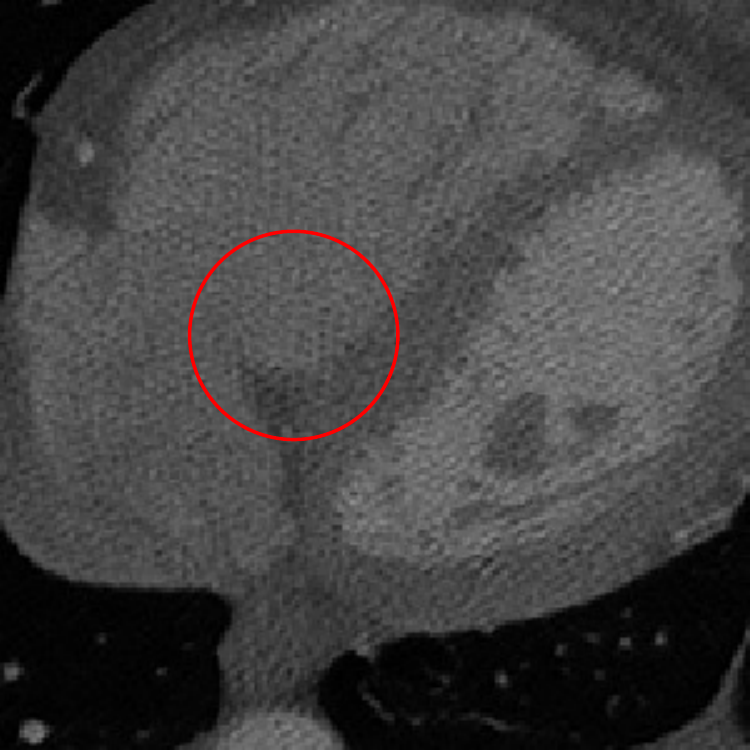} &
    \subfigimg[width=\linewidth]{}{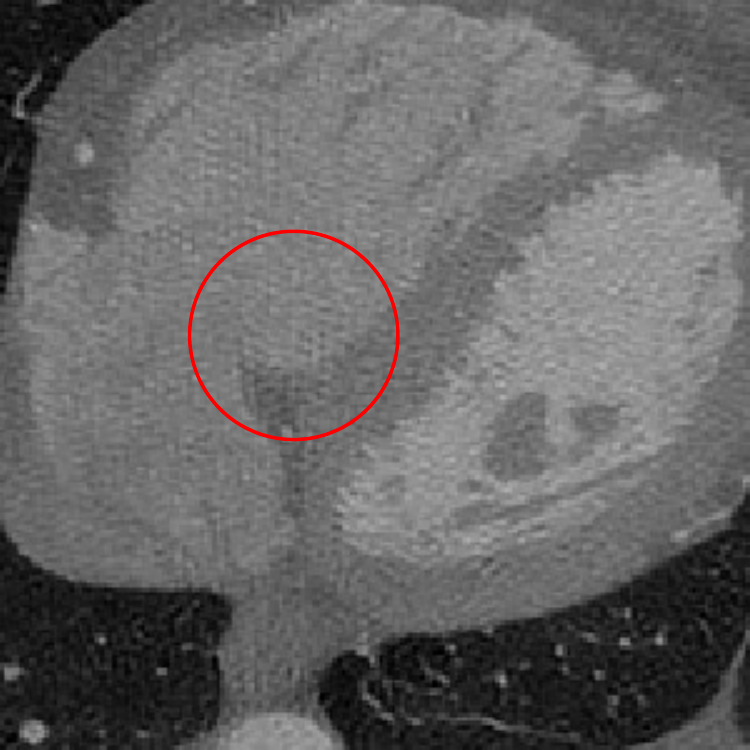} &
    \subfigimg[width=\linewidth]{}{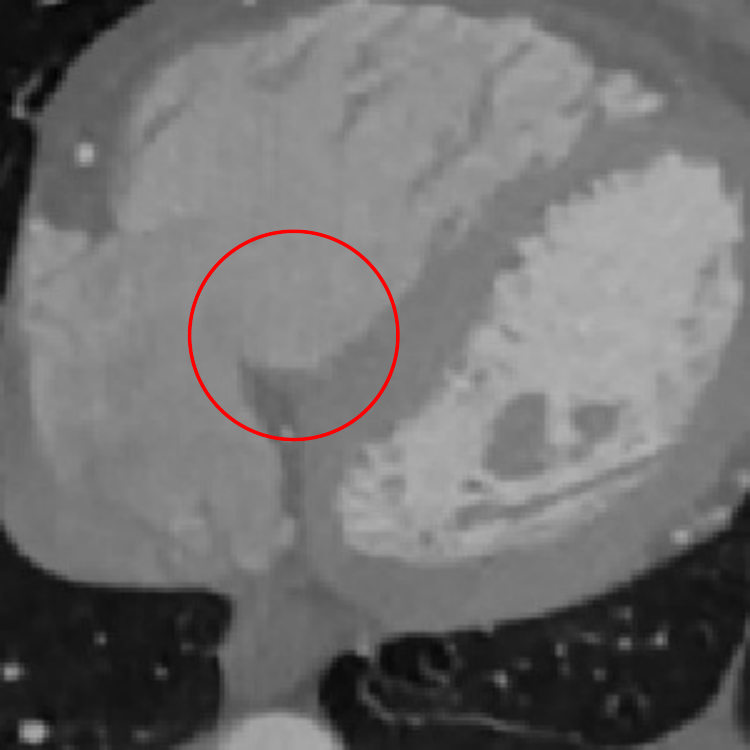} \\
    
    \subfigimg[width=\linewidth]{}{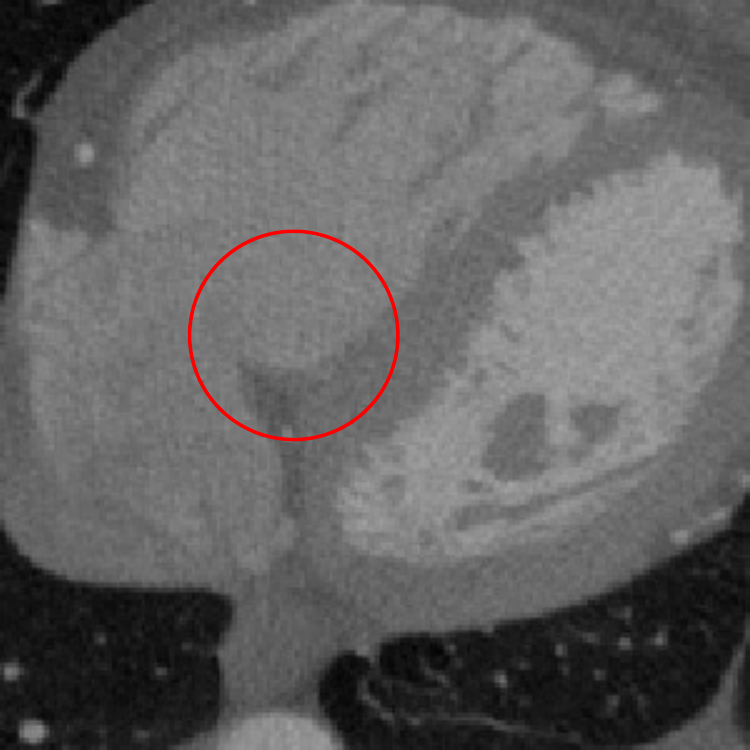}&
    \subfigimg[width=\linewidth]{}{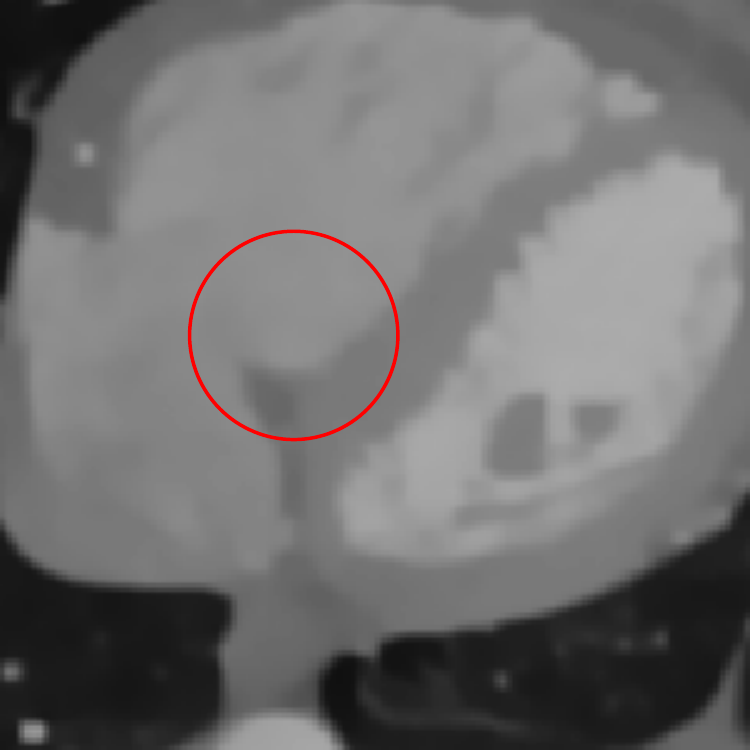} &
    \subfigimg[width=\linewidth]{}{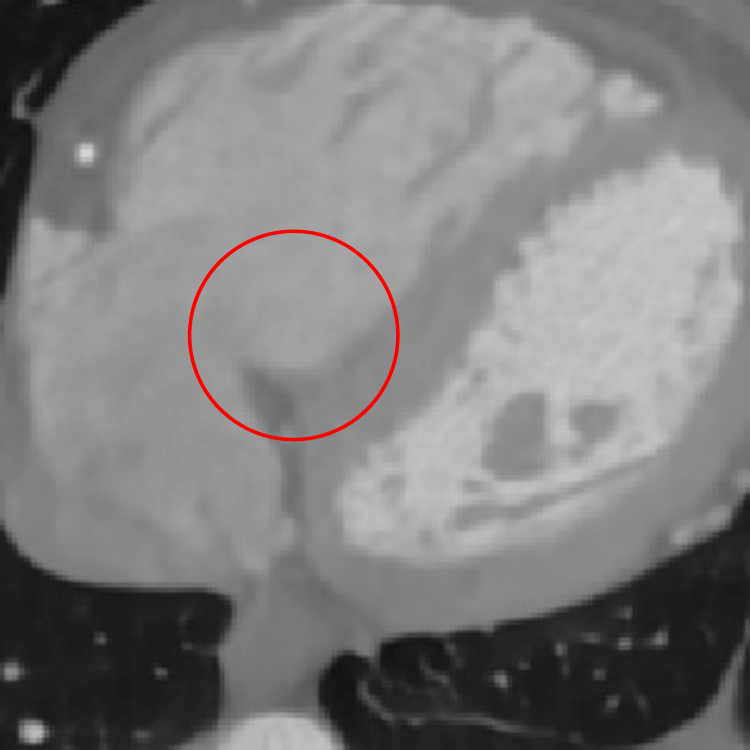} &
    \subfigimg[width=\linewidth]{}{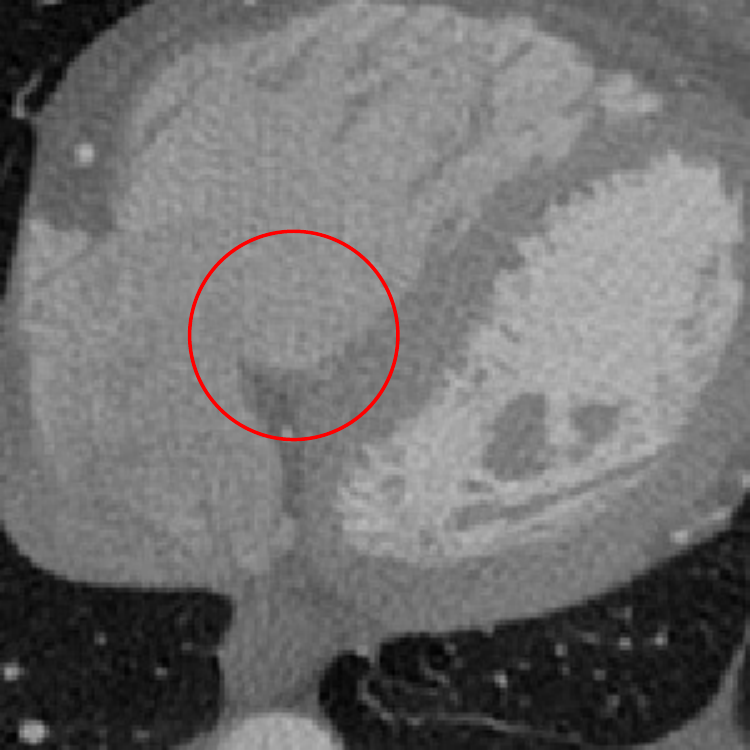} &
    \subfigimg[width=\linewidth]{}{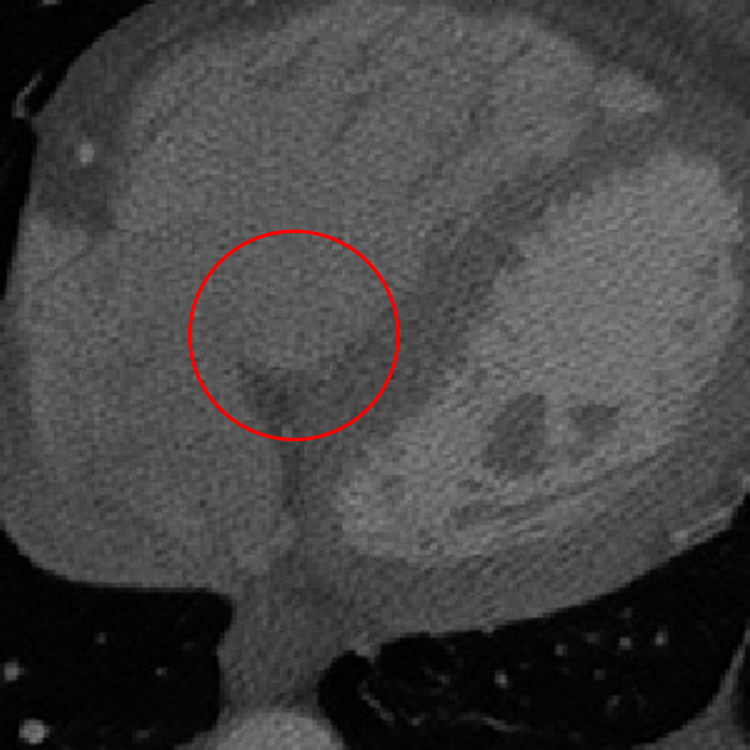} &
    \subfigimg[width=\linewidth]{}{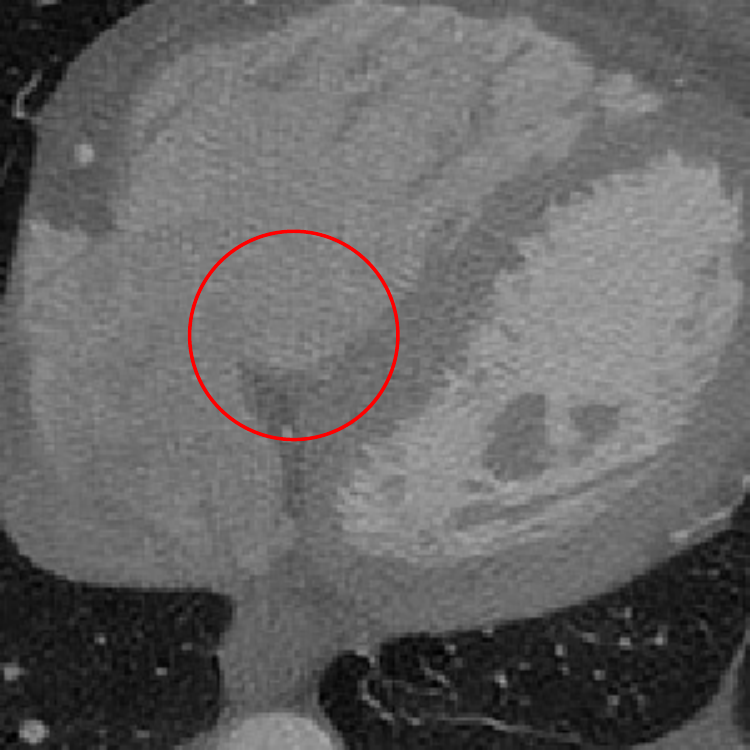} &
    \subfigimg[width=\linewidth]{}{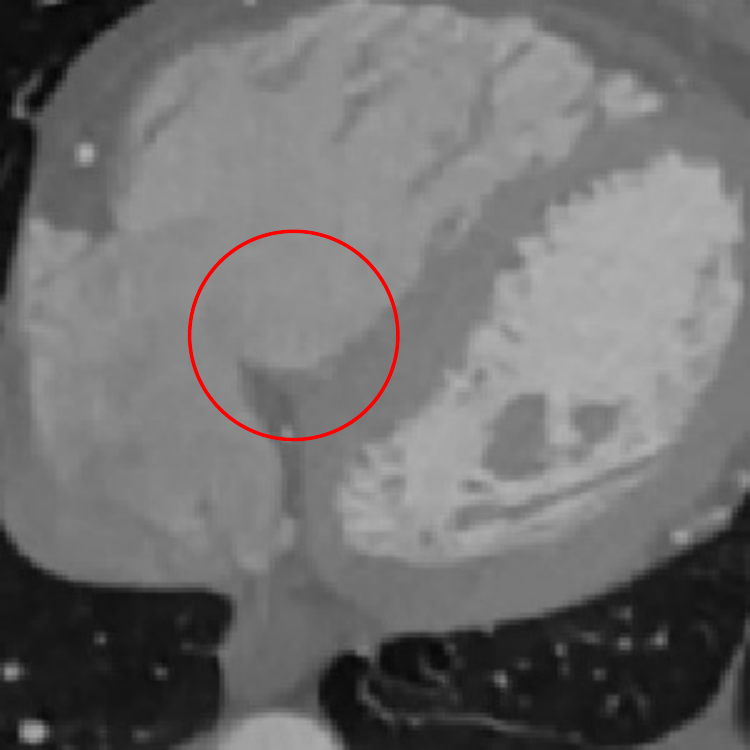} \\
    
    \subfigimg[width=\linewidth]{}{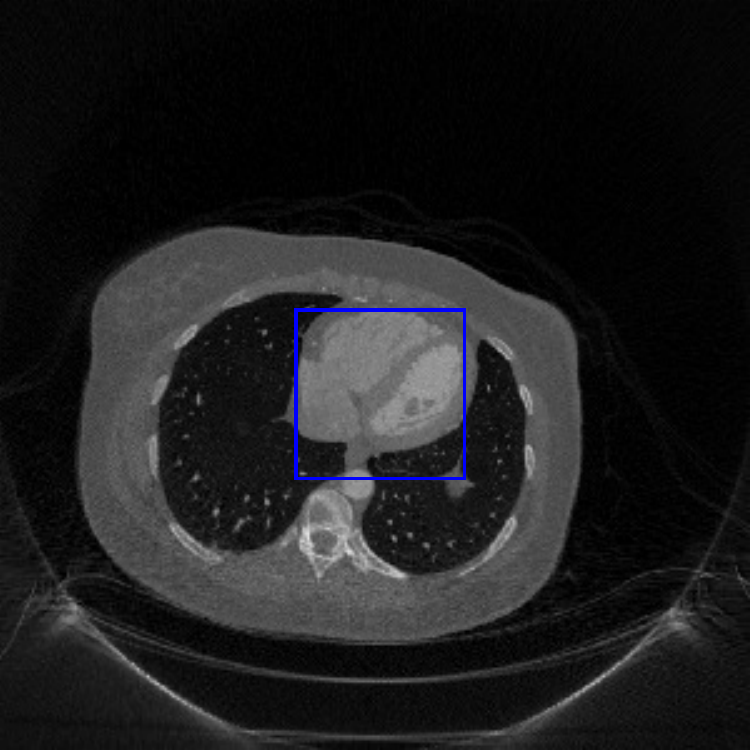}&
    \subfigimg[width=\linewidth]{}{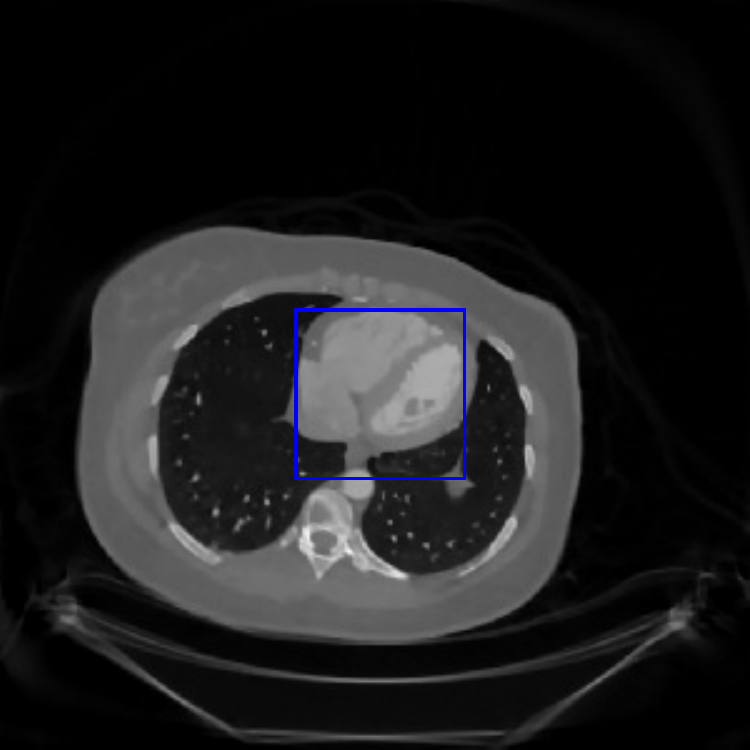} &
    \subfigimg[width=\linewidth]{}{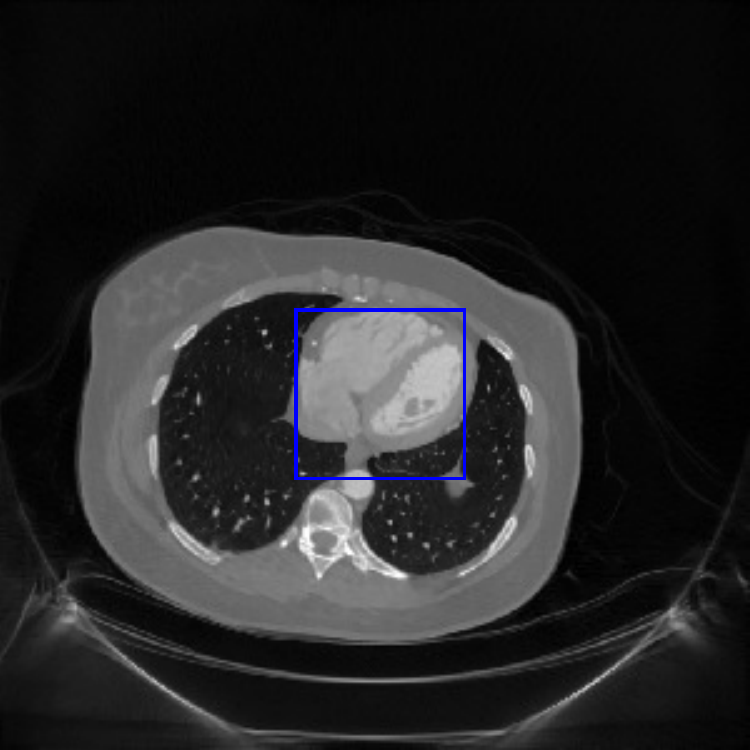} &
    \subfigimg[width=\linewidth]{}{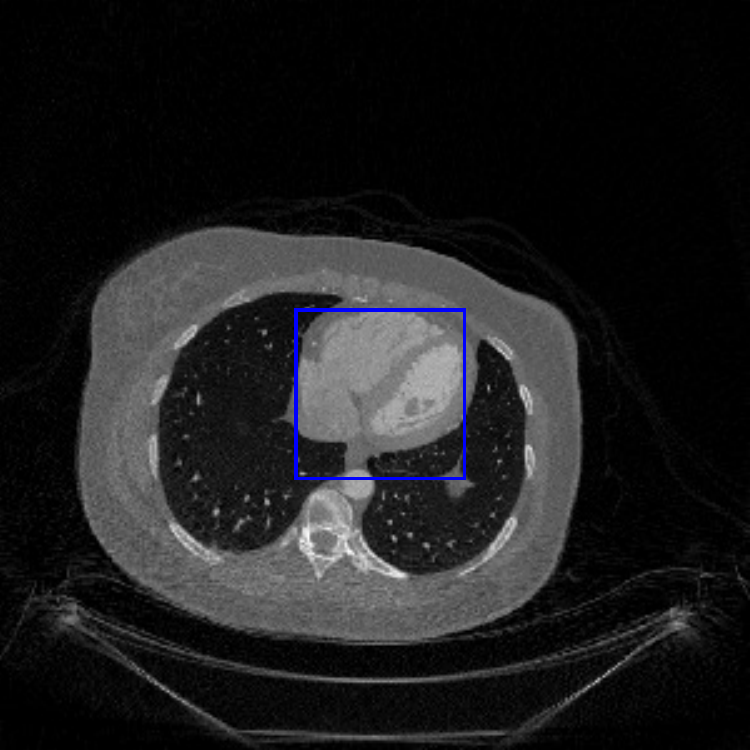} &
    \subfigimg[width=\linewidth]{}{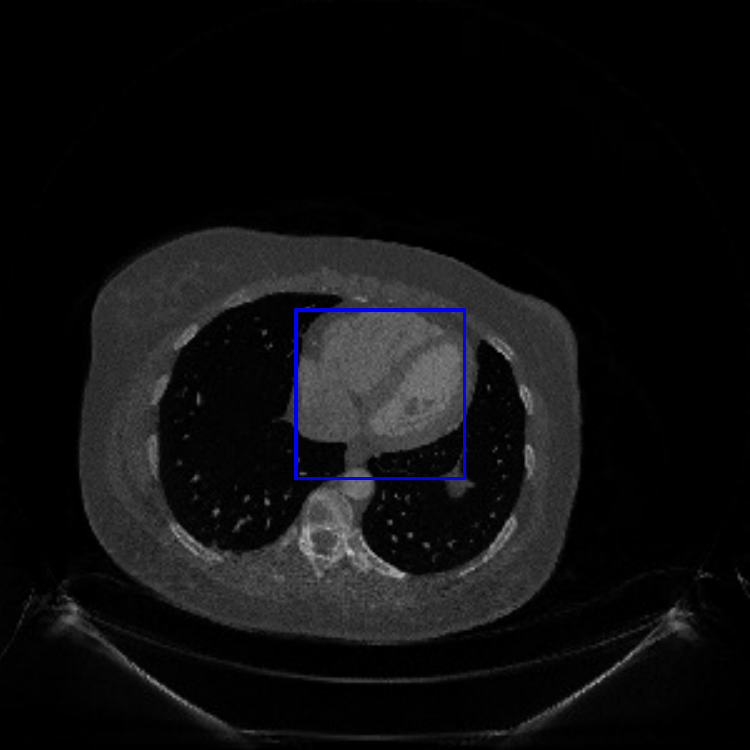} &
    \subfigimg[width=\linewidth]{}{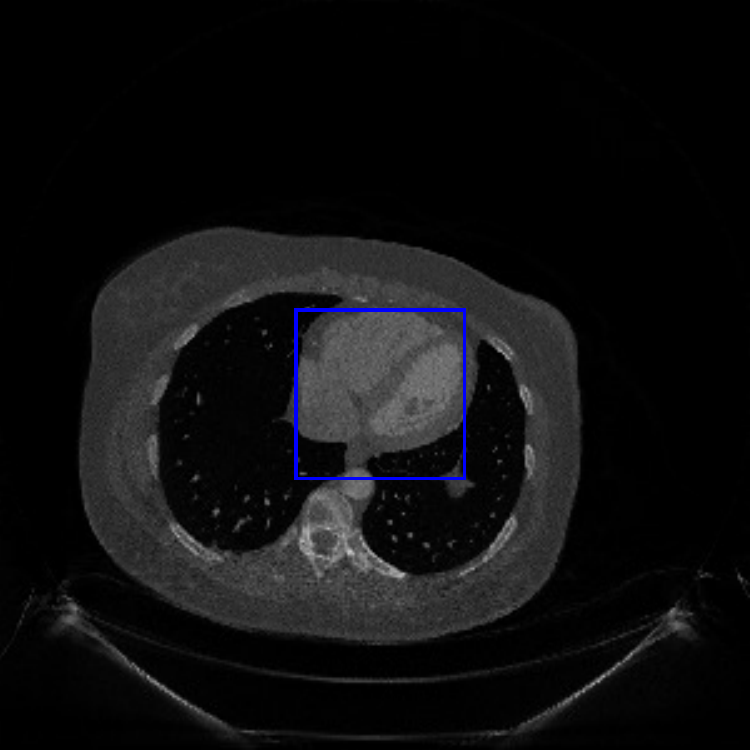} &
    \subfigimg[width=\linewidth]{}{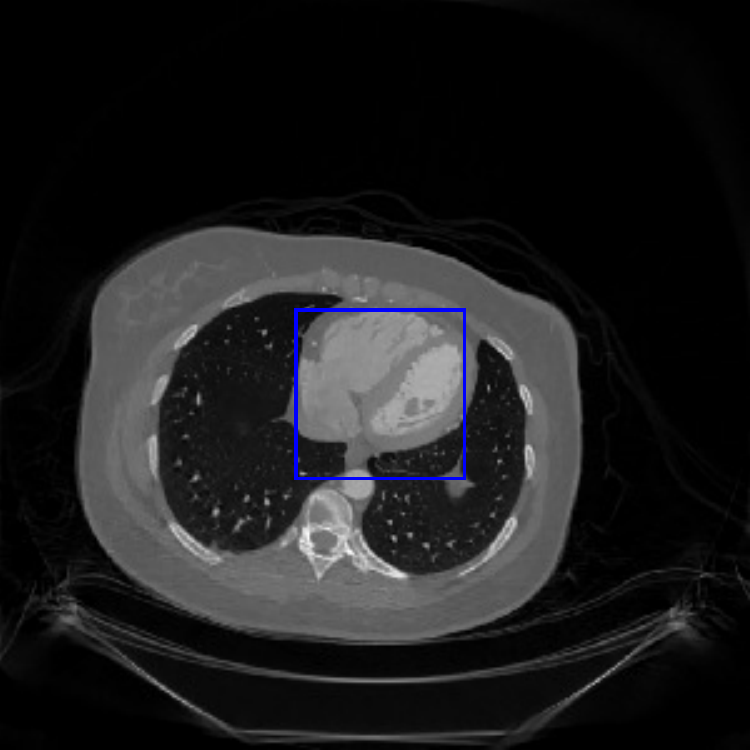} \\
  \end{tabular}
  \caption{Real data reconstruction results. The $1^{\mbox{st}}$ to $7^{\mbox{th}}$ columns correspond to FBP, HQS-CG, FBPConv-Net, PFBS-AIR, PD-Net, JSR-Net, and MetaInv-Net (H). The $1^{\mbox{st}}$ to $5^{\mbox{th}}$ rows correspond to numbers of sparse views $400, 800, 1600, 2200$, and the reference/full view images. The ROI and full view images are displayed with window center 0HU and width 2000HU. From $1^{\mbox{st}}$ to $4^{\mbox{th}}$ rows show the ROI marked by the blue rectangle in the fifth row. Red circles indicate the interested region in the zoomed in images. }
\label{fig:real-data-visual-compare}
\end{figure*}

\begin{figure}[ht]
  \centering
  \begin{tabular}{@{}p{0.5\linewidth}@{}p{0.5\linewidth}}
    \subfigimg[width=\linewidth]{}{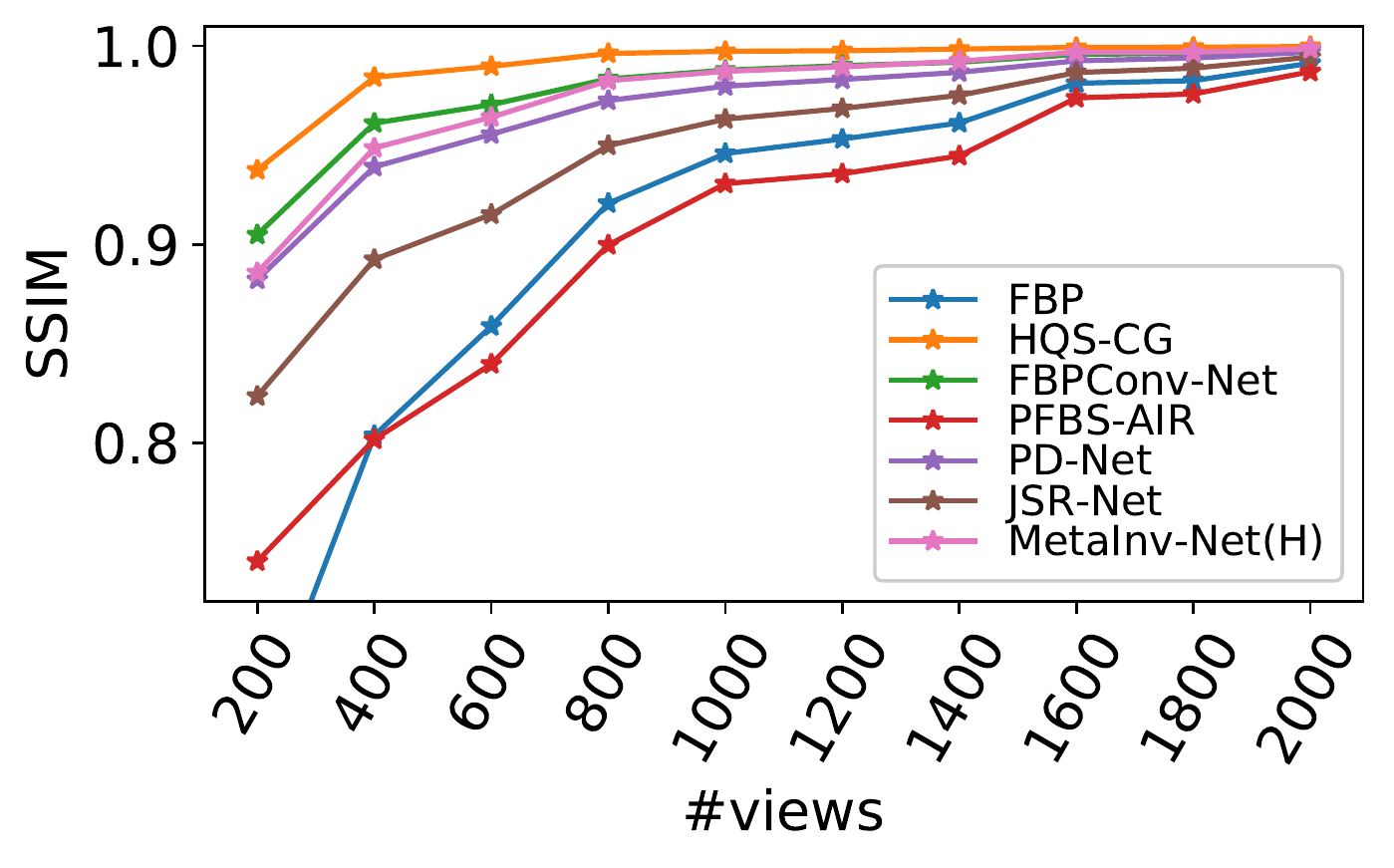} &
    \subfigimg[width=\linewidth]{}{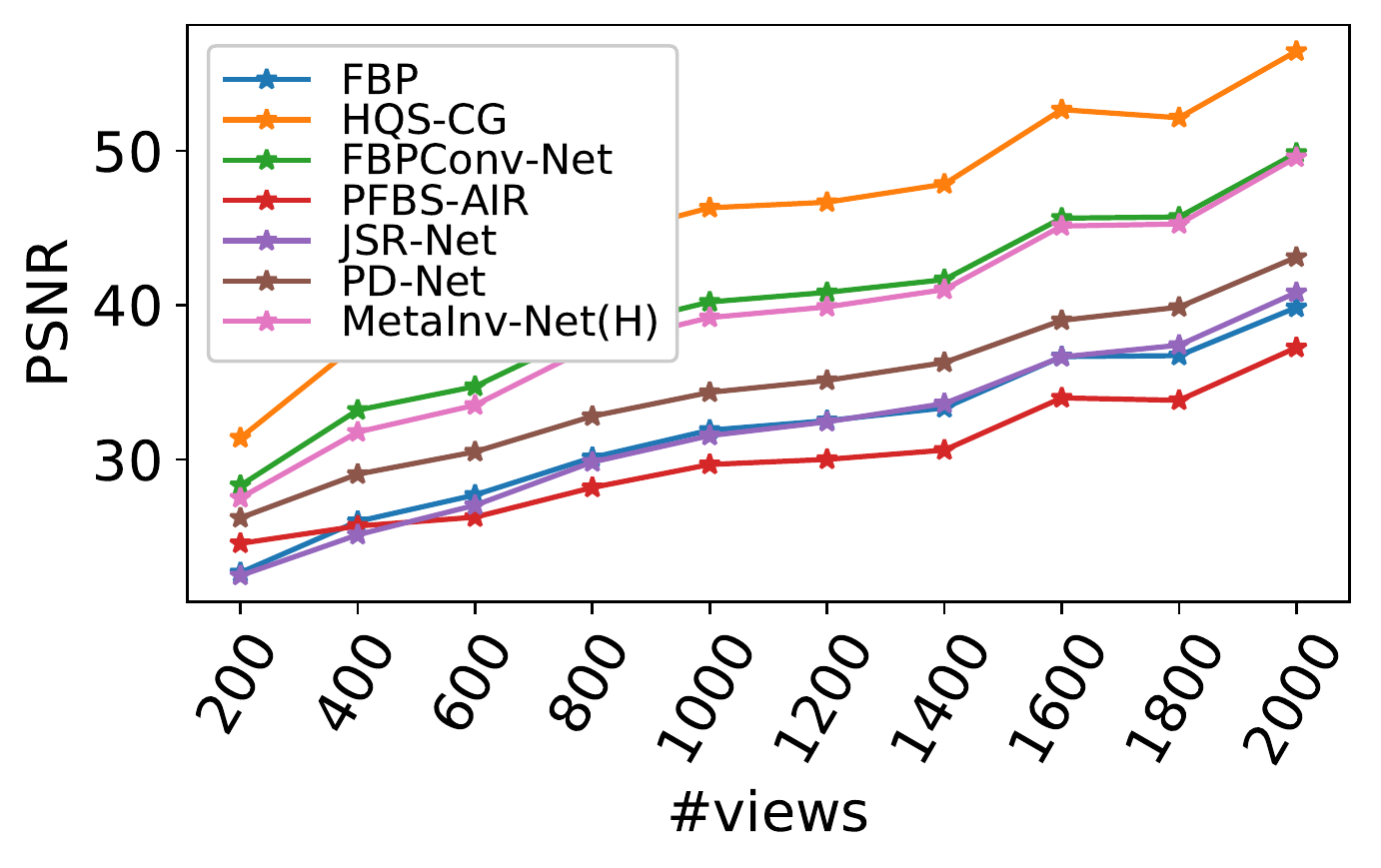} \\
  \end{tabular}
  \caption{MetaInv-Net (H) for sparse view CT image reconstruction on real data. The compared methods are FBP, HQS-CG, FBPConv-Net, PFBS-AIR, JSR-Net, and PD-Net. The left and right curve corresponds to SSIM and PSNR versus number of view (\#view), respectively.}
\label{fig:real-data-ssim-psnr-curve}
\end{figure}

\subsection{Test Setting}
We test the well-trained deep models on different scanning settings, noise levels, and data sets.

We select the rest four patients' data from AAPM-Mayo data as a test set, which comes from the same distribution of training set but not used during training. Ten slices from each patient's data are adopted with equal intervals to form our first test set, denoted by the AAPM-Test set. Another test set is five patient's Pancreas CT \cite{roth2015deeporgan} containing 50 slices (ten slices from each patient), denoted by the Pancreas-Test set. It has a different distribution to the training set. Quantitative results are measured by the multi-scale SSIM index (with level=5, Gaussian kernel size=11, and standard deviation=1.5) \cite{wang2003multiscale,Wang2004} and PSNR.

\subsection{Quantitative Measure at Different Scanning Settings}
In this experiment, we evaluate the performance of well-trained MetaInv-Net and compared models. Table \ref{table:sparse-view-CT-Poisson-AAPM} and \ref{table:sparse-view-CT-Poisson-Panc} report the quantitative results (averaged on the test set) of compared approaches on AAPM-Test and Pancreas-Test sets, respectively. These results are the generalization at different scanning settings (\#views=15, 30, 60, 120, and 180), noise levels (from low to high noise as Noise-1 (Low),  Noise-2,  Noise-3, and Noise-4 (High) corresponding to incident photon intensity $I_0=1\times 10^7, 5\times 10^6, 5\times 10^5$, and $1\times 10^5$, respectively), and test sets. Note that deep models are only trained at Noise-2 and \#views=180. The contents in the bracket indicate the standard deviation.

We observe that the MetaInv-Net (H) has better performance than MetaInv-Net (L) for all cases and is comparable to MetaInv-Net (P). Deep models are superior to HQS-CG when the test settings are similar to the training setting, i.e., \#views=60, 120, and 180. For more challenging tasks (i.e., a smaller number of scanning views and higher noise levels), MetaInv-Net (H) and MetaInv-Net (P) have better generalization performance than FBPConv-Net, PD-Net, and JSR-Net. When the number of scanning views is smaller,  MetaInv-Net (P) is comparable to the iterative algorithms HQS-CG and PWLS-CSCGR, and has better generalization performance than MetaInv-Net (H). PFBS-AIR fails to generalize to smaller numbers of scanning view cases (\#views=15, 30, 60, and 120). FBPConv-Net and MetaInv-Net (L) are even worse than FBP when the noise level is higher and \#views=15. The MetaInv-Net (P) obtains a big margin quality improvement when compared to the post-processing model, FBPConv-Net, at the smaller number of scanning view cases. When the number of scanning views is increased, the performance of MetaInv-Net becomes closer but still better than PD-Net and JSR-Net.

Even though HQS-CG is comparable or better than MetaInv-Net (H) and MetaInv-Net (P) with smaller numbers of scanning view, HQS-CG has a notably higher computation cost (requires more than 40 iterations to reach the stopping relative error $3\times 10^{-4}$) than MetaInv-Net. On the other hand, the hyperparameters of HQS-CG are fine-tuned on each test case. One can expect to obtain improved quantitative results whenever these deep models are fine-tuned on the test setting.

We note that all the settings shown in Table \ref{table:sparse-view-CT-Poisson-AAPM} and \ref{table:sparse-view-CT-Poisson-Panc} are different from the training setting for all deep learning methods. In other words, both tables present the generalization of these methods. This is why HQS-CG outperforms FBPConv-Net, PFBS-AIR, PD-Net, and JSR-Net. Under the training setting, deep learning methods outperform traditional iterative algorithms by a big margin. In contrast, the proposed MetaInv-Net generalizes as well as the none learning based methods HQS-CG and PWLS-CSCGR, which demonstrates the flexibility and robustness of MetaInv-Net. In the following experiments, we mainly focus on the MetaInv-Net (H) and compare it to other iterative algorithms and deep models.

\subsection{Qualitative Comparisons}
Fig. \ref{fig:rec-img-AAPM-Noise-2} shows the region of interest (ROI), marked by red rectangle in Fig. \ref{fig:GT-aapm-pancreas} GT1, in one slice of the reconstructed image from AAPM-Test. Another slice from Pancreas-Test, GT2 in Fig. \ref{fig:GT-aapm-pancreas}, is shown in Fig. \ref{fig:rec-img-Panc-Noise-2}. The visual quality of the proposed MetaInv-Net (H) has comparable performance as HQS-CG and PWLS-CSCGR, and it outperforms the compared deep learning models, especially with fewer scanning views. More specifically, FBP induces serious streak artifacts across the image when the number of views is small, i.e., \#views=15, 30, 60. The post-processing model FBPConv-Net produces reconstructed images with higher quality than FBP. The PFBS-AIR model fails to reconstruct visually meaningful images when generalized to smaller numbers of scanning views (i.e., \#views=15, 30, 60, and 120). The reconstructed images from PD-Net and JSR-Net at \#views=15 and 30 are severely degraded. However, MetaInv-Net (H) produces higher quality images than both PD-Net and JSR-Net at \#views=60, 120, and 180. One can notice the difference of each compared approach at the region indicated by the blue arrow in Fig. \ref{fig:rec-img-AAPM-Noise-2} and \ref{fig:rec-img-Panc-Noise-2}.

\subsection{Computation Time}
We compute the per-image time cost at the inference phase with the image size $512\times 512$ and different sinogram size. The average inference time for HQS-CG, PWLS-CSCGR, FBPConv-Net, PFBS-AIR, PD-Net, JSR-Net, MetaInv-Net (P), MetaInv-Net (L), and MetaInv-Net (H) on AAPM-Test set and Pancreas-Test set are reported at the last column of Table \ref{table:sparse-view-CT-Poisson-AAPM} and \ref{table:sparse-view-CT-Poisson-Panc}, respectively. MetaInv-Net (P), MetaInv-Net (L), and MetaInv-Net (H) require a bit more inference time than other compared deep models. The iterative algorithms HQS-CG and PWLS-CSCGR cost much more time than deep models. Note that the HQS-CG algorithm (Algorithm 1) spends more time at smaller number of views. This is mainly due to the worsened ill-posedness of the inverse problem. Due to the heavy computation burden of PWLS-CSCGR, we shall omit its test results in the remaining numerical experiments.

\subsection{More Experiments}
In this section, we conduct some ablation studies for the MetaInv-Net with different learned components. The first group of experiments are learning the hyperparameters in MetaInv-Net, i.e., $\lambda$, $\bm{\gamma}$, and $\bm{W}$. 
The second group of experiments is comparing the different designs of the denoising layer. 
All these models are trained in the same setting as the previous section. The trained models are tested on the AAPM-Test set.

\subsubsection{Learning Different Components in MetaInv-Net}
There are several variants of MetaInv-Net, which learned: 1) $\lambda$ and $\bm{\gamma}$, 2) $\bm{W}$, 3) CG-Init, $\lambda$ and $\bm{\gamma}$, 4) CG-Init and $\bm{W}$, and 5) CG-Init, $\lambda$, $\bm{\gamma}$, and $\bm{W}$, from data. The hyperparameters $\lambda$ and $\bm{\gamma}$ in MetaInv-Net can be predicted by fully connected neural network (FCN) with the input adopted from each layer's state variables. For $\lambda$ in each layer, the $\ell_{2}$-norm of highpass components of piecewise linear tight wavelet frame coefficients $\bm{W}\bm{u}$ is computed channel-wisely and used as the input of an FCN. For $\bm{\gamma}$ in each layer, the $\ell_{2}$-norm of highpass components of $\bm{W}\bm{u}-\bm{z}, \bm{W}^{\top}\bm{W}\bm{u}, \bm{W}^{\top}\bm{W}\bm{z}$, and $\bm{P}\bm{u}-\bm{Y}$ are concatenated and used as the input of a FCN. Both the predicted  $\lambda$ and $\bm{\gamma}$ have equal dimension as the number of highpass channels of the wavelet frame transform. The learning of the image representation $\bm{W}$ is implemented by adopting the standard error backpropagation during the training phase.

Fig. \ref{fig:learn-diff-hyper-parameters-metainv-net} presents the curves of SSIM/PSNR versus different Poisson noise level from each variant of MetaInv-Net (H). The learnable components in the unrolled HQS-CG include the initialization of CG algorithm denoted as ``CG-Init'' (which is the version of MetaInv-Net (H) that we proposed), ``$\lambda$, $\bm{\gamma}$'', and ``$\bm{W}$'' and a combination of them. Learning adaptive hyperparameters $\lambda$ and $\bm{\gamma}$ in MetaInv-Net (H) brings little benefit when the number of scanning view is larger. When the number of scanning views is 15, learning $\bm{W}$ provides more benefits than the adaptive hyperparameters $\lambda$ and $\bm{\gamma}$. Learning CG-Init along with $\lambda$, $\bm{\gamma}$, $\bm{W}$ does not bring much benefit, while it inevitably introduces more trainable parameters and may bring difficulty during training. Therefore, learning only the initialization of CG while fixing other components is overall the best option.

\subsubsection{Learning Different Denoiser}
In this experiment, we investigate the design of the image denoiser module in MetaInv-Net (i.e. the $\bm{z}$-subproblem). Three unrolled models are created by adopting the same U-Net based CG-Init module and different denoiser modules. This includes 1) Soft-thresholding (ST) with tight wavelet frame transform  (W) of $\bm{u}$-layer reconstruction as input (denoted as CG-Init+ST); 2) DnCNN based denoiser \cite{zhang2017learning} with tight wavelet frame coefficients of $\bm{u}$-layer reconstruction as multi-channel input (denoted as CG-Init+DnCNN-M); and 3) DnCNN based denoiser with $\bm{u}$-layer reconstruction as single-channel input (denoted as CG-Init+DnCNN-S). Note that the CG-Init+ST model is our proposed model, ``MetaInv-Net (H)", which is not retrained in this experiment. Both CG-Init+DnCNN-M and CG-Init+DnCNN-S are trained with the same setting as MetaInv-Net (H) described in the former section. The well trained deep models are evaluated at different Poisson noise levels, i.e., the incident photon intensity $I_{0}=1\times 10^{5}, 5\times 10^{5}, 1 \times 10^{6}, 5 \times 10^{6}, 1 \times 10^{7}, \mbox{and}\; 5\times 10^{7}$, on AAPM-Test set. The tested numbers of scanning view are $30, 60, \mbox{and}\; 120$. Fig. \ref{fig:diff-denoiser-metainv-net} shows the SSIM/PSNR versus Noise level at different scanning view cases.

The proposed MetaInv-Net (H) has a clear advantage when number of views is relatively large while noise level is higher. We observe that the CG-Init+DnCNN-M model performs better than CG-Init+DnCNN-S and the CG-Init+ST when the number of views is small, i.e., \#views=30. However, this quality improvement leads to more trainable parameters in the MetaInv-Net and the difficulty of training the deep model.

\subsection{Different Training Set Size}
In this experiment, we retrain the deep models (FBPConv-Net, PD-Net, JSR-Net, and MetaInv-Net (H)) with the same training setting (except training epochs=20) at different data set sizes (e.g., $1\%, 5\%, 10\%, 20\%, 30\%, 40\%, \mbox{and}\; 50\%$  of the full training set). These smaller training sets are randomly chosen from the full training set and then fixed for all the compared deep models. The training and testing noise level are Noise-2. Since the PFBS-AIR model can not be generalized to a different number of scanning views cases than the training setting, we omit its results in this experiment. Fig. \ref{fig:diff-train-set-size-aapm-test-set} shows the SSIM and PSNR evaluated on the AAPM-Test set.  We observe that 1) When more training data are utilized, the compared deep models have comparable performance to MetaInv-Net (H). 2) The post-processing approach FBPConv-Net has the worst performance among the compared deep models. 3) MetaInv-Net (H) is much less demanding in training data than FBPConv-Net, PD-Net, and JSR-Net.

\subsection{Real Data Study}

In electrocardiographic (ECG)-gated cardiac multiphase 4D imaging, the reconstruction of each phase is a sparse view CT imaging problem. The current setting is a realistic simulation with ground truth. In this section, the proposed MetaInv-Net is tested on real data, which is scanned with the same protocol to \cite{hengyongyu-2011}. The original helical geometry is converted to a fan-beam geometry. There are a total of 2200 scanning views equally distributed $360^\circ$. The number of detector elements is 888, and its length is $95.93\; \mbox{cm}$. The reconstructed image has $800\times 800$ pixels around $24.3\times 24.3\; \mbox{cm}^2$. The distance of source to rotation center is $ 53.85\;\mbox{cm}$ and detector to rotation center is $49.83\;\mbox{cm}$.

Each compared method adopts the complete measured data to reconstruct its reference (Full View in Fig. \ref{fig:real-data-visual-compare}). The sinogram is downsampled to $400, 800$, and $1600$ views to simulate the sparse view CT imaging. Fig. \ref{fig:real-data-visual-compare} shows the reconstructed images at different sparse view settings.
When the number of sparse views is smaller, MetaInv-Net (H) produces an image with better visual quality than FBP, PFBS-AIR, PD-Net, and JSR-Net. HQS-CG and FBPConv-Net produce the image with much fewer artifacts than MetaInv-Net (H). However, the details and edges in HQS-CG and FBPConv-Net reconstruction are smoother than MetaInv-Net (H). JSR-Net and PD-Net have a contrast loss problem on this real data test.

To compare the quality perturbation at different sparse views, we downsample the complete data to $200, 400, ..., 2000$-views. The quantitative measure of each approach is computed with respect to its reference (bottom row in Fig. \ref{fig:real-data-visual-compare}). Fig.\ref{fig:real-data-ssim-psnr-curve} shows the SSIM/PSNR versus number of sparse views. For SSIM, HQS-CG only needs $400$ views to obtain stable value, whereas FBPConv-Net, PD-Net, and MetaInv-Net (H) require at least $800$ views. JSR-Net is inferior to PD-Net when the number of scanning views is smaller. PFBS-AIR fails to reconstruct an image from sparse view data due to its worse than the initialization FBP. For PSNR, all the compared methods gain quality improvement along with the increasing number of available scanning views. MetaInv-Net (H) and FBPConv-Net have comparable performance while are better than other deep models and FBP. However, all the deep models are inferior to HQS-CG in terms of PSNR. Note that the quantitative comparison here is not exactly fair and only serves as a reference due to the lack of ground truth.

\section{Conclusion}\label{sec:conclusion}
In this work, we propose a new deep model, MetaInv-Net, for sparse view CT. Two variants contain much less (MetaInv-Net (L)) and much more (MetaInv-Net (H)) trainable parameters than some existing deep models, while all these deep models reconstruct images with higher quality at various test settings. We observed that most of the backbone model (HQS-CG) components can stay unchanged and we only need to use properly designed CNN to generate good initialization for the CG algorithm module in the HQS-CG algorithm. Extensive numerical experiments on both synthetic and real data show that MetaInv-Net (both light and heavy version) generalizes well beyond the training setting, e.g., to different scanning settings (number of views), noise levels, and data set. As part of the future works, we will explore a better strategy to select hyperparameters and design more powerful data-adaptive image priors in the MetaInv-Net. Since the modifications of the MetaInv-Net to the backbone model is minor, convergence analysis of the MetaInv-Net is also interesting and meaningful.

\bibliographystyle{IEEEtran}

\end{document}